\documentclass[a4paper,11pt]{article}
\usepackage{jheppub} 
\usepackage{lineno}

\usepackage{cleveref}
\usepackage{lipsum}
\usepackage{bm}
\usepackage{physics}
\usepackage{subcaption}
\usepackage{dsfont}
\usepackage{comment}
\usepackage{tikz-feynman}
\usepackage{tikz}
\usetikzlibrary{positioning,calc}
\usetikzlibrary{arrows.meta}

\newcommand{\Y}{\mathcal{Y}}
\newcommand{\Ss}{\mathcal{S}}

\newcommand{\daga}{{}^\dagger}
\newcommand{\eps}{\epsilon}
\newcommand{\Sym}{\text{Sym}}

\title{Hierarchies from Higher Flavor Spin}

\author[]{Admir Greljo}
\author[]{and Alessandro Valenti}
\affiliation[]{Department of Physics, University of Basel\\
Klingelbergstrasse 82, CH-4056 Basel, 
Switzerland}

\emailAdd{admir.greljo@unibas.ch}
\emailAdd{alessandro.valenti@unibas.ch}

\abstract{
We develop a framework in which Yukawa hierarchies arise from powers of fully anarchic spurions transforming in higher representations of the flavor symmetry group $SU(2)^{n_2}\times SU(3)^{n_3}$. The core mechanism is the progressive lifting of Yukawa ranks through successive outer products of composite doublets and triplets. We formulate the general construction in detail and build explicit models realizing it. We then investigate whether renormalizable scalar potentials for higher $SU(2)$ representations can dynamically generate anarchic spurions with non-vanishing composites. The framework predicts distinctive patterns in flavor-changing neutral currents and potentially observable stochastic gravitational-wave backgrounds.
}

\begin{document}
\maketitle

\section{Introduction}
\label{sec:intro}

The origin of flavor is one of the central open questions in particle physics. In the Standard Model (SM), fermion masses and mixings arise from Yukawa interactions with a single Higgs field, $(Y_{f})_{ij} \bar f_{L,i} H f_{R,j}$.  Despite entering the theory in the same way, the observed structure is highly non-generic yet similar across different fermion species. In the absence of an organizing principle, as in the SM, one would naively expect Yukawa matrices $Y_f$ with singular values of comparable size and the large CKM mixing from the product of a priori unrelated $\mathcal{O}(1)$ rotation matrices. Instead, quark and charged lepton spectra are strongly hierarchical, with masses separated by one to two orders of magnitude between successive generations, while the CKM matrix is nearly a unit matrix with hierarchically suppressed off-diagonal elements. These hierarchies constitute \textit{the SM flavor puzzle} and point to new short-distance dynamics beyond the SM yet to be discovered. The neutrino sector further deepens the puzzle, exhibiting a qualitatively distinct pattern: the PMNS matrix features large mixing angles with no evident hierarchy.

Flavor model building is a mature yet evolving field devoted to addressing this puzzle by constructing extensions of the SM that reproduce the observed Yukawa hierarchies and yield testable implications for flavor observables; see reviews~\cite{Altmannshofer:2024hmr, Feruglio:2015jfa}. A prototypical mechanism is the Froggatt–Nielsen (FN) construction~\cite{Froggatt:1978nt, Leurer:1992wg, Leurer:1993gy}, based on a \emph{single} Abelian $U(1)$ flavor symmetry under which SM fermions carry generation-dependent charges, and a \emph{single}, small symmetry breaking spurion $\epsilon$ carrying a unit $U(1)$ charge. Treating the spurion as a background field and enforcing symmetry covariance, the SM Yukawa couplings acquire a hierarchical structure governed by positive powers of the spurion, $(Y_f)_{ij} \sim \epsilon^{|q_{f_{L,i}}-q_{f_{R,j}}|}$. In a UV completion, the spurion $\epsilon \equiv \langle \phi \rangle / M$ is identified with the ratio of the flavon vacuum expectation value (vev) $\langle \phi \rangle$ to the common mass scale $M$ of heavy vector-like fermions (VLF), whose dimensionless couplings are taken to be $\mathcal{O}(1)$.\footnote{In constructions with VLF chains, the scale $\epsilon M$ controls the symmetry-breaking mass mixings along the chain. The associated heavy dynamics is generally required by flavor-changing neutral current constraints to lie well above the TeV scale, although in special symmetry-protected scenarios it may be $\mathcal{O}($TeV)~\cite{Arkani-Hamed:2026wwy, Greljo:2025mwj}.}

The choice of $U(1)$ charges is highly non-unique and involves a trade-off~\cite{Fedele:2020fvh, Cornella:2023zme, Greljo:2024evt}. Restricting to integer charges, assignments with a smaller charge range have limited resolving power, typically requiring $\epsilon \lesssim 10^{-2}$ and yielding a poorer quantitative description of the observed hierarchies, but admit short VLF chains with fewer new states. In contrast, larger maximal charges allow for a more accurate description with $\epsilon \gtrsim 0.1$, at the cost of longer VLF chains. In this regime, the quality of the fit becomes sensitive to the statistical properties of the underlying $\mathcal{O}(1)$ couplings, notably their variance, whose impact is amplified along longer chains.

Non-Abelian flavor symmetries, such as $SU(3)$ or $SU(2)$, offer a more constrained starting point: fermions are assigned to fixed representations, for example, triplets in $SU(3)$ or doublet plus singlet in $SU(2)$. This reduces the arbitrariness present in Abelian constructions. The remaining freedom shifts to the symmetry-breaking sector, which typically requires multiple spurions, with possible hierarchies among them or within their components. Even in minimally broken $U(3)^5$ (MFV)~\cite{DAmbrosio:2002vsn} or $U(2)^5$~\cite{Barbieri:2011ci}, where spurions are bifundamental representations, it is well known that reproducing the observed hierarchies within a given spurion from a flavon potential dynamically is challenging~\cite{Alonso:2011yg, Alonso:2013nca}.

An interesting proposal was recently put forth in Ref.~\cite{Banks:2025baf}, in which a {single} spurion $\mathcal{Y}$ is introduced in a higher-dimensional (non-minimal) irreducible representation of $SU(3)_{f_L}\times SU(3)_{f_R}$, specifically $\mathcal{Y} \sim (\bm{6}_{f_L}, \bar{\bm{6}}_{f_R})$. The effective Yukawa, $(\bm{ 3}_{f_L}, \bar{\bm{3}} _{f_R})$, is then generated through tensor decompositions of powers of $\Y$. In this construction, for a carefully chosen set of non-zero entries in $\Y$ assumed to emerge from some UV dynamics, the leading contribution arises at order $\mathcal{Y}^3$, followed by $\mathcal{Y}^4$ and $\mathcal{Y}^5$, with each successive term \emph{increasing the rank} of the effective Yukawa matrices by one unit, and therefore explaining the fermion mass hierarchies despite the absence of hierarchies in the non-zero elements of $\Y$. The mechanism crucially depends on the assumption of the specific choice of zero and non-zero elements in $\Y$ in a given flavor basis.

While conceptually appealing, this setup struggles to accommodate the top quark Yukawa, which is generated at order $\mathcal{Y}^3$ and is therefore parametrically suppressed, in tension with the observed $y_t \sim 1$. Furthermore, the dynamics generating the assumed textures of zeros in $\Y$ remains unspecified, and cannot be realized through a renormalizable potential of a single field. In this work, we explore higher-representation breaking of $U(2)$ flavor symmetries~\cite{Barbieri:1995uv, Barbieri:1996ww, Barbieri:1997tu, Barbieri:2011ci, Linster:2018avp, Greljo:2023bix, Antusch:2023shi, Greljo:2024zrj, Greljo:2025mwj}, which organize some or all SM fermion representations into a doublet, associated with the light generations, and a singlet, associated with the third generation. The most general framework can then be written as $SU(2)^{n_2} \times SU(3)^{n_3} \times G_{\rm Abelian}$, with $1 \leq n_2 + n_3 \leq 5$, depending on how many SM fermion species are charged under independent $SU(2)$ and $SU(3)$ factors. Representative examples with $n_2=1$ and $n_3=0$ include $U(2)_{\rm all}$~\cite{Barbieri:1995uv,Barbieri:1996ww, Barbieri:1997tu}, $U(2)_{\psi_L=q+\ell}$~\cite{Greljo:2023bix, Greljo:2024zrj}, $U(2)_{q+e^{(c)}}$ and $U(2)_{\bm{10}} \times Z_2$~\cite{Antusch:2023shi}, as well as constructions such as ${\rm MFP}=SU(2)_q \times U(1)_X$~\cite{Greljo:2025mwj} and $SU(2)_{\rm all} \times U(1)_F$~\cite{Linster:2018avp}. At the other extreme, the fully factorized case $U(2)^5$~\cite{Barbieri:2011ci} assigns an independent $U(2)$ symmetry to each SM fermion sector.

Within $U(2)$, a rank-1 Yukawa structure is already present at zeroth order, allowing the top quark Yukawa to be unsuppressed, while the lighter generations arise from higher powers of a single spurion $\Ss$ with higher \emph{flavor spin} $j_\Ss \ge 3/2 $, half-integer. As shown in \cref{sec:framework}, with a single $U(2)$ acting on a definite chirality, the second- and first-generation Yukawas are generated at orders $\epsilon^3$ and $\epsilon^5$ where $\epsilon \sim |\Ss|/M$, respectively, and independently of the choice of $j_\Ss$. Importantly, all entries of the spurion $\Ss$ are assumed to be non-zero and of the same order. The hierarchies arise from the successive appearance of independent composite doublet representations at different orders. The Yukawa matrices are then built progressively as sums of \emph{outer products}, with each new composite lifting the rank at a parametrically higher order. The resulting structure of Yukawa eigenvalues is therefore
\begin{equation}\label{eq:u2hierarchy}
U(2)\text{ + single higher-spin $j_\Ss \ge 3/2$ breaking}: \quad \{y_{1},\, y_{2},\, y_{3}\} \sim \{\epsilon^5,\, \epsilon^3,\, 1\}\,.
\end{equation}
The novelty of our mechanism with respect to~\cite{Banks:2025baf} is precisely that it only relies on the properties of the tensor decomposition of successive spurion powers, and not on the assumed zero texture within the spurion.

Thus, within $U(2)$ flavor symmetries, the tensor products of a {single} non-minimal representation $\Ss$ serve to lift the rank of the Yukawa matrices within the light-family sector. This complements the standard $U(2)$ mechanism, which distinguishes the third generation from the light ones already in the exact symmetry limit. The two mechanisms therefore operate in a coherent and mutually reinforcing way. As in FN models, the expansion is controlled by a small parameter with no internal hierarchy. Unlike in Abelian constructions, however, the hierarchies are not imposed through arbitrary charge assignments, but follow from the rigid tensor product structure of the symmetry. A similar mechanism can also arise in $SU(3)$, as we show later, with the difference that the third-generation Yukawa is then generated by the higher-irrep spurion as well.

Let us contrast this with existing implementations of $U(2)$ breaking~\cite{Barbieri:1995uv, Barbieri:1996ww, Barbieri:1997tu, Barbieri:2011ci, Linster:2018avp, Greljo:2023bix, Antusch:2023shi, Greljo:2024zrj, Greljo:2025mwj}, which generate the flavor structure of the light families. These typically rely on two or more spurions in minimal representations (doublets, bidoublets, singlets, or triplets) to split the first and second generations. In many cases, this requires introducing additional hierarchies either among different spurions or within the entries of a given spurion, as in the bifundamentals $\Delta_{u,d,e}$ of~\cite{Barbieri:2011ci}.\footnote{Notable exceptions include~\cite{Greljo:2023bix}, where the $1$--$2$ hierarchy is generated radiatively, and~\cite{Greljo:2025mwj}, where multiple spurions are present but remain of comparable size.} By contrast, employing a single higher flavor spin spurion $\Ss$ leads to a more constrained and predictive pattern of flavor symmetry breaking.

The paper is organized as follows. In \cref{sec:framework} we introduce the framework in full generality. Then, in \cref{sec:models}, we present its concrete implementation in explicit models, including several realistic constructions with different symmetry structures and vector-like chain completions. In \cref{sec:pot}, we analyze the scalar potential of the higher-representation field required for the mechanism to operate. In \cref{sec:pheno}, we discuss the phenomenological implications, comparing the resulting flavor selection rules with those of established benchmarks. We summarize our results and conclude in \cref{sec:concl}. Technical details of the $SU(2)$ representation theory and of the potential minimization are collected in the appendices.

\section{Framework}
\label{sec:framework}

We assume a flavor symmetry of the form
\begin{align}
    G_\text{flavor}
    =
    SU(2)^{n_2}
    \times
    SU(3)^{n_3}
    \times
    G_\text{Abelian}.
    \label{eq:Gflavor}
\end{align}
under which the SM fermions are charged. The integers $n_2$ and $n_3$ satisfy $n_2+n_3\leq 5$, corresponding to the five SM chiral fermion species $q,u,d,\ell,e$, while $G_\text{Abelian}$ is bounded by the maximal number of independent Abelian charge assignments.

For each SM chiral multiplet $f_i=q_i,u_i,d_i,\ell_i,e_i$, we allow the three generations to transform, with respect to the non-Abelian flavor factors, either as a triplet, as a doublet plus singlet, or as three singlets:
\begin{align}
    f_i \longrightarrow \bm{3}_{2,3}
    \qquad \text{or} \qquad
    f_i \longrightarrow \bm{2}_2 \oplus \bm{1}
    \qquad \text{or} \qquad
    f_i \longrightarrow \bm{1}\oplus\bm{1}\oplus\bm{1}.
\end{align}
Here, the triplet may be either of $SU(2)$ or $SU(3)$, as indicated by the subscript, while doublets arise only for $SU(2)$. Each of these multiplets may, in addition, carry charges under the Abelian factors.

With these assignments, the spurion structure of the Yukawa matrices is strongly constrained by representation theory. Let us first illustrate this in the up quark sector. As discussed in \cref{sec:intro}, the fact that the top Yukawa $y_t \approx 1 $ strongly suggests that $q$ and $u$ should transform at most as $SU(2)$ doublets plus singlets, since an $SU(3)$ assignment would generically require a spurion expansion even for the top Yukawa. One is then led to three qualitatively distinct possibilities:
\newcommand{\SsuBigCell}{\begin{array}{c}
\bm{1}_2\oplus\bm{3}_2\\
\text{or }(\bm{2}_2^L,\bm{2}^R _2)
\end{array}}
\begin{align}
\begin{array}{c@{\qquad}c@{\qquad}c@{\qquad}c}
\raisebox{-3.75em}{$Y_u:$}
&
\begin{array}{c}
\\
SU(2)_L \\[0.2cm]
q \sim \bm{2}_2 \oplus \bm{1}\\[0.15cm]
u \sim \bm{1} \oplus \bm{1} \oplus \bm{1}\\[0.35cm]
\left(
\begin{array}{@{\hskip 0.35em}c|c@{\hskip 0.35em}|c@{\hskip 0.35em}}
\bm{2}_2 & \bm{2}_2 & \bm{2}_2 \\
\hline
\bm{1} & \bm{1}& \bm{1}
\end{array}
\right)
\vphantom{\left(
\begin{array}{c|c}
\SsuBigCell & \bm{2}_2 \\
\hline
\bm{2}_2 & \bm{1}
\end{array}
\right)}
\end{array}
&
\begin{array}{c}
SU(2)_{L+R} \text{ or }\\
SU(2)_L \times SU(2)_R \\[0.2cm]
q \sim \bm{2}_2 \oplus \bm{1}\\[0.15cm]
u \sim \bm{2}_2 \oplus \bm{1}\\[0.35cm]
\left(
\begin{array}{c|c}
\SsuBigCell & \bm{2}_2 \\
\hline
\bm{2}_2 & \bm{1}
\end{array}
\right)
\end{array}
&
\begin{array}{c}
\\
SU(2)_R 
\\[0.2cm]
q \sim \bm{1} \oplus \bm{1} \oplus \bm{1}\\[0.15cm]
u \sim \bm{2}_2 \oplus \bm{1}\\[0.35cm]
\left(
\begin{array}{@{\hskip 0.35em}c|c@{\hskip 0.35em}}
\bm{2}_2 & \bm{1} \\
\hline
\bm{2}_2 & \bm{1} \\
\hline
\bm{2}_2 & \bm{1}
\end{array}
\right)
\end{array}
\end{array}
\label{eq:YfStructure}
\end{align}
where we do not display the purely $U(1)$ Froggatt-Nielsen possibility in which all fields are singlets of the non-Abelian factors. We stress that here and in the following $L$ and $R$ label the flavor group, orthogonal to the SM gauge group $SU(3)_c \times SU(2)_{\rm L} \times U(1)_{\rm Y}$.

These structures can be understood quite easily. The Yukawa is built from lined up or stacked doublets when only one chirality transforms as an $SU(2)$ doublet. If instead both chiralities are doublets, two possibilities arise. If the same $SU(2)$ is shared between $q$ and $u$, namely $SU(2)_{L+R}$, then $\bar q  u \sim \bm{1}_2\oplus\bm{3}_2$, so the corresponding spurions in the light $2\times2$ block must transform as singlets or triplets. If the left- and right-handed sectors transform under different $SU(2)$ factors, as in $U(2)^5$, the same block is instead described by a bifundamental spurion. The off-diagonal $1\times 2$ and $2\times 1$ blocks connect the light generations to the third one, and therefore transform as doublets whenever the corresponding light generations form an $SU(2)$ doublet, and as singlets otherwise. Finally, the bottom-right entry is always a singlet.

Since $y_{b,\tau}\sim 10^{-2}$, the down quarks and leptons have more possibilities, since $d$, $\ell$, and $e$ may also transform as triplets. In the down quark sector, in addition to the structures in \cref{eq:YfStructure}, one can also have
\newcommand{\SsdBigCell}{\begin{array}{c}
\bm{2}_2 \oplus \bm{4}_2\\
\text{or }(\bm{2}_2 ^L,\bar{\bm{3}}_{2,3} ^R)
\\[0.1cm]
\end{array}}
\newcommand{\SsdBigMatrix}{\left(
\begin{array}{c}
\SsdBigCell\\
\hline\\[-0.5cm]
\bar{\bm{3}}_{2,3}
\end{array}
\right)}
\begin{align}
\begin{array}{c@{\qquad}c@{\qquad\qquad}c}
\raisebox{-3.75em}{$Y_d:$}
&
\begin{array}{c}
SU(2)_{L+R} \text{ or}\\
SU(2)_L \times SU(2 \text { or } 3)_R\\[0.2cm]
q \sim \bm{2}_2 \oplus \bm{1}\\[0.15cm]
d \sim \bm{3}_{2,3}\\[0.35cm]
\SsdBigMatrix
\end{array}
&
\begin{array}{c}
\\
SU(2 \text { or } 3)_R\\[0.2cm]
q \sim \bm{1}\oplus\bm{1}\oplus\bm{1}\\[0.15cm]
d \sim \bm{3}_{2,3}\\[0.35cm]
\left(
\begin{array}{@{\hskip 0.35em}c@{\hskip 0.35em}}
\bar{\bm{3}}_{2,3}\\
\hline
\\[-0.5cm]
\bar{\bm{3}}_{2,3}\\
\hline
\\[-0.5cm]
\bar{\bm{3}}_{2,3}
\end{array}
\right)
\end{array}
\end{array}
\label{eq:YdStructure}
\end{align}
where the $\bm{2}_2\oplus\bm{4}_2$ option arises for $SU(2)_{L+R}$, while in the remaining cases the left and right flavor structures factorize.
In the lepton sector one may also have both $\ell$ and $e$ transforming as triplets. Hence, in addition to the possibilities in \cref{eq:YfStructure}, \cref{eq:YdStructure}, and their transposes, this gives rise also to the pattern
\begin{align}
\begin{array}{c@{\quad}c}
\raisebox{-3.25em}{$Y_e:$}
&
\begin{array}{c}
SU(2 \text{ or } 3 )_{L+R} \text{ or}\\
SU(2 \text{ or } 3 )_{L} \times SU(2 \text{ or } 3 )_{R}\\[0.2cm]
\begin{aligned}
\ell &\sim \bm{3}_{2,3}\\
e &\sim \bm{3}_{2,3}
\end{aligned}
\\[0.5cm]
\left(
\begin{array}{c}
\begin{array}{c}
\bm{1}_2 \oplus \bm{3}_2 \oplus \bm{5}_2\\
\text{or } \bm{1}_3 \oplus \bm{8}_3\\
\text{or } (\bm{3}_{2,3} ^L,\bar{\bm{3}}_{2,3} ^R)
\end{array}
\end{array}
\right).
\end{array}
\end{array}
\label{eq:YeStructure}
\end{align}
It follows that the spurions entering $Y_u$, $Y_d$, and $Y_e$ can only transform in a limited set of representations, reflecting the limited number of consistent flavor assignments of the SM fields. In particular, they can be at most $SU(2)$ quintuplets or $SU(3)$ octets. If all such spurions are treated as independent, one is led to a proliferation of flavor-breaking parameters with hierarchical sizes. Our aim is instead to generate these effective spurions from a smaller set of elementary spurions transforming in higher representations of the flavor group. These are assumed to be completely anarchic, with no zero textures and with entries of comparable size. The effective spurions then arise as composite objects, with the observed flavor hierarchies generated by the non-Abelian selection rules themselves.

\paragraph{Composite spurions.} 
Let us illustrate this mechanism explicitly in the $SU(2)$ case. As discussed above, $SU(2)$ plays a special role, since it is the largest non-Abelian flavor symmetry that can still accommodate an unsuppressed top Yukawa, while additional $SU(3)$ factors, if present, can only act on a reduced set of SM fermions. We then consider a single elementary spurion, which we denote by $\Ss$, with unit mass dimension, transforming in a higher-spin representation of $SU(2)$:
\begin{align}
    \Ss \sim (\bm{2j_\Ss+1}), \qquad j_\Ss > 1/2.
\end{align}
If $j_\Ss$ were integer, products of $\Ss$ could never generate half-integer representations, and so never generate doublets. We therefore take $j_\Ss$ to be half-integer. We assume the entries of $\Ss$ to be \emph{completely generic}, meaning with comparable sizes and without accidental symmetries. The effective spurions that can then be built from $\Ss$ are simply all those allowed by $SU(2)$ representation theory, which schematically take the form
\begin{align}
    \begin{aligned}
        \bm{1}_2:& \quad  (\Ss \bar \Ss)_0, \quad (\Ss^4)_{0,i}, (\Ss^3 \bar \Ss)_{0,i}, (\Ss^2 \bar \Ss^2)_{0,i}, \dots \\
        \bm{2}_2:& \quad (\Ss^2 \bar \Ss)_{1/2}, \quad (\Ss^5)_{1/2,i}, (\Ss^4 \bar \Ss)_{1/2,i}, (\Ss^3 \bar \Ss^2)_{1/2,i}, \dots \\
        \bm{3}_2:& \quad (\Ss^2)_1, (\Ss \bar \Ss)_{1}, \quad (\Ss^3 \bar \Ss)_{1}, (\Ss^2 \bar \Ss ^2)_{1}, \dots
    \end{aligned}
    \label{eq:Ydecomp}
\end{align}
where $\bar\Ss$ denotes the $SU(2)$ conjugate of $\Ss$,\footnote{$(\bar \Ss)_m = (-1)^{j_\Ss-m} \Ss_{-m}^*$, with $m$ denoting the $SU(2)$ magnetic quantum number.} and we do not display separately the conjugate composite representations. The subscript denotes the flavor spin of the composite representation, with an index $i$ denoting the fact that the tensor product decomposition includes multiple independent representations (see \cref{app:countingIrreps}).

From the EFT point of view, this is
an intriguing situation. If the effective theory is organized in powers of a heavy scale $M$, and we define
\begin{align}
    \eps \equiv \frac{|\Ss|}{M},
\end{align}
then the leading singlets and triplets scale as $\mathcal{O}(\eps^2)$, while the leading doublets scale as $\mathcal{O}(\eps^3)$. Higher-order composites are further suppressed by additional powers of $\eps^2$. This immediately points toward a dynamical origin for the observed pattern of SM masses and mixings: the elementary spurion itself carries no finite internal hierarchy, while the flavor hierarchies arise from the non-Abelian tensor decomposition. The idea is analogous in spirit to the mechanism proposed in \cite{Banks:2025baf} in the context of $SU(3)^5$, and can be viewed as a non-Abelian analogue of the Froggatt-Nielsen mechanism. It differs from more standard $U(2)$ spurion constructions, such as the $\Delta_{u,d,e}$ of $U(2)^5$, where non-trivial hierarchies are already encoded in the entries of the effective spurions.
Of course, when $\Ss$ is realized as a scalar field, the genericity of its vev is not automatically guaranteed dynamically. We discuss in detail in \cref{sec:pot} the minimization of the tree-level renormalizable scalar potential for an elementary $\Ss$.

\paragraph{Outer products.} A crucial observation is that the bosonic nature of $\Ss$ strongly constrains the representations that can appear. Let us consider the important example of doublets. The would-be doublet $(\Ss^3)_{1/2}$ vanishes identically, while $(\Ss^2\bar\Ss)_{1/2}$ contains \emph{exactly one} independent spin-$1/2$ representation, as proven in \cref{app:countingIrreps}.
This is highly suggestive for model building. Suppose, for instance, that the $\bm{2}_2$, $\bm{1}_2\oplus\bm{3}_2$, or $(\bm{2}_2,\bm{2}_2)$ structures appearing in the $2\times 2$ light block of the Yukawas in \cref{eq:YfStructure} are built out of the leading composites above (two distinct ones for the bifundamental). If the conjugate structure is forbidden, for instance by an additional $U(1)$ symmetry, the uniqueness of the leading order (LO) composite immediately enforces a rank-1 structure.
To show this explicitly, let us focus on the case in which only one chirality, say the left one, transforms as a doublet under the relevant $SU(2)_L$ factor. The light block then necessarily factorizes as an \emph{outer product}:
\begin{align}
    (Y_f^{2\times 2})_{\alpha i}^{\rm LO}
    \sim
    \eps^3\, v_\alpha w_i,
    \qquad
    \Longrightarrow
    \qquad
    \rank(Y_f^{2\times 2})^{\rm LO}=1,
\end{align}
where $v_\alpha$ is the normalized composite doublet and $w_i$ is a vector of $\mathcal{O}(1)$ coefficients. By a flavor rotation one can always choose a basis in which $v_\alpha\sim(0,1)$, making it manifest that only one light generation receives a mass at this order.

At next-to-leading order (NLO), new spin-$1/2$ composites appear, for example from $(\Ss^3\bar\Ss^2)_{1/2}$, with an additional suppression $\mathcal{O}(\eps^2)$ relative to the leading doublet. These give new vectors $v'_\alpha$, so that in an EFT expansion one has effectively a sum of outer products,
\begin{align}
    (Y_f^{2\times 2})_{\alpha i}  ^{\rm NLO}
    \sim
    \eps^3\, v_\alpha w_i
    +
    \eps^5\, v'_\alpha w'_i
    +\dots  \qquad
    \Longrightarrow
    \qquad
    \rank(Y_f^{2\times 2})^{\rm NLO}=2.
\end{align}
For generic $v', w'$ misaligned from $v,w$, this lifts the rank. The light block, in the aforementioned basis, then has the schematic form
\begin{align}
    Y_f^{2\times 2}
    \sim
    \begin{pmatrix}
        \eps^5 & \eps^5\\
        \eps^3 & \eps^3
    \end{pmatrix},
\end{align}
up to $\mathcal{O}(1)$ coefficients. Thus, the hierarchy between second- and first-generation masses follows directly from the $SU(2)$ tensor structure: the leading composite doublet gives one massive light generation, while the next independent one lifts the rank at order $\eps^2$ relative to it. This rank-lifting mechanism by outer products of doublets is the core of our construction and is illustrated schematically in \cref{fig:ranklifting}.
\begin{figure}[!htb]
\centering
\begin{tikzpicture}[x=1cm,y=1cm,>=latex]

\tikzset{
    smallball/.style={circle,fill=black,inner sep=1.6pt},
    bigball/.style={circle,fill=black,inner sep=2.4pt}
}

\node at (1.5,0) {$\mathcal S \sim$};

\node at (2.9,0)
{$
\left(
\begin{array}{c}
\vdots\\
\vdots \\
\vdots
\end{array}
\right)
_{2j_{\mathcal S}+1}
$};

\coordinate (B) at (3.7,0);

\draw[thick] (B) -- (5.4, 1.9);
\draw[thick] (B) -- (5.6, 0.0);
\draw[thick] (B) -- (5.4,-1.9);

\def\xtext{5.9}
\def\xvec{8.1}
\def\xY{10.5}

\node[anchor=west] at (\xtext,1.95)
{$(\mathcal S^2\bar{\mathcal S})_{1/2}$};
\node[anchor=west] at (7.7,1.95)
{$\sim$};

\node[anchor=west] at (\xvec,1.95)
{$
\left(
\begin{array}{c}
0\\
\epsilon^3
\end{array}
\right)
$};

\node[anchor=west] at (\xtext,-0.1)
{$(\mathcal S^3\bar{\mathcal S}^{\,2})_{1/2,i}$};
\node[anchor=west] at (7.7,0.05)
{$\sim$};

\node[anchor=west] at (\xvec,0.05)
{$
\left(
\begin{array}{c}
\epsilon^5\\
\epsilon^5
\end{array}
\right)
$};

\node[anchor=west] at (\xtext+0.35,-0.60) {$\cdots$};
\node[anchor=west] at (\xtext+0.35,-1.9) {$\cdots$};

\node at (\xY+0.85,2.95) {$Y_f^{2\times 2}$};

\begin{scope}[shift={(\xY,1.85)}]
    \node at (0,0) {$\left(\vphantom{\begin{array}{cc}a&a\\a&a\end{array}}\right.$};
    \node at (1.35,0) {$\left.\vphantom{\begin{array}{cc}a&a\\a&a\end{array}}\right)$};
    \node[bigball] at (0.42,-0.25) {};
    \node[bigball] at (0.93,-0.25) {};

    \node at (2.7,0.0275) {Rank-1};
\end{scope}

\begin{scope}[shift={(\xY,-0.05)}]
    \node at (0,0) {$\left(\vphantom{\begin{array}{cc}a&a\\a&a\end{array}}\right.$};
    \node at (1.35,0) {$\left.\vphantom{\begin{array}{cc}a&a\\a&a\end{array}}\right)$};

    \node[smallball] at (0.42, 0.30) {};
    \node[smallball] at (0.93, 0.30) {};
    \node[bigball] at (0.42,-0.25) {};
    \node[bigball]   at (0.93,-0.25) {};

    \node at (2.7,0.0275) {Rank-2};
\end{scope}

\end{tikzpicture}
\caption{Schematic illustration of the rank lifting mechanism via outer products in the light $2\times 2$ Yukawa block.}
\label{fig:ranklifting}
\end{figure}

The mechanism works in an analogous way for the $(\bm{2}_2^L, \bm{2}_2^R)$ case, but is actually problematic for the $\bm{1}_2 \oplus \bm{3}_2$ case. This matters when the light generation left- and right-handed fermions transform under a shared $SU(2)_{L+R}$ as doublets, see \cref{eq:YfStructure}. To show why, suppose that the light $2\times 2$ Yukawa block requires a spurion with Abelian integer charge $q$, even, and let us conventionally assign charge $+1$ to $\Ss$. Then, since $\Ss^q$ always contains a singlet or a triplet,\footnote{Often not both because of the commuting nature of $\Ss$, see \cref{app:countingIrreps}. This does not change the conclusion here.} it can be inserted as a spurion and for generic $\Ss$ would induce common first and second generation masses of $\mathcal{O}(\eps^{|q|})$, together with either a very small Cabibbo angle, in the singlet case, or an $\mathcal{O}(1)$ mixing angle, in the triplet case. Neither possibility is satisfactory.
One may wonder whether the leading light $2\times 2$ block could instead be approximately rank-1, for example, because it is generated by the outer product of two doublets as suggested before, with the singlet or triplet entering only at subleading order and lifting the rank to generate the first generation masses. This mechanism does not work, though. If the two doublets have charges $q_1$ and $q_2$, then the overall suppression is $\mathcal{O}(\epsilon^{|q_1|+|q_2|})$, while the total charge is $q=q_1+q_2$. Since
\begin{align}
    |q_1|+|q_2|\ge |q| \, ,
\end{align}
the two doublet construction can never appear parametrically earlier than the direct singlet or triplet operator of charge $q$, which already arises at $\mathcal{O}(\epsilon^{|q|})$. Therefore, the rank-1 structure built from two doublets cannot generically precede the direct singlet/triplet contribution. Note that $q_1$ and $q_2$ must be odd, so $q=q_1+q_2$ is necessarily even, always consistent with the fact that the singlet or triplet carries even charge.

This example was formulated for the structures in \cref{eq:YfStructure}, but the conclusion is in fact more general and constraining. Including also the possibilities of \cref{eq:YdStructure} and \cref{eq:YeStructure}, one is led to the same lesson: spurion insertions of $SU(2 \text{ or } 3)_{L+R}$ that directly couple left- and right-handed fields, namely $\bm{1} \oplus \bm{3}_2$ in \cref{eq:YfStructure}, $\bm{2}_2 \oplus \bm{4}_2$ in \cref{eq:YdStructure} and  $\bm{3}_2 \oplus \bm{3}_2 \oplus \bm{5}_2$ or $\bm{1} \oplus \bm{8}_3$ in \cref{eq:YeStructure}, generically induce full rank matrices and thus comparable masses at the first available spurion insertion, if these are themselves assumed to arise as composites of more elementary higher-representation fields with anarchic entries.

There are then essentially two ways forward. One possibility is to abandon the assumption that the elementary higher-representation spurions are generic and instead assume that they already carry some non-trivial internal structure, for instance in the form of strategic zeros, which is then inherited by the composite spurions, as in \cite{Banks:2025baf}. This is, however, basis-dependent and does not necessarily follow from an underlying dynamical principle, such as minimizing a scalar potential. The other, which is the path we follow here, is to exploit the mechanism illustrated above and construct the Yukawas through sequential outer products, so that the rank is lifted step by step by successive insertions of composite doublets or triplets. For a Yukawa interaction of the form 
\begin{align}
\bar f_L H f_R,
\end{align}
only two classes of structures among those in \cref{eq:YfStructure}, \eqref{eq:YdStructure}, and \eqref{eq:YeStructure} thus remain viable: 
\begin{itemize}
    \item $ G_\text{flavor} \supset SU(2\text{ or }3)_{L/R} $: a single non-Abelian factor acts on either $f_L$ or $f_R$, but not both, with the corresponding light fermions transforming as a doublet or triplet, possibly supplemented by Abelian charges of $G_\text{Abelian}$ \eqref{eq:Gflavor}:
    \begin{align}
    \begin{gathered}
    SU(2 \text{ or } 3)_L\\
    \begin{aligned}
        f_{L} &\sim \bm{2}_{2} \oplus  \bm{1} \text{ or } \bm{3}_{2,3}  \\
        f_{R} &\sim \bm{1} \oplus \bm{1} \oplus \bm{1}
    \end{aligned}
    \end{gathered}
    \qquad \text{ or } \qquad
    \begin{gathered}
    SU(2 \text{ or } 3)_R\\
    \begin{aligned}
    f_{L} &\sim \bm{1} \oplus \bm{1} \oplus \bm{1} \\
        f_{R} &\sim \bm{2}_{2} \oplus  \bm{1} \text{ or } \bm{3}_{2,3}
    \end{aligned}
    \\
    \end{gathered}
    \end{align}
    \item $G_\text{flavor} \supset  SU(2\text{ or }3)_L \times SU(2\text{ or }3)_R $: two different non-Abelian factors act separately on $f_L$ and $f_R$, again possibly supplemented by additional Abelian charges:
    \begin{align}
    \begin{gathered}
    SU(2 \text{ or } 3)_L \times SU(2 \text{ or } 3)_R\\
    \begin{aligned}
        f_{L} &\sim \bm{2}_{2} \oplus  \bm{1} \text{ or } \bm{3}_{2,3} \\
        f_{R} &\sim \bm{2}_{2} \oplus  \bm{1} \text{ or } \bm{3}_{2,3} \\
    \end{aligned}.
    \end{gathered}
    \end{align}
\end{itemize}
A given $SU(2)$ or $SU(3)$ factor may be shared among fields that do not couple directly through the same Yukawa interaction, for example among $q,\ell$ or $u,d,e$.

Finally, we briefly discuss how neutrino masses can be accommodated within this framework. If neutrinos are Majorana particles, the leading effective operator responsible for their masses is the Weinberg operator, $ O_5=(\bar{\tilde \ell} H)(\tilde H \daga \ell)$. 
The spurion structures entering the neutrino mass matrix $M_\nu$ are then immediately fixed by the representation properties of $\ell$:
\newcommand{\MnuTripletCell}{\begin{array}{c}
\bm{1}_2\oplus\bm{5}_2\\
\text{or } \bm{6}_3
\end{array}}
\newcommand{\MnuTripletMatrix}{\left(
\begin{array}{c}
\MnuTripletCell
\end{array}
\right)}
\begin{gather}
\begin{gathered}
\hspace{1.75cm}SU(2 \text{ or } 3)_L\\
\begin{array}{c@{\qquad}c@{\qquad}c}
\raisebox{-1.4em}{$M_\nu:$}
&
\begin{array}{c}
\ell \sim \bm{3}_{2,3}\\[0.5cm]
\MnuTripletMatrix
\end{array}
&
\begin{array}{c}
\ell \sim \bm{2}_{2} \oplus \bm{1}\\[0.5cm]
\left(
\begin{array}{c|c}
\bm{3}_2 & \bm{2}_2 \\
\hline
\bm{2}_2 & \bm{1}
\end{array}
\right)
\vphantom{\MnuTripletMatrix}
\end{array}
\end{array}
\end{gathered}
\label{eq:MnuStructure}
\end{gather}
where again we omitted the pure $U(1)_\text{FN}$ structure, and display only the representations allowed by the symmetric nature of $M_\nu$. The first possibility, $\ell \sim \bm{3}_{2,3}$, naturally leads to anarchic neutrino masses and PMNS angles, if the singlet $\bm{1}_2$ is excluded by some additional selection rule, such as a $U(1)$. The second instead predicts at best a modest splitting between the third generation and the first two of order $\mathcal{O}(\eps^2)$, see \cref{eq:Ydecomp}, together with anarchic PMNS angles among the light generations, and PMNS angles involving the third generation of at least order a doublet insertion $\sim \mathcal{O}(\eps^3)$. In view of the observed neutrino mass pattern and PMNS mixings \cite{Esteban:2024eli}, this appears disfavored, although not completely excluded if the bottom-right entry could somehow be engineered to be of order $\mathcal{O}(\eps^2)$. Selection rules associated with the possible $U(1)^m$ symmetry factors may help in this direction.
If neutrinos are instead Dirac particles, the Yukawa interaction between the SM lepton doublet and the right-handed neutrinos decomposes as in \cref{eq:YfStructure}, \cref{eq:YdStructure}, \cref{eq:YeStructure}. Similar considerations apply.

To summarize, in these setups, the effective spurions entering the Yukawas can genuinely arise as composite objects built from more elementary spurions that transform in higher representations and have generic entries of comparable size. The central mechanism by which hierarchical Yukawa structures emerge is that the matrices are built sequentially through outer products, so that the rank is filled one step at a time.  The resulting hierarchies are then encoded in the tensor decomposition of higher products of the same elementary spurions. The crucial ingredients for this construction are composite $SU(2)$ doublets, required by the large top Yukawa, and possibly $SU(2)$ or $SU(3)$ triplets for the down-quark and charged-lepton masses.

\section{Models}
\label{sec:models}

In this section, we present explicit models that implement the framework in \cref{sec:framework}. We start with the simplest realization and progressively move to more elaborate constructions that better fit the observed pattern of fermion masses and mixings.

\subsection{Model I: $U(2)_{q+\ell}$} 
\label{sec:U2ql}

As a first example, we consider the model based on the $U(2)_{q+\ell}$ flavor symmetry of \cite{Greljo:2023bix, Greljo:2024zrj}. In this framework, the first two generations of left-handed quarks and leptons transform as doublets of an $SU(2)$ flavor symmetry and carry a common $U(1)$ charge,
\begin{align}
    q,\ell \sim \bm{2}_{x} \, ,
\end{align}
while all other fields are neutral. This setup can be embedded in a Pati-Salam unification \cite{Greljo:2024zrj}, and its $SU(2)$ factor is anomaly-free.\footnote{An alternative variation charges $e$ instead of $\ell$~\cite{Antusch:2023shi}, motivated by the large PMNS mixing. The fit to the charged fermion masses and the CKM mixing is identical, but the $SU(2)$ symmetry is then anomalous.} Among the possibilities discussed above, this is one of the most minimal and already illustrates clearly the logic of our approach.

The observed pattern of SM masses and mixings can be reproduced by introducing two effective spurion doublets,
\begin{align}
\phi_{1,2} \sim \bm{2}_x \, .
\end{align}
The effective dimension-4 Yukawa interactions then take the form
\begin{align}
    \mathcal{L}_\text{Yuk} \supset 
    x^{d}_{i} \,  \bar q_3 H d_i +  x^{e}_{i} \, \bar \ell_3 H e_i +  \bar q H (\phi_1 z^{d}_{i} + \phi_2 y^{d}_{i}) d_i + 
    \bar \ell H (\phi_1 z^{e}_{i} + \phi_2 y^{e}_{i}) e_i 
    + u\leftrightarrow d + \text{h.c.}
\end{align}
where $z^f_i$, $y^f_i$, and $x^f_i$ are $\mathcal{O}(1)$ coefficients. This structure yields a rank-1 Yukawa matrix at LO, which is then promoted to rank 2 and rank 3 through the insertion of the two spurions. In a basis where $\phi_{2}\sim(0,a)$ and $\phi_1 \sim (b,b)$ with $b \ll a$, the Yukawa matrices take the generic form
\begin{align}
    Y_{u,d,e} \sim 
    \begin{pmatrix}
        b & b & b\\
        a & a & a \\
        1 & 1 & 1
    \end{pmatrix}
    \label{eq:U2qlYuk}
\end{align}
which leads to the mass and CKM angles predictions
\begin{gather}
\begin{aligned}
y_{u,d,e}&\sim b,  \\
y_{c,s,\mu}&\sim a,  \\
y_{t,b,\tau}&\sim 1  \\
\end{aligned}\\
\theta_{12} \sim b/a,\quad 
\theta_{23} \sim a, \quad \theta_{13} \sim b.
\label{eq:predictions_qe}
\end{gather}
This provides an acceptable fit to the observed data for $a \approx 3 \cdot 10^{-3}$ and $b \approx 5 \cdot 10^{-5}$ \cite{Greljo:2023bix,Antusch:2023shi}, and is the minimal model capable of partially addressing the origin of the mass and CKM hierarchies, at the price of leaving some residual hierarchies unexplained, such as the splitting between $y_t$ and $y_{b,\tau}$.

Let us now replace the two effective doublets by a single spurion $\Ss$ transforming in an $SU(2)$ representation of half-integer spin-$j_\Ss \geq 3/2$, with generic entries of comparable size.
As discussed around \cref{eq:Ydecomp}, the first spin-$1/2$ representation that can be constructed from $\Ss$ is $(\Ss^2 \bar \Ss)_{1/2}$, and is unique. At the next order, the structures $\Ss^3 \bar \Ss^2$, $\Ss^4 \bar \Ss$, and $\Ss^5$ contain several spin-$1/2$ components, linearly independent from the leading $\mathcal{O}(|\Ss|^3)$ one. Hence, momentarily dropping the $U(1)\subset U(2)_{q+\ell}$, the Yukawa interactions can be written as
\begin{align}
\begin{aligned}
    \mathcal{L}_\text{Yuk} \supset &   \; x_i ^d \,   \bar q_3 H d_i  \\
    & + \bar q  H \left( \frac{(\Ss^2 \bar \Ss)_{1/2}}{M^3} y_{i} ^{d} + 
    \frac{(\Ss \bar \Ss^2)_{1/2}}{M^3} \bar y_{i} ^{d}  + \frac{(\Ss^3 \bar \Ss^2 )_{1/2,k}}{M^5} z_{k,i}^{d}  + \frac{(\Ss^4 \bar \Ss)_{1/2,k}}{M^5} z_{k,i}^{\prime \, d} + \dots \right)d_i \\
    &+ \{u\} + \{\text{leptons}\} + \text{h.c.}
\end{aligned}
\label{eq:U2qlLagr}
\end{align}
for some common UV scale $M$ and $\mathcal{O}(1)$ coefficients.

We immediately see that the full $3\times 3$ Yukawa is naturally rank-1 at $\mathcal{O}(|\Ss|^0)$, rank-2 at $\mathcal{O}(|\Ss|^3)$ and full rank at $\mathcal{O}(|\Ss|^5)$, as schematically depicted in \cref{fig:U2qlCartoon}, provided the two leading doublet contributions from $(\Ss^2\bar\Ss)_{1/2}$ and $(\Ss\bar\Ss^2)_{1/2}$ are either aligned or one of them is forbidden. This can be achieved in two simple ways. The first is to re-introduce the common $U(1)$ factor, thus recovering a $U(2)_{q+\ell}$ setup, in which, say, $\Ss$ carries charge conjugate to that of $q$ and $\ell$, so the term $(\Ss \bar \Ss^2)_{1/2}$ is absent. The second, more economical, possibility is to impose a $Z_2$ symmetry exchanging $\Ss \leftrightarrow \bar \Ss$, which enforces $y^f=\bar y^f$. In that case, the leading light block takes the form
\begin{align}
    Y_{f,\alpha i}^{2\times 2}=
    \Big ( \frac{(\Ss^2 \bar \Ss)_{1/2}}{M^3}+\frac{(\Ss \bar \Ss^2)_{1/2}}{M^3}\Big )_\alpha y_i^{f}  ,
    \label{eq:Z2yuk}
\end{align}
and is manifestly rank-1. As we discuss later in \cref{sec:UV}, such a symmetry can arise accidentally.\footnote{The symmetry could also be identified as CP. However, even allowing for $\mathcal{O}(1)$ CP-violating phases in $\Ss$, one can prove that the structure in \cref{eq:U2qlYuk} cannot generate an $\mathcal{O}(1)$ CP-violating phase in $V_\text{CKM}$ without fine-tunings.}

\begin{figure}[!htb]
\centering
\begin{tikzpicture}[scale=1.0, baseline=(current bounding box.center)]
\tikzset{
  dotS/.style={circle, fill=black, inner sep=1.pt},
  dotM/.style={circle, fill=black, inner sep=1.8pt},
  dotL/.style={circle, fill=black, inner sep=2.5pt},
  arrow/.style={->, thick},
  mat/.style={thick}
}
\node at (0,1.8) {$\mathrm{Rank}\text{-}1$};
\node at (4.6,1.8) {$\mathrm{Rank}\text{-}2$};
\node at (9.2,1.8) {$\mathrm{Rank}\text{-}3$};
\node at (-2.35,0) {$Y_{u,d,e}: \quad $};
\draw[mat] (-1.0,1.0) .. controls (-1.2,0.4) and (-1.2,-0.4) .. (-1.0,-1.0);
\draw[mat] (1.0,1.0) .. controls (1.2,0.4) and (1.2,-0.4) .. (1.0,-1.0);

\node[dotL] at (-0.5,-0.6) {};
\node[dotL] at (0,-0.6) {};
\node[dotL] at (0.5,-0.6) {};

\draw[arrow] (1.55,0) -- (3.05,0);
\node at (2.3,0.45) {$\left(\Ss^2\bar \Ss\right)_{1/2}$};
\node at (2.3,-0.45) {$\mathcal{O}(\epsilon^3)$};
\draw[mat] (3.6,1.0) .. controls (3.4,0.4) and (3.4,-0.4) .. (3.6,-1.0);
\draw[mat] (5.6,1.0) .. controls (5.8,0.4) and (5.8,-0.4) .. (5.6,-1.0);

\node[dotM] at (4.1,0.00) {};
\node[dotM] at (4.6,0.00) {};
\node[dotM] at (5.1,0.00) {};

\node[dotL] at (4.1,-0.6) {};
\node[dotL] at (4.6,-0.6) {};
\node[dotL] at (5.1,-0.6) {};
\draw[arrow] (6.15,0) -- (7.65,0);
\node at (6.9,0.45) {$\left(\Ss^3\bar \Ss^2\right)_{1/2}$};
\node at (6.9,-0.45) {$\mathcal{O}(\epsilon^5)$};
\draw[mat] (8.2,1.0) .. controls (8.0,0.4) and (8.0,-0.4) .. (8.2,-1.0);
\draw[mat] (10.2,1.0) .. controls (10.4,0.4) and (10.4,-0.4) .. (10.2,-1.0);

\node[dotS] at (8.7,0.6) {};
\node[dotS] at (9.2,0.6) {};
\node[dotS] at (9.7,0.6) {};

\node[dotM] at (8.7,0.00) {};
\node[dotM] at (9.2,0.00) {};
\node[dotM] at (9.7,0.00) {};

\node[dotL] at (8.7,-0.6) {};
\node[dotL] at (9.2,-0.6) {};
\node[dotL] at (9.7,-0.6) {};
\end{tikzpicture}
\caption{Schematic illustration of the generation of hierarchical Yukawa structures in the $U(2)_{q+\ell}$ model. Higher-order insertions of composite doublet spurions progressively lift the Yukawa matrix from rank-1 to full rank.}
\label{fig:U2qlCartoon}
\end{figure}

We then see that the $U(2)_{q+\ell}$ pattern can be fully reproduced in terms of a single higher flavor spin spurion $\Ss$. The matching is remarkably successful: choosing the overall size
\begin{align}
    \frac{|\Ss|}{M} = \eps \approx 0.1
\end{align}
reproduces the observed hierarchies obtained in the $\phi_1, \phi_2$ framework. Crucially, \emph{the spurion $\Ss$ is completely generic}: it does not require any special alignment or internal hierarchy among its components. The most generic spin-$j_\Ss$ spurion naturally generates two independent and hierarchical spin-$1/2$ structures; the observed flavor hierarchies and mixings arise from the tensor decomposition properties of a higher $SU(2)$ flavor spin.

Moving beyond the renormalizable Lagrangian, for the Weinberg operator, consistently with \cref{eq:MnuStructure}, we can schematically write
\begin{align}
    \frac{(\bar \Ss^2)_{1}}{M^2} (\bar{\tilde \ell} H)(\tilde H \daga \ell)  + \frac{(\Ss \bar\Ss^2)_{1/2}}{M^3} (\bar{\tilde \ell} H)(\tilde H \daga \ell_3)  + (\bar{\tilde \ell}_3 H)(\tilde H \daga \ell_3)+\left\{ \Ss \leftrightarrow \bar \Ss \right\} + \text{h.c.} .
\end{align}
The resulting neutrino mass matrix then takes the parametric form already anticipated in \cref{eq:MnuStructure},
\begin{align}
    M_\nu \sim 
    \begin{pmatrix}
        \eps^2 & \eps^2 & \eps^3\\
        \eps^2 & \eps^2 & \eps^3\\
        \eps^3 & \eps^3 & 1
    \end{pmatrix},
\end{align}
which qualitatively leads to non-hierarchical masses and anarchic PMNS mixings, provided the bottom right entry is accidentally of order $\eps^2$, as already happens for $y_{b,\tau}$. At the lowest order, this gives an acceptable fit to neutrino data \cite{Esteban:2024eli}.
This is an important difference with respect to the construction based on two effective doublets $\phi_{1,2}$. In that case, the light $2\times 2$ block would be of the form $\sim \phi_2 \phi_2 ^T + \phi_1 \phi_1^T$, leading generically to overly hierarchical neutrino masses and mixings, unless extra sources of symmetry breaking are introduced~\cite{Greljo:2023bix}, or one moves to a $U(2)_{q+e}$ setup \cite{Antusch:2023shi}.

Other higher-dimensional operators are only weakly constrained by the symmetry and spurion structure, and their phenomenological implications are discussed in \cref{sec:pheno}.

\paragraph{Non-Abelian origin.}
The Yukawa pattern illustrated schematically in \cref{fig:U2qlCartoon} relies crucially on the non-Abelian tensor product structure of our construction. To show this, let us retain only the maximal Abelian subgroup of the flavor symmetry $U(2)_{q+\ell}$,
\begin{align}
    U(1)_q \times U(1)_{T^3},
\end{align}
where $U(1)_{T^3}\subset SU(2)_{q+\ell}$ is associated to the diagonal generator $T^3$. Taking the quark Yukawa matrices as a concrete example, the $SU(2)$ doublet $q$ decomposes into two fields carrying the same $U(1)_q$ charge and $T^3$ charges $m=\pm 1/2$:
\begin{align}
    q \sim
    \begin{pmatrix}
        q_{-1/2} \\
        q_{+1/2}
    \end{pmatrix}.
\end{align}
Similarly, the higher-irrep spurion $\Ss$ decomposes into $2j_\Ss+1$ fields $s_m$, that carry the same $U(1)_q$ charge as $q$ and $T^3$ charges $m=-j_\Ss,-j_\Ss+1,\ldots,j_\Ss$:
\begin{align}
    \Ss \sim
    \begin{pmatrix}
        s_{-j_\Ss} \\
        s_{-j_\Ss+1} \\
        \vdots \\
        s_{j_\Ss}
    \end{pmatrix}.
\end{align}

Retaining only the Abelian symmetry, operators involving different $T^3$
components are no longer related, and therefore appear with independent
coefficient vectors. In particular, the most general Lagrangian allowed by
$U(1)_q \times U(1)_{T^3}$ contains the terms
\begin{align}
\begin{aligned}
    \mathcal{L}_\text{Yuk}
    &\supset
    \bar q_{-1/2} H
    \left(
        \frac{s_{-1/2}}{M}\,y_i^d
        +
        \frac{s_{-1/2}s_{3/2}s_{3/2}^*}{M^3}\,z_i^d
        + \ldots
    \right)d_i
    \\
    &\quad+
    \bar q_{+1/2} H
    \left(
        \frac{s_{+1/2}}{M}\,y_i^{\prime d}
        +
        \frac{s_{+1/2}s_{3/2}s_{3/2}^*}{M^3}\,z_i^{\prime d}
        + \ldots
    \right)d_i
    + \ldots \, .
\end{aligned}
\end{align}
The linear terms already generate two independent light-family rows at
$\mathcal{O}(\eps)$, rendering the Yukawa
matrices full rank and delivering the qualitative pattern
\begin{align}
    Y_{u,d},Y_e^t
    \sim
    \begin{pmatrix}
        \eps & \eps & \eps \\
        \eps & \eps & \eps \\
        1    & 1    & 1
    \end{pmatrix}.
\end{align}
This is undesirable and should be contrasted with the pattern arising from the
$U(2)_{q+\ell}$ mechanism \cref{fig:U2qlCartoon},
\begin{align}
    Y_{u,d},\,Y_e^t \sim
    \begin{pmatrix}
        \eps^5 & \eps^5 & \eps^5 \\
        \eps^3 & \eps^3 & \eps^3 \\
        1 & 1 & 1
    \end{pmatrix}.
\end{align}
which delivers hierarchical properties much more compatible with those observed
in Nature. Ultimately, it arises from the constraints imposed by the non-Abelian flavor symmetry,
which enforce the first contribution involving the light families to appear at
$\mathcal{O}(\eps^3)$, while the Yukawa matrix becomes full rank only at
$\mathcal{O}(\eps^5)$.

This shows that our mechanism relies crucially on the non-Abelian nature of the flavor symmetry group. We wish to stress, however, that we are not claiming that this is the only possible way to obtain such a Yukawa pattern. The crucial point is that, once the higher-irrep spurion $\Ss$ is decomposed under the maximal Abelian subgroup, its components $s_m$ behave as several Abelian spurions with different charges. Taken as independent Abelian fields, these allow many different contractions and, as shown above, generically fill the Yukawa rank already at low order. The full non-Abelian symmetry instead forces all these components to belong to a single multiplet and to enter only in specific combinations, which is what produces the hierarchical rank lifting. A genuine $U(1)$ Froggatt-Nielsen construction can evade this simply by retaining only a restricted subset of such Abelian spurions, so that the Abelian selection rules alone can enforce a hierarchy. For instance, a simple $U(1)_{\rm FN}$ model with a single scalar field $\Phi$ of unit charge and charge assignments $[q_1]=[\ell_1]=5$, $[q_2]=[\ell_2]=3$ reproduces the same row hierarchy. Our claim is therefore that, with the full field content and charge assignments inherited from $\Ss$, the Yukawa structure cannot be accounted for by the maximal Abelian subgroup alone. The same logic extends straightforwardly to the other models introduced in the following subsections.

\subsection{Model II: $U(2)_{q+\ell} \times U(1)_R$}
\label{sec:LRmodel}

We next consider a more structured model in which the hierarchy of the SM Yukawa couplings factorizes into a \emph{row hierarchy} and a \emph{column hierarchy}. The former is controlled by the same left-handed non-Abelian factor as before, broken by a single higher flavor spin spurion $\Ss$ with $j_\Ss \geq 3/2$, while the latter is controlled by an Abelian Froggatt-Nielsen symmetry, broken by a spurion $\phi$. The flavor symmetry is
\begin{equation}
G_\text{flavor} = U(2)_{q+\ell} \times U(1)_R\,.
\end{equation}
The first two left-handed generations transform as a doublet of $U(2)_L$, while the third generation is a singlet,
$(q,\ell) \sim {\bf 2}_{1}\oplus{\bf 1}_0$.
All right-handed fields are singlets of $U(2)_{q+\ell}$ and carry generation-dependent charges under $U(1)_R$, as shown in \cref{tab:chargesLRmodel}. This structure reproduces the observed flavor hierarchies up to $\mathcal{O}(1)$ coefficients, as we now show. We stress, though, that the assignment of FN charges is not unique, and that other choices may work equally well or even better. This simply reflects the freedom, already mentioned in the introduction, that is intrinsic to FN models. The charges chosen here are motivated by a possible quark-lepton unification based on
$SU(4) \times SU(2)_L \times U(1)_R$~\cite{Greljo:2024zrj}, under which the five SM fermion representations of a single generation unify into three. In particular, the left chiral quark and lepton doublets are combined into a single $(\bm{4},\bm{2},0)$ multiplet, while the right-chiral fields are organized into $(\bm{4},\bm{1}, \pm 1/2)$ multiplets. 

\begin{table}[!ht]
\centering
\renewcommand{\arraystretch}{1.15}
\begin{tabular}{c|ccccc|cc}
 & $q_L,\ell_L$ & $q_L^3,\ell_L^3$ & $u_R^i$ & $d_R^i,e_R^i$ & $H$ & $\Ss$ & $\phi$ \\
\hline
$U(2)_{q+\ell}$ 
& $\mathbf{2}_1$ 
& $\cdot$
& $\cdot$
& $\cdot$
& $\cdot$
& $(\bm{2j_\Ss+1})_1$ 
& $\cdot$ \\

$U(1)_R$ 
& $\cdot$
& $\cdot$
& $(4,\,2,\,0)$ 
& $(4,\,3,\,2)$ 
& $\cdot$ 
& $\cdot$
& $-1$
\end{tabular}
\caption{Flavor charges of matter fields and spurions under $U(2)_{q+\ell}\times U(1)_R$. Here $i=1,2,3$, and entries denoted with $\cdot$ refer to total singlets. $\Ss$ transforms in a higher-spin irrep of the non-Abelian factor, with half-integer $j_\Ss \geq 3/2$, while $\phi$ is the Abelian Froggatt--Nielsen spurion. See \cref{sec:LRmodel} for details.}
\label{tab:chargesLRmodel}
\end{table}

Denoting the flavor-breaking expansion parameters with
\begin{equation}
\epsilon_{q+\ell} \equiv \frac{| \Ss |}{M_\Ss}\,,\qquad
\epsilon_R \equiv \frac{| \phi|}{M_\phi}\,,
\end{equation}
the Yukawa matrices acquire a factorized form
\begin{equation}
Y_u \sim
{
\renewcommand{\arraystretch}{1.3}
\setlength{\arraycolsep}{6pt} 
\begin{pmatrix}
\epsilon_{q+\ell}^5 \epsilon_R^4 & \epsilon_{q+\ell}^5 \epsilon_R^2 & \epsilon_{q+\ell}^5 \\
\epsilon_{q+\ell}^3 \epsilon_R^4 & \epsilon_{q+\ell}^3 \epsilon_R^2 & \epsilon_{q+\ell}^3 \\
\epsilon_R^4 & \epsilon_R^2 & 1
\end{pmatrix}
},
\qquad
Y_{d,e} \sim
{
\renewcommand{\arraystretch}{1.3}
\setlength{\arraycolsep}{6pt} 
\begin{pmatrix}
\epsilon_{q+\ell}^5 \epsilon_R^4 & \epsilon_{q+\ell}^5 \epsilon_R^3 & \epsilon_{q+\ell}^5 \epsilon_R^2 \\
\epsilon_{q+\ell}^3 \epsilon_R^4 & \epsilon_{q+\ell}^3 \epsilon_R^3 & \epsilon_{q+\ell}^3 \epsilon_R^2 \\
\epsilon_R^4 & \epsilon_R^3 & \epsilon_R^2
\end{pmatrix}
}.
\label{eq:textures}
\end{equation}
The hierarchy between generations arises from two conceptually distinct sources: the full flavor pattern is organized by a simple factorization principle:
\begin{equation}
\text{rows} \leftrightarrow U(2)_{q+\ell},
\qquad
\text{columns} \leftrightarrow U(1)_R.
\end{equation}
After perturbative singular value decomposition of Eq.~\eqref{eq:textures}, we obtain the parametric predictions
\begin{gather}
\begin{aligned}
y_{u,c,t}&\sim \left(\epsilon_{q+\ell}^5\epsilon_R^4,\;\epsilon_{q+\ell}^3\epsilon_R^2,\;1\right), \\
y_{d,s,b}\sim y_{e,\mu,\tau} &\sim \left(\epsilon_{q+\ell}^5\epsilon_R^4,\;\epsilon_{q+\ell}^3\epsilon_R^3,\;\epsilon_R^2\right),
\end{aligned}\\
\theta_{12} \sim \epsilon_{q+\ell}^2,\quad \theta_{23} \sim \epsilon_{q+\ell}^3,\quad  \theta_{13} \sim \epsilon_{q+\ell}^5 .
\label{eq:predictions}
\end{gather}
These relations should be interpreted up to unspecified coefficients of order unity. Taking numerical values from~\cite{Huang:2020hdv, Greljo:2023bix} evaluated at the renormalization scale $\mu_R=100$ TeV, we find a good fit for
\begin{align}
\epsilon_{q+\ell} \simeq 0.35, \quad \epsilon_R \simeq 0.2
\end{align} with $\mathcal{O}(1)$ fit parameters scattered in the range of $[0.26,1.8]$.

\subsection{Model III: $U(2)_{q+e} \times U(2)_{u} \times G$}
\label{sec:10model}

\subsubsection{$G=Z_2$ ($\tan \beta$)}

We now turn to examples that go beyond the simple benchmark model of \cref{sec:U2ql}, providing a better fit while still avoiding the freedom introduced by a Froggatt-Nielsen $U(1)$, as in \cref{sec:LRmodel}. We begin with a model based on the flavor symmetry
\begin{align}
    U(2)^2 = U(2)_{q+e} \times U(2)_{u},
\end{align}
with universal $U(1)$ factors and an additional $Z_2$.\footnote{
We distinguish the universal $U(1)$ associated with $U(2)$ from genuine $U(1)_\text{FN}$ assignments, which correspond to qualitatively different assumptions. The former only provides an additional selection rule tied to the doublet insertions and can, in fact, always be replaced by a discrete $Z_2$/CP symmetry of the type discussed in \cref{sec:U2ql}. Such a symmetry may arise accidentally in explicit UV completions, as for $Z_2$ in the model of \cref{sec:UV}, or can be a well-motivated UV assumption, in the case of CP \cite{Choi:1992xp, Dine:1992ya}. The latter instead relies on generation-dependent charge assignments and is therefore a genuine high-level input.
}

\begin{table}[!ht]
\centering
\renewcommand{\arraystretch}{1.15}
\begin{tabular}{c|cccccc|ccc}
 & $q,e$ & $q^3, e^3$ & $u$ & $u^3$ & $d^i,\ell^i$ & $H$ & $\mathcal{S}_{q+e}$ & $\mathcal{S}_{u}$ & $\cot_\beta $\\
\hline
$U(2)_{q+e}$ 
& $\bm{2}_1$ 
& $\cdot$
& $\cdot$
& $\cdot$
& $\cdot$
& $\cdot$
& $(\bm{2 j_{q+e}+1})_1$ 
& $\cdot$
& $\cdot$\\

$U(2)_{u}$ 
& $\cdot$
& $\cdot$
& $\bm{2}_1$ 
& $\cdot$
& $\cdot$
& $\cdot$
& $\cdot$
& $(\bm{2 j_{u}+1})_{-1}$ 
& $\cdot$ \\

$Z_2$ 
& $\bm{+}$
& $\bm{+}$
& $\bm{+}$
& $\bm{+}$
& $\bm{-}$
& $\bm{+}$
& $\bm{+}$
& $\bm{+}$
& $\bm{-}$\\
\end{tabular}
\caption{Flavor charges of matter fields and spurions under $U(2)_{q+e}\times U(2)_{u} \times Z_2$. Here $i=1,2,3$, and entries denoted with $\cdot$ refer to total singlets. $\mathcal{S}_{q+e}, \mathcal{S}_{u}$ transforms in a higher-spin irrep of the non-Abelian factors, with half-integer $j_\Ss \geq 3/2$.}
\label{tab:chargesU22model}
\end{table}

The charges of the various SM fields are reported in \cref{tab:chargesU22model}. As the notation for the two $U(2)$ factors suggests, one factor acts on $q$ and $e$, while the other acts only on $u$. This structure is inspired by the $U(2)_{10}$ setup of \cite{Antusch:2023shi}, albeit with an important modification: in the present framework, the flavor symmetry cannot be shared between $q$ and $ u$, because this would reintroduce the triplet problem discussed in \cref{sec:framework}. Flavor breaking is driven by the two spurions $\mathcal{S}_{q+e}$ and $\mathcal{S}_{u}$,
\begin{equation}
\epsilon_{q+e} \equiv \frac{| \mathcal{S}_{q+e}|}{M_{q+e}}\,,\qquad
\epsilon_{ u} \equiv \frac{| \mathcal{S}_{ u} |}{M_u}\,.
\end{equation}
The $Z_2$ factor is introduced to suppress the overall size of the down quark and charged lepton Yukawa matrices, playing a role somewhat analogous to that of $\cot \beta = (\tan \beta)^{-1}$ in 2HDM models, hence the notation. 

With these fields and charge assignments, the Yukawa matrices receive contributions from the LO and NLO spin-1/2 composites, yielding the structures
\begin{equation}
Y_u \sim
{
\renewcommand{\arraystretch}{1.3}
\setlength{\arraycolsep}{6pt} 
\begin{pmatrix}
\epsilon_{q+ e}^5  \epsilon_{u}^5 & \epsilon_{q+ e}^5  \epsilon_{u}^3 & \epsilon_{q+e}^5 \\
\epsilon_{q+e}^3  \epsilon_{u}^5 & \epsilon_{q+e}^3  \epsilon_{ u}^3 & \epsilon_{q+e}^3 \\
\epsilon_{u}^5 & \epsilon_{u}^3 & 1
\end{pmatrix}
},
\qquad
Y_{d}, Y_e^T \sim \cot_\beta
{
\renewcommand{\arraystretch}{1.3}
\setlength{\arraycolsep}{6pt} 
\begin{pmatrix}
\epsilon_{q+ e}^5 & \epsilon_{q+ e}^5 & \epsilon_{q+ e}^5 \\
\epsilon_{q+ e}^3 & \epsilon_{q+ e}^3 &\epsilon_{q+ e}^3 \\
1 & 1 & 1
\end{pmatrix}
},
\label{eq:U22Yuk}
\end{equation}
which leads to 
\begin{gather}
\begin{aligned}
y_{u,c,t}&\sim \left(\epsilon_{q+e}^5\epsilon_u^5,\;\epsilon_{q+e}^3\epsilon_u^3,\;1\right), \\
y_{d,s,b}\sim y_{e,\mu,\tau} &\sim \cot_\beta \left(\epsilon_{q+e}^5, \; \epsilon_{q+e}^3,\;1\right),
\end{aligned}\\
\theta_{12} \sim \epsilon_{q+e}^2,\quad \theta_{23} \sim \epsilon_{q+e}^3,\quad \theta_{13} \sim \epsilon_{q+e}^5 .
\label{eq:predictions_tbeta}
\end{gather}
The pattern in the down quark and charged lepton sectors is essentially the same as in the $U(2)_{q+\ell}$ model, up to the overall suppression by $\cot_\beta$. In this way, $\mathcal{S}_{q+ e}$ can account for the observed ratios of down quark and charged lepton masses and CKM angles. The more stretched hierarchies of the up sector, together with an $\mathcal{O}(1)$ top Yukawa, are instead obtained through a double spurion insertion in the light block, effectively building the \emph{composite bifundamental}
\begin{align}
\begin{aligned}
\Delta_{u}\sim (\bm{2}_{q+e}, \bm{2}_u) &\sim \frac{(\Ss_{q+ e}^2 \bar \Ss_{q+ e})_{1/2}}{M_{q+e}^3} \otimes \frac{(\Ss_{ u}^2 \bar \Ss_{ u})_{1/2}}{M_{u}^3} \; + \; 
\frac{(\Ss_{q+ e}^2 \bar \Ss_{q+ e})_{1/2}}{M_{q+e}^3} \otimes 
\frac{(\Ss_{ u}^3 \bar \Ss_{ u}^2)_{1/2}}{M_{u}^5}\\
&+\frac{(\Ss_{q+ e}^3 \bar \Ss_{q+ e}^2)_{1/2}}{M_{q+e}^5} \otimes \frac{(\Ss_{ u}^2 \bar \Ss_{ u})_{1/2}}{M_{u}^3} \; + \; 
\frac{(\Ss_{q+ e}^3 \bar \Ss_{q+ e}^2)_{1/2}}{M_{q+e}^5} \otimes 
\frac{(\Ss_{ u}^3 \bar \Ss_{ u}^2)_{1/2}}{M_{u}^5} + \dots\\
& \sim
{
\renewcommand{\arraystretch}{1.3}
\setlength{\arraycolsep}{6pt} 
\begin{pmatrix}
\epsilon_{q+ e}^5  \epsilon_{u}^5 & \epsilon_{q+ e}^5  \epsilon_{u}^3 \\
\epsilon_{q+e}^3  \epsilon_{u}^5 & \epsilon_{q+e}^3  \epsilon_{ u}^3
\end{pmatrix},
}
\end{aligned}
\label{eq:compBifund}
\end{align}
which, by construction, is rank-1 at LO and full rank at NLO, naturally implementing the required inner hierarchies precisely through the sequential outer product structure introduced in \cref{sec:framework}.
A good fit to the measured parameters is obtained for 
\begin{equation}
\epsilon_{q+e} \simeq 0.3,
\qquad
\epsilon_{ u} \simeq 0.32, \qquad \cot_\beta \simeq 8.7 \times 10^{-3}
\end{equation}
with $\mathcal{O}(1)$ coefficients in \cref{eq:U22Yuk} in the range $[0.13,2.5]$. The Weinberg operator has no selection rules, leading to completely anarchic neutrino masses and mixings.

\subsubsection{$G=U(2)_5$ or $SU(3)_5$}
\label{sec:U3_5}

An interesting variation on the previous idea is to let the right-handed down quarks and the left-handed leptons play a more active role in generating flavor hierarchies, rather than merely providing the overall suppression previously encoded by the $Z_2$.

As already discussed in \cref{sec:intro}, the need for an $SU(2)$ symmetry in the up sector, rather than $SU(3)$, is tied to the fact that the top Yukawa is $\mathcal{O}(1)$, so that a spurion expansion would not be well defined there. This forces $q$ and $u$ to be charged at most under a $U(2)$. By contrast, the hierarchies in the right-handed down quark and lepton sectors begin already at the level of $\mathcal{O}(10^{-2})$, opening the possibility that $d$ and $\ell$ could instead be unified into a single triplet, either of $SU(2)$ or of $SU(3)$. In the present context, one may therefore consider $d$ and $\ell$ as triplets of a common $SU(2)_5$ or $SU(3)_5$, where the subscript 5 refers to their natural embedding into a single $SU(5)$ GUT multiplet.

The possibility that $d$ and $\ell$ transform as triplets of $SU(2)_5$ is interesting, but also problematic. Indeed, without an additional selection rule, the Weinberg operator $(\bar{\tilde \ell} \tilde H)(\tilde H^\dagger \ell)$ already contains an $SU(2)$ singlet. As a result, independently of which spurion is responsible for breaking $SU(2)_5$, one would generically predict exactly degenerate neutrino masses at LO. This is in tension with neutrino oscillation data and cosmological bounds, which together indicate that neutrino mass splittings should be of the same order as the masses themselves \cite{Esteban:2024eli}. At the very least, one therefore needs a common $U(1)\subset U(2)_5$ factor, so that the Weinberg operator carries a non-trivial charge and is pushed to higher order. In that case, for example, an integer $SU(2)_5$ spin spurion $\Ss_5\sim (\bm{2j_5+1})_1$ would generate anarchic neutrino masses through the spin-2 irrep $(\Ss_5^2)_2$, while the effective triplets would arise from structures such as $(\Ss_5^2\bar \Ss_5)_1$, $(\Ss_5^3\bar \Ss_5^2)_1$, and so on.

A more attractive possibility is instead to take $d$ and $\ell$ to transform as triplets of $SU(3)_5$. In this case, both the effective triplets and the neutrino masses are generated only at higher order, without the need for any additional $U(1)$ factor, through a mechanism entirely analogous to the $SU(2)$ one, as hinted in \cref{sec:framework}. For example, taking a spurion 
\begin{align}
\Ss_5\sim \bm{24}  \in  SU(3)_5,
\end{align}
one finds that the first relevant composites are:\footnote{We verified this with \texttt{GroupMath} \cite{Fonseca:2020vke} and explicitly constructed the triplets to check their independence.}
\begin{itemize}
    \item one $\bar{\bm{6}}$ in $\Ss_5^2$;
    \item two independent $\bar{\bm 3}$'s in $\Ss_5^2\bar\Ss_5$, and several more in higher structures such as $\Ss_5^4$ and $\Ss_5\bar\Ss_5^3$.
\end{itemize}
The $\bar{\bm{6}}$ can be used directly to generate effectively anarchic neutrino masses and mixings. The $\bar{\bm 3}$'s instead provide the effective triplet spurions entering the down quark and charged lepton Yukawas. Defining
$
    \eps_5 \equiv | \Ss_5|/M_5,
$
the two independent $\bar{\bm 3}$'s generated at order $\eps_5^3$, together with higher-order triplets starting at order $\eps_5^4$, induce a column hierarchy in $Y_d$ and $Y_e^T$. Combining this with the row hierarchy from the $U(2)_{q+e}$ doublets, in analogy with the model of \cref{sec:LRmodel}, one obtains
\begin{align}
    Y_{d}, Y_e^T \sim
    {
    \renewcommand{\arraystretch}{1.3}
    \setlength{\arraycolsep}{6pt} 
    \begin{pmatrix}
        \eps_{q+ e}^5 \eps_5 ^4 & \eps_{q+ e}^5 \eps_5 ^3  & \eps_{q+ e}^5 \eps_5 ^3 \\
        \eps_{q+ e}^3 \eps_5 ^4 & \eps_{q+ e}^3 \eps_5 ^3  & \eps_{q+ e}^3 \eps_5 ^3 \\
        \eps_5 ^4 &  \eps_5 ^3  & \eps_5 ^3 
    \end{pmatrix}
    }.
\end{align}
This leads to the parametric predictions
\begin{gather}
\begin{aligned}
y_{u,c,t}&\sim \left(\eps_{q+ e}^5\eps_u^5,\;\eps_{q+ e}^3\eps_u^3,\;1\right), \\
y_{d,s,b}\sim y_{e,\mu,\tau} &\sim  \left(\eps_{q+ e}^5 \eps_5 ^4, \; \eps_{q+ e}^3 \eps_5 ^3 ,\; \eps_5^3 \right),
\end{aligned}\\
\theta_{12} \sim \eps_{q+ e}^2,\quad \theta_{23} \sim \eps_{q+ e}^3,\quad \theta_{13}\sim \eps_{q+ e}^5 ,
\end{gather}
which are in excellent agreement with the observed pattern. Compared with the previous models, the main improvement is that the relation $y_{d,e}/y_{s,\mu} \sim \theta_{12} \sim \eps_{q+ e}^2$ is relaxed, since the first generation Yukawas receive an additional suppression from the $SU(3)_5$ sector. A good fit is obtained for
\begin{align}
\eps_{q+e} \approx 0.38, \qquad \eps_u \approx 0.29, \qquad \eps_5 \approx 0.24,
\end{align}
with all remaining coefficients in the Yukawa matrices lying within the range $[0.26,1.64]$.

An important point is that we treat each independent composite $\bar{\bm 3}$ arising at a given order as carrying an independent coefficient. For example, in the down quark Lagrangian one may write
\begin{align}
    \mathcal{L} &\supset \bar q H 
    \Bigl [ \frac{(\Ss_{q+ e}^2 \bar\Ss_{q+ e})_{1/2}}{M_{q+e}^3}
    \Bigl ( a \frac{(\Ss_5^2\bar\Ss_5)_{\bar{\bm{3}},1}}{M_5^3} + b \frac{(\Ss_5^2\bar\Ss_5)_{\bar{\bm{3}},2}}{M_5^3} + c \frac{(\Ss_5^4)_{\bar{\bm{3}}}}{M_5^4} + \dots  \Bigr ) \nonumber \\
    &\qquad \quad + 
    \frac{(\Ss_{q+e}^3 \bar\Ss_{q+ e}^2)_{1/2}}{M_{q+e}^5}
    \Bigl ( a' \frac{(\Ss_5^2\bar\Ss_5)_{\bar{\bm{3}},1}}{M_5^3} + b' \frac{(\Ss_5^2\bar\Ss_5)_{\bar{\bm{3}},2}}{M_5^3} + c' \frac{(\Ss_5^4)_{\bar{\bm{3}}}}{M_5^4}  + \dots  \Bigr )  \Bigr] d \nonumber \\
    &+ \bar q_3 H \Bigl [ a'' \frac{(\Ss_5^2\bar\Ss_5)_{\bar{\bm{3}},1}}{M_5^3} + b'' \frac{(\Ss_5^2\bar\Ss_5)_{\bar{\bm{3}},2}}{M_5^3} + c'' \frac{(\Ss_5^4)_{\bar{\bm{3}}}} {M_5^4} + \dots 
    \Bigr ] d .
\end{align}
with $(a,b,c,\dots), (a',b',c',\dots), \dots$ taken as independent coefficient vectors. This is crucial: otherwise, only one or two fixed linear combinations of $\bar{\bm 3}$'s would appear, and the Yukawa matrix would fail to be of full rank. This assumption is well motivated. In explicit UV completions with VLF chains, for instance, different choices of VLF representations along the chain generate different combinations of effective triplets, and therefore naturally lead to independent coefficients, as we argue in \cref{sec:UV}. Note that for $SU(2)$ the same logic applies in principle as well, with the simplification that at a given order, namely $\eps^3$ or $\eps^5$, a single doublet combination is already enough; so one only cares about relative coefficients at different orders, which even more naturally differ.

\subsection{Comments on $U(2)^5$}
\label{sec:U25model}

Finally, we comment on possible embeddings of the popular $U(2)^5$ framework within our construction. In this setup, each SM fermion representation is split into a doublet and a singlet under its own $U(2)$ factor,
\begin{align}
    f_i \sim \bm{2}_f \oplus \bm{1}.
\end{align}
The effective spurions traditionally used in this context are \cite{Barbieri:2011ci, Barbieri:2012uh, Fuentes-Martin:2019mun}
\begin{align}
     V_q \sim \bm{2}_q, \qquad \Delta_{u,d} \sim (\bm{2}_q, \bm{2}_{u,d}), \qquad \Delta_e \sim (\bm{2}_\ell, \bm{2}_{e}).
\end{align}
The doublet $V_q$ is of order $|V_{cb}|$, while the diagonal entries of the bifundamentals $\Delta_{u,d,e}$ are essentially set by the Yukawas of the light generations. This is precisely the kind of hierarchical structure that, in our framework, can be understood in terms of generic spurions of comparable size.
Concretely, one may introduce higher flavor spin spurions $\mathcal{S}_f$ for each SM fermion representation, and construct the effective doublet and bifundamentals by sequential outer products, as in \cref{eq:compBifund},
\begin{align}
\begin{aligned}
    V_q &\sim \frac{(\Ss_q^2 \bar \Ss_q)_{1/2}}{M_q^3} + \dots,\\
    \Delta_{u,d} &\sim 
    \frac{(\Ss_q^2 \bar \Ss_q)_{1/2}}{M_{q}^3} \otimes \frac{(\Ss_{u,d}^2 \bar \Ss_{u,d})_{1/2}}{M_{u,d}^3}
    \; + \;
    \frac{(\Ss_q^2 \bar \Ss_q)_{1/2}}{M_q^3} \otimes \frac{(\Ss_{u,d}^3 \bar \Ss_{u,d}^2)_{1/2}}{M_{u,d}^5} + \dots,\\
    \Delta_e &\sim
    \frac{(\Ss_\ell^2 \bar \Ss_\ell)_{1/2}}{M_\ell^3}\otimes \frac{(\Ss_e^2 \bar \Ss_e)_{1/2}}{M_e^3}
    \; + \;
    \frac{(\Ss_\ell^2 \bar \Ss_\ell)_{1/2}}{M_\ell^3} \otimes \frac{(\Ss_e^3 \bar \Ss_e^2)_{1/2}}{M_e^5} +\dots .
\end{aligned}
\end{align}
This works qualitatively well; for example, for $\eps_{q,\ell} \approx 0.35$, $\eps_u \approx 0.29$, and $\eps_{d,e}\approx 0.24$ one obtains a good fit, again with $\mathcal{O}(1)$ coefficients varying within roughly one order of magnitude. Note that this embedding holds only at the level of reproducing the SM Yukawas. Selection rules on new physics coupled to the SM are very different, as discussed in \cref{sec:pheno}.

It is worth stressing that the introduction of one $\mathcal S_f$ for each SM fermion representation appears to be the most natural implementation of our philosophy. Constructions based on spurions charged under more than one $U(2)$ factor seem less promising. For instance, consider a doubly-charged $\Delta \sim ( \bm{2j_q+1}, \bm{2j_{u}+1})$ with half-integer $j_q, j_u$, introduced to generate $\Delta_u$ as a composite object. Then $\Delta^3, \Delta^2 \bar \Delta$ would indeed contain a bidoublet $(\bm{2}_q,\bm{2}_u)$ of order $\eps_\Delta^3$, but each of these structures would not generically be rank-1, and therefore would not naturally account for the observed splitting between first and second generation fermions.\footnote{For example, $\Delta^3$ does not project only onto $\Sym^3 V_q \otimes \Sym^3 V_u$ with $V_q,V_u$ the $SU(2)_{q,u}$ spaces; one has the decomposition
$
\Sym^3(V_q\otimes V_u)
\sim
\Sym^3 V_q \otimes \Sym^3 V_u
\oplus
S_{(2,1)}V_q \otimes S_{(2,1)} V_u
\oplus
S_{(0,3)} V_q \otimes S_{(0,3)}  V_u
$.
Hence, the $(1/2,1/2)$ component can also receive contributions from the mixed symmetry sector $S_{(2,1)}V_q \otimes S_{(2,1)}V_u$, not only from the fully symmetric piece. This renders $(1/2,1/2)$ not vanishing and full rank in general. In other words, from a fixed $SU(2)$ factor point of view, the other $SU(2)$ factor indices behave as external flavor indices that break the single irrep counting of \cref{app:countingIrreps}. This is precisely why the doubly charged $SU(3)$ spurions of \cite{Banks:2025baf} require additional assumptions, namely strategic zeros, in order to avoid generating generic full-rank structures too early.}
Moreover, such a field would not allow one to isolate a single $U(2)_q$ factor in order to generate $V_q$.

One could also imagine a hybrid setup involving both singly- and doubly-charged fields, for example, generating $\Delta_{u,d,e}$ at LO via outer products of doublets and then lifting the rank via composite bifundamentals. However, this tends to reintroduce the very type of hierarchy we are trying to avoid. For instance, second generation masses would arise at $\mathcal{O}(\eps_q^3 \eps_u ^3)$, while first generation masses could already be generated at $\mathcal{O}(\eps_q \eps_\Delta \eps_u )$ through structures such as $\mathcal S_q \Delta \mathcal S_{\bar u}$. Note that one cannot invert the roles of these two contributions, again because the composite bifundamental is generically full rank. Reproducing the observed pattern would then require a significant hierarchy between $\eps_{q,u}$ and $\eps_\Delta$, running against the basic motivation of the construction.

\subsection{UV completion}
\label{sec:UV}

In this subsection, we discuss how the EFT structures identified in the models above can be realized from renormalizable UV completions.

\paragraph{$\bm{U(2)_{q+\ell}}$.} To illustrate the general logic, we start by focusing on the $U(2)_{q+\ell}$ model of \cref{sec:U2ql}.
In a weakly coupled theory, at tree level, there are essentially two ways to build such a UV completion: one based purely on chains of VLFs, and another involving both VLFs and additional scalars.

An example completion using only VLF chains requires, in the quark sector, the introduction of three partners of $q,\ell$:
\begin{align}
     Q_{0}, L_0 \sim \bm{1}_0, \qquad  Q_{j_\Ss}, L_{j_\Ss} \sim (\bm{2j_\Ss+1})_{1}, \qquad  Q_{J}, L_J \sim (\bm{2J+1})_{2}
     \label{eq:U2qlVLF}
\end{align}
with (spin decomposition) 
\begin{align}
     J \equiv j_\Ss \pm \frac{1}{2}.
\end{align}
Here, the subscript on $Q,L$ denotes the $SU(2)$ flavor spin, while the subscript appended to the dimension of the $SU(2)$ representation denotes the $U(1)$ charge. These VLFs are all assumed to live at the common mass scale $ M_{q+\ell}$.

\begin{figure}[!htb]
  \centering

\begin{tikzpicture}
\begin{feynman}
\def\xscale{0.5} 
\def\yscale{0.5} 
\vertex (i0) at (0*\xscale, 0*\yscale) {\(\bar q\)};
\vertex (i1) at (3*\xscale, 4*\yscale) {\(\bar \Ss\)};
\vertex (i2) at (7*\xscale, 4*\yscale) {\( \Ss\)};
\vertex (i3) at (11*\xscale, 4*\yscale)  {\(\Ss\)};
\vertex (i4) at (15*\xscale, 4*\yscale)  {\(H\)};
\vertex (i5) at (18*\xscale, 0*\yscale)  {\(d\)};
\vertex (i6) at (3*\xscale, 0*\yscale);
\vertex (i7) at (7*\xscale, 0*\yscale);
\vertex (i8) at (11*\xscale, 0*\yscale);
\vertex (i9) at (15*\xscale, 0*\yscale);
\diagram* {
(i0) -- [plain, thick] (i6) -- [plain, edge label = \(Q_{J}\), thick] (i7) -- [plain, edge label = \(Q_{j_\Ss}\), thick] (i8) -- [plain, edge label = \(Q_0\), thick] (i9) -- [plain, thick] (i5),
(i6) -- [scalar, thick] (i1),
(i7) -- [scalar, thick] (i2),
(i8) -- [scalar, thick] (i3),
(i9) -- [scalar, thick] (i4)
};
\end{feynman}
\end{tikzpicture}

\begin{tikzpicture}
\begin{feynman}
\def\xscale{0.5} 
\def\yscale{0.5} 
\vertex (i0) at (0*\xscale, 0*\yscale) {\(\bar q\)};
\vertex (i1) at (3*\xscale, 4*\yscale) {\(\bar \Ss\)};
\vertex (i2) at (7*\xscale, 4*\yscale) {\( \Ss\)};
\vertex (i3) at (11*\xscale, 4*\yscale)  {\(\bar \Ss\)};
\vertex (i4) at (15*\xscale, 4*\yscale)  {\( \Ss\)};
\vertex (i6) at (3*\xscale, 0*\yscale);
\vertex (i7) at (7*\xscale, 0*\yscale);
\vertex (i8) at (11*\xscale, 0*\yscale);
\vertex (i9) at (15*\xscale, 0*\yscale);
\vertex (i10) at (19*\xscale, 0*\yscale);
\vertex (i10u) at (19*\xscale, 4*\yscale)  {\(\Ss\)};
\vertex (i11) at (23*\xscale, 0*\yscale);
\vertex (i12) at (27*\xscale, 0*\yscale) {\(d\)};
\vertex (i13) at (23*\xscale, 4*\yscale)  {\(H\)};
\diagram* {
(i0) -- [plain, thick] (i6) -- [plain, edge label = \(Q_{J}\), thick] (i7) -- [plain, edge label = \(Q_{j_\Ss}\), thick] (i8) -- [plain, edge label = \(Q_{J}\), thick] (i9) -- [plain, edge label = \(Q_{j_\Ss}\), thick] (i10) -- [plain, edge label = \(Q_{0}\), thick] (i11) -- [plain, thick] (i12),
(i6) -- [scalar, thick] (i1),
(i7) -- [scalar, thick] (i2),
(i8) -- [scalar, thick] (i3),
(i9) -- [scalar, thick] (i4),
(i10) -- [scalar, thick] (i10u),
(i11) -- [scalar, thick] (i13)
};
\end{feynman}
\end{tikzpicture}

\caption{Yukawa generation at LO and NLO through VLF chains in the $U(2)_{q+\ell}$ model.}
\label{fig:UVql}
\end{figure}

With these fields, the chains generating the leading and next-to-leading spurion insertions of \cref{eq:U2qlLagr} are shown in \cref{fig:UVql} for the down-type Yukawa sector; the up quark and lepton sectors are entirely analogous. Following the chain, the effective doublets are explicitly built as
\begin{align}
\begin{aligned}
    (\Ss^2 \bar \Ss)_{1/2}&: [(\Ss \otimes   \Ss)_{J} \otimes \bar \Ss ]_{1/2} \\
    (\Ss^3 \bar \Ss^2)_{1/2}&: 
    [ ( [(\Ss \otimes \Ss)_{J} \otimes \bar \Ss ]_{j_\Ss} \otimes  \Ss)_{J} \otimes \bar  \Ss]_{1/2}
    \label{eq:Y3Y5UV}
\end{aligned}
\end{align}
where the nested products involve intermediate representations that are tied to the VLF associated with the propagation.
A few important points should be emphasized:
\begin{itemize}
    \item The VLF with flavor spin-$0$, $j_\Ss$, $J$ form the minimal set of flavor representations needed to connect the left-handed SM fermions ($SU(2)$ doublets) to the right-handed ones ($SU(2)$ singlets). In particular, since $(\Ss \otimes \Ss)_J$ is not identically null only for odd $J$ by the bosonic nature of $\Ss$ (see \cref{app:countingIrreps}), it follows that among $J = j_\Ss \pm 1/2$ only
    \begin{align}
        J = j_\Ss + \frac{1}{2} (-1)^{j_\Ss - 1/2}
    \end{align}
    is viable, namely the odd $J$. For example, for $j_\Ss =3/2$ it is $J=1$, for $j_\Ss = 5/2$ it is $J=3$, and so on.

    \item The same set of VLFs is also sufficient to generate the NLO chain and render
    the Yukawa matrix full rank. No additional mediators are required beyond
    those already present at LO.

    This is a priori not obvious in a weakly coupled UV completion. The NLO
    correction lifts the rank only if the mediator chain generates a
    spin-$1/2$ composite doublet that is genuinely independent of the LO one.
    If the chain simply repeats the singlet mediator $Q_0$, for example, this does not happen. In that case, flavor is not propagated through new non-trivial $SU(2)$ contractions, and the NLO effective doublet is exactly aligned with the LO one, up to an overall factor. The light $2\times 2$ Yukawa block then takes the form
    \begin{align}
        Y_{d,\alpha i}^{2\times 2} \sim 
        \frac{((\Ss^2 \bar \Ss)_{1/2})_\alpha }{M_{q+\ell}^3} \left(a + b \eps^2+\dots\right) y_i^{d}
    \end{align}
    with $a,b$ given by products of $\mathcal{O}(1)$ couplings. This remains manifestly rank-1 at every order.
    
    By contrast, repeating the non-trivial mediators $Q_{j_\Ss}$ and $Q_{J}$ opens up additional spin-$1/2$ structures at NLO. In this case $(\Ss^3 \bar \Ss^2)_{1/2}$ is not forced to be aligned with the LO doublet, and the rank can be lifted.
    A check of this can be done via recoupling identities. The LO and NLO chains of \cref{eq:Y3Y5UV} can be rewritten as
    \begin{gather}
    \begin{aligned}
        (\Ss^2 \bar \Ss)_{1/2} &= [(\Ss \otimes  \Ss)_J \otimes \bar \Ss ]_{1/2},\\
        (\Ss^3 \bar \Ss^2)_{1/2} &=
        \sum_{K,L}
        C_{KL}\,
        \left[
        (\Ss\otimes \bar \Ss)_K
        \otimes
        \bigl((\Ss\otimes \Ss)_J\otimes \bar \Ss\bigr)_L
        \right]_{1/2},
    \end{aligned} \nonumber \\
        C_{KL}
        =
        (-1)^{J+3j_\Ss+1/2}\,
        d_{j_\Ss} d_K d_J d_L
        \begin{Bmatrix}
        J & j_\Ss & j_\Ss\\
        j_\Ss & J & K
        \end{Bmatrix}
        \begin{Bmatrix}
        K & J & J\\
        j_\Ss & \frac12 & L
        \end{Bmatrix},
    \end{gather}
    where $d_{j_\Ss}\equiv \sqrt{2j_\Ss+1}$, etc., and the parentheses denote Wigner $6j$ symbols.
    From these expressions, one sees immediately that the $K=0$ contribution in the NLO chain is aligned with the LO contraction, up to an overall coefficient $\sim |\Ss|^2$. Therefore, LO and NLO can be independent only if some $K\neq 0$ term survives, namely if $C^{KL}\neq 0$ for some $K \neq 0$. In that case, the NLO doublet contains additional structures with coefficients that differ from those of the LO one; the two are generically independent.
    We find that this condition is typically satisfied. For example, for $j_\Ss=3/2$, in which $J=1$, $j_\Ss =5/2$, in which $J=3$, and $j_\Ss =7/2$, in which again $J=3$, the NLO doublet is not aligned with the LO one.

    \item The VLF must come in at least two internal flavors. This can be understood from the fact that, if they carried no internal flavor structure, the light $2\times 2$ Yukawa block would read
    \begin{align}
    Y_{f,\alpha i}^{2\times 2} \sim 
    \Bigr (\frac{(\Ss^2 \bar \Ss)_{1/2}}{M_{q+\ell}^3}+ \frac{(\Ss^3 \bar \Ss^2)_{1/2}}{M_{q+\ell}^5} + \dots  \Bigr )_\alpha y_i^{d}  ,
    \end{align}
    at any order, where $y_i^d$ is the coupling appearing in the vertex $y_i^d \,\bar Q_0 H d_i$. This would again remain rank-1 to all orders. If instead the VLF carries a non-trivial flavor structure, one obtains
    \begin{align}
    Y_{f,\alpha i}^{2\times 2} \sim 
    \Bigr ( \frac{(\Ss^2 \bar \Ss)_{1/2}}{M_{q+\ell}^3} \Bigr )_\alpha y_i ^d  + \Bigr ( \frac{ (\Ss^3 \bar \Ss^2)_{1/2}}{M_{q+\ell}^5}  \Bigr )_\alpha z_i ^d  + \dots
    \end{align}
    where $y_i^d$ and $z_i^d$ are now genuinely independent, thanks to the non-trivial internal flavor contractions of the vertices entering the diagram.
\end{itemize}

The last point also makes explicit why this model strictly requires the $U(1)\subset U(2)_{q+\ell}$ assumption, or at least a $Z_2$ symmetry, as discussed around \cref{eq:Z2yuk}. Since $(\Ss^2 \bar \Ss)_{1/2}$ and $(\Ss \bar \Ss^2)_{1/2}$ are independent two-dimensional vectors, and the VLF must come in two or more flavors, then $y_i^f$ and $\bar y_i^f$ are independent vectors, obtained by different VLF flavor contractions. Dropping the $U(1)$ would generically make the light Yukawa block rank-2 already at LO, unless the $Z_2$ is imposed.

The need for a $U(1)$ or $Z_2$ makes the model somewhat less economical, especially given the aim of avoiding the introduction of global symmetries by gauging the flavor group. The $U(1)$ factor has mixed anomalies with $SU(2)$ and the SM gauge group. We have checked that it is possible to gauge it by introducing suitable anomalons, but this, in turn, requires an ad hoc mechanism to generate their masses and, importantly, to allow them to decay, since some of them must be charged under the SM group. The proposed $Z_2$ acting solely on $\Ss$, on the other hand, is part of the $\mathcal{O}(4j_\Ss+2)$ symmetry left unbroken by the kinetic term of $\Ss$. It could therefore arise accidentally from a larger scale 
UV completion, but this would still require an additional layer of explanation.

An alternative completion that avoids the need for any additional $U(1)$ or $Z_2$ can be constructed. In this case, the flavor symmetry is simply taken to be $SU(2)_{q+\ell}$, which is anomaly-free and can be gauged directly. Relative to the field content of \cref{eq:U2qlVLF}, this requires the addition of a scalar with flavor spin-$j_\Delta$ and associated VLFs
\begin{align}
\Delta \sim \left (\bm{1}_c, \bm{2}_{\rm L}, - \left(1/2 \right)_{\rm Y}, \bm{2 j_\Delta +1 }\right), \qquad Q_{j_\Delta}, L_{j_\Delta} \sim (\bm{2j_\Delta +1} ).
\end{align}
The scalar is charged under the SM gauge group in such a way as to allow a quartic coupling with one SM Higgs. The diagram generating the LO doublet remains the same as in \cref{fig:UVql}, while the next-to-leading doublet is now generated through the diagram shown in \cref{fig:UVscalarNLO}.

\begin{figure}[!htb]
  \centering

\begin{tikzpicture}
\begin{feynman}
\def\xscale{0.5} 
\def\yscale{0.5} 
\vertex (i0) at (0*\xscale, 0*\yscale) {\(\bar q\)};
\vertex (i1) at (3*\xscale, 4*\yscale) {\(\Ss\)};
\vertex (i2) at (7*\xscale, 4*\yscale) {\(\Ss\)};
\vertex (i3) at (11*\xscale, 4*\yscale)  {\(\Ss\)};
\vertex (i4) at (15*\xscale, 4*\yscale);
\vertex (i4uL) at (12.5*\xscale, 6*\yscale)  {\( \Ss\)};
\vertex (i4uC) at (15*\xscale, 6*\yscale)  {\( \Ss\)};
\vertex (i4uR) at (17.5*\xscale, 6*\yscale)  {\( H\)};
\vertex (i6) at (3*\xscale, 0*\yscale);
\vertex (i7) at (7*\xscale, 0*\yscale);
\vertex (i8) at (11*\xscale, 0*\yscale);
\vertex (i9) at (15*\xscale, 0*\yscale);
\vertex (i10) at (19*\xscale, 0*\yscale) {\(d\)};
\diagram* {
(i0) -- [plain, thick] (i6) -- [plain, edge label = \(Q_{J}\), thick] (i7) -- [plain, edge label = \(Q_{j_\Ss}\), thick] (i8) -- [plain, edge label = \(Q_{j_\Delta}\), thick] (i9) -- [plain, thick] (i10),
(i6) -- [scalar, thick] (i1),
(i7) -- [scalar, thick] (i2),
(i8) -- [scalar, thick] (i3),
(i9) -- [scalar, thick, edge label = \(\Delta\)] (i4),
(i4) -- [scalar, thick] (i4uL),
(i4) -- [scalar, thick] (i4uC),
(i4) -- [scalar, thick] (i4uR),
};
\end{feynman}
\end{tikzpicture}

\caption{Down Yukawa generation at NLO for the VLF+scalar UV completion.}
\label{fig:UVscalarNLO}
\end{figure}

In this setup, the couplings $y^d_i \bar Q_0 H d_i$ and $z_i^d \bar Q_{j_\Delta} \Delta d_i$ involve two independent vectors, $y_i^d$ and $z_i^d$. As a result, already a single VLF family is sufficient to generate independent LO and NLO contributions to the fermion masses. At the same time, precisely because only one VLF family is present, the leading contribution automatically has the structure of \cref{eq:Z2yuk}, effectively making the required alignment accidental at LO. Hence, the model reproduces the desired flavor structure without introducing symmetries beyond the gauged $SU(2)_{q+\ell}$.  This, however, requires the important structural condition 
\begin{align}
   |j_\Delta -j_\Ss |> 1/2
\end{align}
which avoids the insertion of $\Delta$ in the LO chain that would lift the rank too soon.

The tradeoff of this model is the presence of an additional elementary scalar, charged under the flavor group and with a mass comparable to that of the VLF, $M_\Delta \sim M_{q+\ell}$. Such a structure could perhaps be motivated by a more complete UV framework, for instance, supersymmetry in which $\Delta, L_{j_\Delta}$ may form a supermultiplet. A fully consistent embedding that does not spoil the flavor structure appears, however, to be non-trivial, and we leave it for future work.

\paragraph{Left-right models.}
Independent of the specific model, the generation of Yukawa structures in which only one of the two fermion fields is charged under a flavor factor proceeds through diagrams and arguments completely analogous to those discussed above. A qualitative difference arises only when both fermion chiralities require spurion insertions, so that the Yukawa block is generated by a genuinely new left-right chain. This occurs in the models of \cref{sec:LRmodel}, \cref{sec:10model}, and \cref{sec:U25model}.

To illustrate the structure, let us focus on the light up-quark Yukawa block in the $U(2)_{q+e}\times U(2)_u \times G$ models. The LO diagram involving only VLF chains is shown in \cref{fig:UVqeu}. The chain is similar to that of \cref{fig:UVql}: for each SM fermion appearing in the diagram, one minimally needs the three VLF representations \cref{eq:U2qlVLF} required to connect that fermion to the $SU(2)$ neutral sector. The main difference is that the Higgs vertex now couples two $SU(2)$-neutral VLFs to each other, rather than coupling directly to one SM fermion chirality. The NLO diagram is obtained as in \cref{fig:UVql}, namely by repeating a non-trivial VLF representation along the chain.

\begin{figure}[!htb]
  \centering

\begin{tikzpicture}
\begin{feynman}
\def\xscale{0.5} 
\def\yscale{0.5} 
\vertex (i0) at (0*\xscale, 0*\yscale) {\(\bar q\)};
\vertex (i1) at (3*\xscale, 4*\yscale) {\(\bar \Ss_{q+e}\)};
\vertex (i2) at (7*\xscale, 4*\yscale) {\( \Ss_{q+e}\)};
\vertex (i3) at (11*\xscale, 4*\yscale)  {\(\Ss_{q+e}\)};
\vertex (i4) at (15*\xscale, 6*\yscale)  {\(H\)};
\vertex (i6) at (3*\xscale, 0*\yscale);
\vertex (i7) at (7*\xscale, 0*\yscale);
\vertex (i8) at (11*\xscale, 0*\yscale);
\vertex (i9) at (15*\xscale, 0*\yscale);
\vertex (i5) at (19*\xscale, 0*\yscale);
\vertex (r2) at (23*\xscale, 0*\yscale);
\vertex (r3) at (27*\xscale, 0*\yscale);
\vertex (r4) at (30*\xscale, 0*\yscale) {\(u\)};
\vertex (u5) at (19*\xscale, 4*\yscale) {$\Ss_{u}$};
\vertex (u6) at (23*\xscale, 4*\yscale) {$\Ss_{u}$};
\vertex (u7) at (27*\xscale, 4*\yscale) {$\bar \Ss_{u}$};

\diagram* {
(i0) -- [plain, thick] (i6) -- [plain, edge label = \(Q_{J_{q+e}}\), thick] (i7) -- [plain, edge label = \(Q_{j_{\Ss_{q+e}}}\), thick] (i8) -- [plain, edge label = \(Q_0\), thick] (i9),
(i6) -- [scalar, thick] (i1),
(i7) -- [scalar, thick] (i2),
(i8) -- [scalar, thick] (i3),
(i9) -- [scalar, thick] (i4),
(i9) -- [plain, thick,  edge label = \(U_0\)] (i5) -- [plain, thick,  edge label = \(U_{j_{\Ss_u}}\)] (r2) -- [plain, thick,  edge label = \(U_{J_{u}}\)] (r3)  -- [plain, thick] (r4), 
(i5) -- [scalar, thick] (u5),
(r2) -- [scalar, thick] (u6),
(r3) -- [scalar, thick] (u7),
};
\end{feynman}
\end{tikzpicture}

\caption{Generation of the light $Y_u$ block at LO through VLF chains in the $U(2)_{q+e} \times U(2)_u \times G$ models.}
\label{fig:UVqeu}
\end{figure}

There are a few interesting points to mention:
\begin{itemize}
    \item A single flavor copy for each VLF is sufficient in this case. The LO and NLO doublets generated by the $\Ss_{q+e}$ and $\Ss_u$ spurions on the left and right sides of the chain are independent, and therefore generate a full-rank light Yukawa block. The only subtlety arises when several independent composites at the same order are required. This happens, for example, for the leading composite triplets in the $U(2)_{q+e}\times U(2)_u\times SU(3)_5$ model of \cref{sec:U3_5}. In that case, one needs more than the minimal three VLF representations in order to generate the required independent composites.

    \item The VLF chains are rather long, involving up to eleven Yukawa vertices in the NLO diagram responsible for the $\mathcal{O}(\eps_{q+e}^5\eps_u^5)$ suppression. This has an interesting consequence. The residual $\mathcal{O}(1)$ coefficients in the SM Yukawas, fixed by the fit and typically spanning about one order of magnitude,\footnote{In the fits above, the typical range of these coefficients is $[0.1,1]$.} can be generated by much smaller variations among the UV couplings. These variations are amplified by the long products of couplings along the chains, potentially becoming an integral part of the hierarchy generation mechanism~\cite{Greljo:2024zrj}. At the same time, this means that the predictions of these models can be sensitive to the assumed distribution of UV couplings.

    \item The presence of a large number of VLFs in several representations may induce early Landau poles for the SM gauge couplings or for the gauged flavor group. This issue is clearly UV-model dependent and should be checked carefully in each explicit realization.
\end{itemize}

\section{Scalar potential and vacuum alignment}
\label{sec:pot}

A crucial ingredient of the models in \cref{sec:models}, and more generally
of the framework introduced in \cref{sec:framework}, is the appearance of
composite $SU(2)$ doublets built from products of a higher-spin spurion
$\Ss$. At the EFT level, $\Ss$ simply parametrizes the breaking of the
flavor symmetry and may arise from the vacuum expectation value of an
elementary scalar field or a composite operator, or even represent the
first allowed explicit symmetry-breaking coupling. In the following, we
study the dynamical realization in which the spurion is generated
by the vacuum expectation value of a scalar field transforming in the
corresponding representation.\footnote{With a slight abuse of notation, we
denote by $\Ss$ both the scalar field and its vacuum expectation value.}

The flavor construction requires this vacuum to be sufficiently generic, so that the spin-$1/2$ composites allowed by representation theory are non-vanishing and independent at successive orders.
We assess the viability of this assumption by first studying the most general renormalizable potential for $\Ss$. The vacuum structure of higher-dimensional $SU(2)$ representations has been investigated previously \cite{Kim:1983mc, Jetzer:1983ij, Hisano:2013sn,Heikinheimo:2017nth,Moultaka:2020dmb,Brummer:2023znr,Jurciukonis:2024bzx,Milagre:2025vcx,Jurciukonis:2025cyy}. We provide a new perspective on this problem by introducing a geometrical description of residual vacuum symmetries and of the corresponding selection rules for composite representations, with the details collected in \cref{app:vacuumSymm}. We then apply these tools to study the two simplest cases relevant to our construction, $j_\Ss=3/2$ and $j_\Ss=5/2$, with results in \cref{sec:treeVacum}.

As we will see, generic renormalizable tree-level potentials tend to select vacua with residual symmetries, which can force the required spin-$1/2$ composites to vanish or become aligned.\footnote{Higher-dimensional representations may also lead to accidental massless moduli. The mechanisms discussed in \cref{sec:additonalContrib} address this issue as well \cite{Brummer:2023znr}.} As a result, obtaining the desired vacuum typically requires going beyond the most naive tree-level analysis and calls for additional steps in the construction, which we analyze in \cref{sec:additonalContrib}. We anticipate that a similar analysis could be carried out for composite $SU(3)$ triplets, employed briefly in \cref{sec:U3_5}. We leave this direction for future work.

\subsection{Leading order vacuum structure}
\label{sec:treeVacum}

Let us consider a scalar field $\Ss$ with half-integer $SU(2)$ spin $j_\Ss$. We separate its radial and angular components as
\begin{align}
    \Ss \equiv |\Ss|\,n,
    \qquad
    n^\dagger n=1.
\end{align}
The quadratic invariant is unique, $(\Ss^\dagger \Ss)=|\Ss|^2$, while no cubic invariant exists. The $U(2)$-invariant quartics can be classified through the quantities
$B_k\equiv |(n\otimes n)_k|^2$. Since $\Ss$ is bosonic, only odd spins
$k=1,3,\ldots,2j_\Ss$ occur, giving $j_\Ss+1/2$ quartic invariants. Moreover,
$\sum_k B_k=1$, so only $j_\Ss-1/2$ combinations contribute non-trivially to the angular dependence.
For the minimization, it is more convenient to use the equivalent basis
\begin{align}
    S_k\equiv |(n\otimes\bar n)_k|^2,
    \qquad
    k=0,1,\ldots,j_\Ss-\frac12,
\end{align}
whose relation to the $B_k$ basis is discussed in \cref{app:potentialBases}. The most general $U(2)$-invariant renormalizable potential can then be written as
\begin{align}
\begin{aligned}
    V_{\rm tree}(\Ss)
    &=
    -m^2|\Ss|^2+|\Ss|^4\lambda_{\rm eff}(n),\\
    \lambda_{\rm eff}(n)
    &=
    \sum_{k=0}^{j_\Ss-1/2}\lambda_k S_k .
\end{aligned}
\label{eq:TLpotential}
\end{align}
The radial minimization gives
\begin{align}
    |\Ss|^2=\frac{m^2}{2\lambda_{\rm eff}(n)},
\end{align}
so the remaining problem is to determine the direction $n$ that minimizes the effective quartic coupling. The invariant
$S_0=1/(2j_\Ss+1)$ is fixed by the normalization of $n$ and therefore does not affect the angular minimization, although its coefficient remains relevant for the boundedness of the potential. The remaining $S_k$ are proportional to the squared rank-$k$ multipole moments of the spin-$j_\Ss$ state $n$, $ 
    S_k
    \propto
    \left|
    \left\langle n\left|
    \left(T^{(a_1}\cdots T^{a_k)}\right)_{\rm traceless}
    \right|n\right\rangle
    \right|^2$,
which provides a useful physical and geometrical interpretation of the variables entering the minimization.

If the symmetry of the potential is reduced from $U(2)$ to $SU(2)$, two additional classes of quartic invariants are allowed,
\begin{align}
    B_{U,i}\equiv(n^3\bar n)_{0,i},
    \qquad
    F_{U,i}\equiv(n^4)_{0,i},
\end{align}
with $\lfloor(2j_\Ss+3)/6\rfloor$ independent invariants of each type, as discussed in \cref{app:countingIrreps}. The effective quartic then generalizes to
\begin{align}
    \lambda_{\rm eff}(n)
    =
    \sum_k\lambda_k S_k
    +
    \frac12\sum_i
    \left(
        \lambda_{B_{U,i}}B_{U,i}
        +
        \lambda_{F_{U,i}}F_{U,i}
        +
        \text{h.c.}
    \right).
    \label{eq:potentialSU2}
\end{align}

The normalized invariants $S_k$, $B_{U,i}$, and $F_{U,i}$ define bounded orbit variables \cite{Abud:1981tf,Abud:1983id}. Since the renormalizable angular potential is linear in these variables, minimizing $\lambda_{\rm eff}(n)$ amounts to minimizing a linear function on the corresponding compact orbit space, and the minimum is therefore reached on its boundary. Boundary points often belong to singular strata with enhanced residual symmetry, in line with the Michel--Radicati conjecture \cite{Michel:1971th,Michel:121951}. In the full orbit space, these strata are characterized by a deficient rank of the Jacobian of the invariant map \cite{Cabibbo:1970rza,Procesi:1985hr,Talamini:2006wd,Feruglio:2019ybq}. For the projected orbit spaces relevant to truncated renormalizable potentials, however, some boundary regions may instead correspond to less symmetric configurations \cite{Kim:1983mc,Kim:1997wf,Kim:1998qy}.

In our context, residual symmetries are potentially dangerous. They are encoded in the stabilizer of a vacuum direction $n$, namely the subgroup
\begin{align}
    H_n
    \equiv
    \left\{
        h\in SU(2)
        \,\middle|\,
        D^{(j_\Ss)}(h)\,n=e^{i\alpha_h}n
    \right\}.
\end{align}
The non-trivial stabilizers relevant for half-integer $j_\Ss$ are a continuous $U(1)_{T^3}\subset SU(2)$ or cyclic and dihedral subgroups $C_p\simeq Z_p$ and $D_p$. For a $U(2)$ model, $H_n$ corresponds to a genuine residual symmetry, since the phase $e^{i\alpha_h}$ can be compensated by an overall $U(1)$ transformation. Even in the $SU(2)$ case, however, the stabilizer imposes selection rules on composite representations. These can force the leading composite doublet
\begin{align}
    v_3\equiv(n^2\bar n)_{1/2}
\end{align}
to vanish, or constrain composite doublets arising at different orders to be aligned, for example $v_5=(n^3\bar n^2)_{1/2}\propto v_3$. Their derivation, together with a geometrical characterization of the stabilizers through the Majorana representation, is presented in \cref{app:vacuumSymm}. We now apply the resulting logic to the two simplest cases relevant to our construction, $j_\Ss=3/2$ and $j_\Ss=5/2$.

\paragraph{$\bm{j_\Ss=3/2}$.}
The case $j_\Ss=3/2$ is the simplest non-trivial example. In the
$U(2)$ case, the angular potential depends only on the dipole invariant
$S_1$,
\begin{align}
    \lambda_{\rm eff} (n)
    =
    \frac{1}{4}\lambda_0+\lambda_1 S_1 (n),
    \qquad
    0\leq S_1\leq \frac{9}{20}.
\end{align}
The minimum is therefore attained at one of the two endpoints:
\begin{align}
\begin{alignedat}{3}
    \lambda_1<0:\qquad
    & n=\ket{3/2,\pm3/2},
    &\qquad
    & S_1=\frac{9}{20},
    &\qquad
    & H_n=U(1)_{3/2},
    \\[2mm]
    \lambda_1>0:\qquad
    & n=\frac{1}{\sqrt{2}}
    \left(
        \ket{3/2,-3/2}
        +
        \ket{3/2,3/2}
    \right),
    &\qquad
    & S_1=0,
    &\qquad
    & H_n=D_3 .
\end{alignedat}
\label{eq:j32U2minima}
\end{align}
The first configuration maximizes the dipole and preserves a continuous
$U(1)$ subgroup of rotations around a fixed axis, which we have chosen to
be the $T^3$ axis, with any other choice being $SU(2)$-equivalent. The
second has vanishing dipole and preserves the dihedral symmetry $D_3$ of
an equilateral triangle.\footnote{The $D_3$ minimum does not correspond to
a maximal subgroup, providing a counterexample to the Michel--Radicati
conjecture \cite{Kim:1983mc}.}
In both cases, the residual symmetry forces the spin-$1/2$ composites
required by our construction to vanish, as show in \cref{app:genericSU2_3/2}. The renormalizable $U(2)$-invariant
potential therefore does not yield a viable vacuum for $j_\Ss=3/2$.

Allowing the additional quartics present in the general $SU(2)$-invariant
potential does not change this conclusion. As shown in
\cref{app:genericSU2_3/2}, its global minimum is generically
attained on a family of vacua preserving a discrete $C_3$ symmetry, for
which the leading composite doublet $v_3$ vanishes. Non-symmetric vacua
with $v_3\neq0$ can occur only at tuned or degenerate points in the
quartic-coupling space. We therefore conclude that neither the $U(2)$ nor
the $SU(2)$-invariant renormalizable potential generically produces a viable $j_\Ss=3/2$ vacuum.

\begin{figure}[!htb]
    \centering
    \includegraphics[width=1\linewidth]{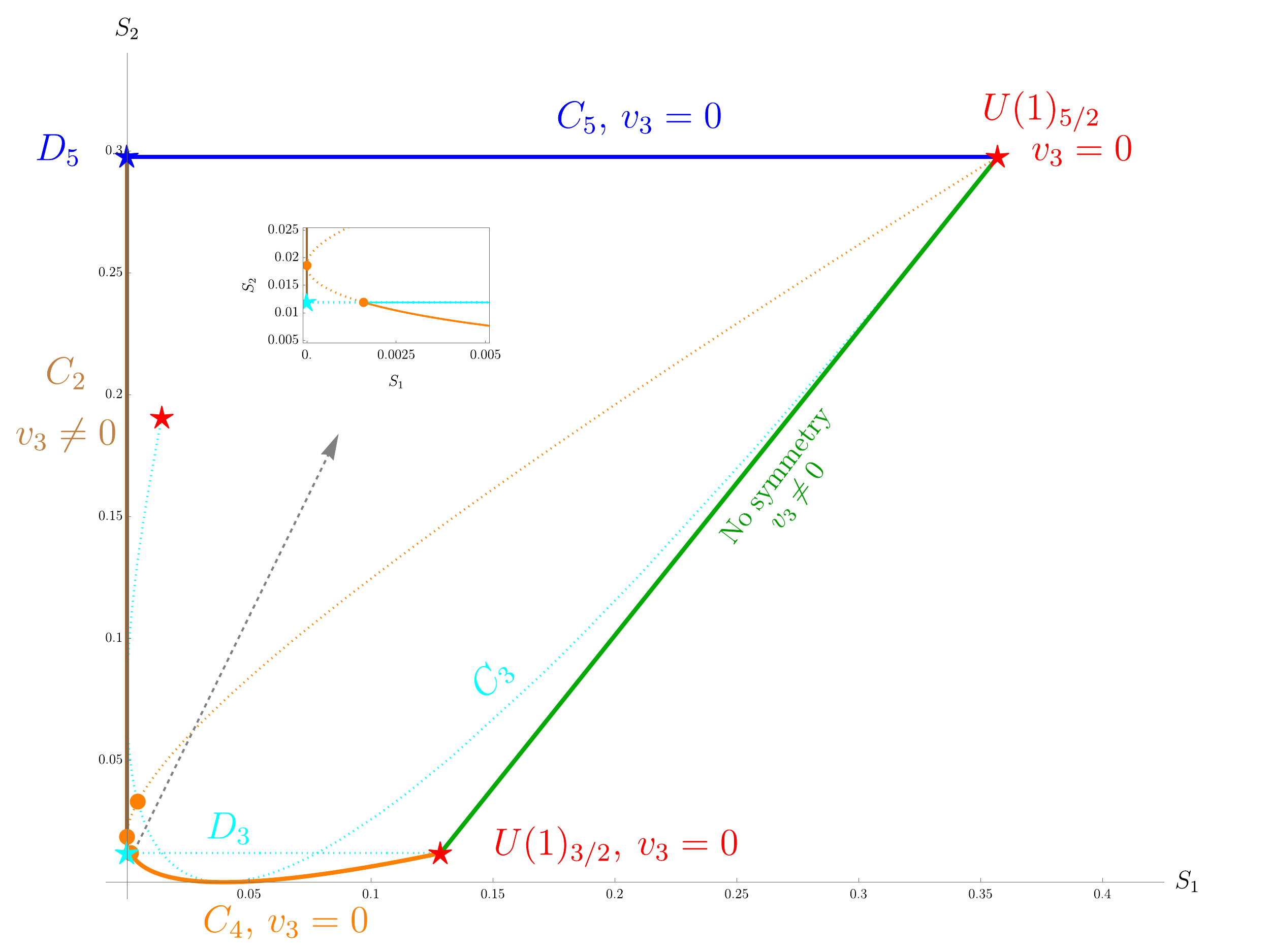}
    \caption{Boundary of the orbit space for $j_\Ss=5/2$ in the $U(2)$-invariant case. Each boundary is labeled by its stabilizer and by whether the spin-$1/2$ composite $v_3$ is identically vanishing. The inset highlights the non-trivial crossing near the $D_3$ limit, while the dotted lines denote symmetric configurations lying in the interior of the orbit space.}
    \label{fig:5/2_domain}
\end{figure}

\paragraph{$\bm{j_\Ss = 5/2}$.}
We now turn to $j_\Ss=5/2$. In the $U(2)$ case, the angular potential depends on the dipole and quadrupole invariants,
\begin{align}
    \lambda_{\rm eff} (n)
    =
    \frac{1}{6}\lambda_0+\lambda_1 S_1 (n)+\lambda_2 S_2(n),
\end{align}
so the orbit space is two-dimensional and can be parametrized by $(S_1,S_2)$. Its boundary, shown in \cref{fig:5/2_domain}, is considerably richer than in the $j_\Ss=3/2$ case and is formed by several branches associated with symmetric configurations. The explicit representative vacua are collected in \cref{app:genericSU2_5/2}, while here we only emphasize the crucial point: most of these branches correspond to vacua for which the leading composite doublet $v_3\equiv(n^2\bar n)_{1/2}$ vanishes. In particular, the most protruding and convex parts of the orbit space are associated with $U(1)_{5/2}$, $D_5$, and $C_4$ configurations, all of which have $v_3=0$. 

Since equipotential surfaces are straight lines in the $(S_1,S_2)$ plane, a generic choice of quartic couplings selects precisely these branches. The only branches along which one can obtain $v_3\neq0$ are either a non-symmetric branch or a $C_2$ branch, and they are selected only for the tuned relations
\begin{align}
\begin{alignedat}{3}
    \lambda_1+\frac{5}{4}\lambda_2
    &=0,
    &\qquad
    \lambda_2
    &>0,
    &\qquad
    &(v_3\neq 0,\ \text{no stabilizer}),
    \\
    \lambda_2
    &=0,
    &\qquad
    \lambda_1
    &\geq 0,
    &\qquad
    &(v_3\neq 0,\ C_2\text{ stabilizer}) .
\end{alignedat}
\end{align}
At this level, any departure from either of these tuned relations tilts the equipotential line toward one of the branches with $v_3=0$. Furthermore, even in the tuned cases, the corresponding vacua are degenerate with other configurations for which $v_3=0$. The renormalizable $U(2)$-invariant potential therefore does not generically yield a viable vacuum for $j_\Ss=5/2$ either.

Allowing the additional quartics present in the general $SU(2)$-invariant potential does not qualitatively change this picture. The corresponding orbit space is five-dimensional, or four-dimensional in the $Z_2$/CP-invariant limit, and is therefore no longer easily visualized. We have studied this case numerically, and we find that the minimum is again always attained on symmetric configurations, in particular on $C_5$ and $C_4$ vacua, for which $v_3=0$. We therefore conclude that also for $j_\Ss=5/2$, neither the $U(2)$- nor the $SU(2)$-invariant renormalizable potential generically produces a viable vacuum.

\subsection{Additional contributions and Coleman-Weinberg potential}
\label{sec:additonalContrib}

From the two explicit examples discussed above, we conclude that the renormalizable potential is very unlikely to select vacuum configurations with non-vanishing spin-$1/2$ composites.\footnote{We also tested numerically the case $j_\Ss=7/2$ for some representative quartics and, unsurprisingly, always found $v_3=0$.} One may in fact suspect that larger values of $j_\Ss$ are even less likely to yield viable vacua, since the number of possible stabilizers increases, while the number of configurations within those stabilizers that admit non-vanishing spin-$1/2$ composites grows only very slowly.\footnote{A simple counting argument shows that among $C_n$ vacua, the ratio of viable to possible configurations scales as $1/j_\Ss$.}

Hence, rather than relying on the hope that some higher value of $j_\Ss$ will accidentally lead to such a vacuum, we discuss robust mechanisms that can guarantee their presence. The mechanisms are essentially three.
\begin{enumerate}

    \item \textbf{Assisted symmetry breaking}. A possibility is to let an additional scalar field, charged under the same flavor symmetry, participate in the breaking. Enlarging the scalar sector can remove the accidental residual symmetries that often arise in the single field case, and can therefore make fully generic vacua much easier to obtain \cite{Espinosa:2012uu, deMedeirosVarzielas:2025byb}. In the EFT setup, however, this option is typically more delicate: unless additional selection rules are imposed, the assisting scalar will itself enter the effective spurion operators and can therefore modify the Yukawa hierarchies. In practice, preserving the desired spurion pattern usually requires extra structure from the UV mediator sector, or more in general a construction in which the assisting field participates in the vacuum alignment but is prevented from entering the mediator chains that generate the effective Yukawas. Assisted breaking is therefore viable, but more model-dependent and UV-sensitive.
    
    \item \textbf{Non-renormalizable operators}. Another possibility is to allow non-renormalizable operators to participate actively in vacuum selection. These operators can tilt the vacuum away from the symmetric tree-level configurations, inducing a small breaking of the stabilizer present at that level. 
    As an example, consider the dimension six operator
    $
    O_6 = |\Ss|^6 |v_3|^2 =|(\Ss^2 \bar \Ss)_{1/2}|^2
    $
    with Wilson coefficient $C_6 \sim c_6/M_6^2 <0$. In the minimization, this operator effectively acts as a Lagrange multiplier favoring non-vanishing values of $v_3$ and the other spin-1/2 composites. If $M_6 \sim M$, the induced size is parametrically
    $
    |v_3| \sim \min \left(c_6 \eps^2/\lambda_4,1\right),
    $
    where $\lambda_4$ denotes the typical size of the renormalizable quartic couplings. 
    
    This modifies the EFT power counting underlying the hierarchy generation, which assumes $|v_3| \sim \mathcal{O}(1)$, unless one takes $\lambda_4 \lesssim c_6 \eps^2 $. Such an assumption is, however, typically not realized: the same UV physics generating $O_6$ will in general also renormalize the marginal couplings, maintaining $\lambda_4/c_6 \sim \mathcal{O}(1)$ for $\mathcal{O}(1)$ UV couplings and so $|v_3| \sim \eps^2$ (and similarly for the higher order spurions). This is especially evident with VLF chains, in which both $O_6$ and the renormalization of the quartic are 1-loop closed chains. Hence, the power counting is altered in a potentially problematic way, possibly rendering the induced flavor hierarchies and mixings no longer viable.
    
    \item \textbf{1-loop Coleman-Weinberg potential}. This is by far the most elegant mechanism. As we argue below, it generically yields a completely generic vacuum with no residual stabilizer, and therefore non-vanishing composites of any allowed spin, including spin-$1/2$. It requires no additional ingredients and is radiatively stable. For this reason, we now discuss it more in detail.
\end{enumerate}

With the Coleman-Weinberg option, at 1-loop the vacuum is determined by the full effective potential,
\begin{equation}
V_{\rm eff}(\Ss)=V_\text{tree}(\Ss)+V_\text{CW}(\Ss).
\end{equation}
For a generic background field configuration $\Ss$, the Coleman-Weinberg contribution \cite{Coleman:1973jx} in the $\overline{\rm MS}$ scheme reads \cite{Martin:2001vx}
\begin{equation}
V_{\rm CW}(\Ss)=
\frac{1}{64\pi^2}
\sum_a (-1)^{F_a} N_a\, m_a^4(\Ss)
\left[
\log \frac{m_a^2(\Ss)}{\mu^2}-c_a
\right],
\end{equation}
where the sum runs over all fluctuating fields with $\Ss$-dependent masses, $F_a=0,1$ for bosons and fermions, $N_a$ is the corresponding multiplicity, and $c_a = \{3/2, 5/6\}$ for scalars/fermions and gauge bosons, and $\mu$ is the renormalisation scale.

In our setup, four classes of fields can acquire $\Ss$-dependent masses: $\Ss$ itself, the gauge bosons of the flavor group (if the symmetry is gauged), the UV completion mediators, and the SM Higgs and fermions. The SM fermions and Higgs are too light to contribute appreciably to the Coleman-Weinberg potential, since in general $|\Ss|\gg v_\text{EW}$ as we will see in \cref{sec:pheno}. The UV mediators, whether scalars or fermions, will in general also receive $\Ss$-dependent contributions to their masses. However, since, by construction, they live at scales $\sim |\Ss|/\eps$, their effects are more appropriately captured in the low-energy SM+$\Ss$ EFT via induced higher-dimensional operators. Indeed, including them explicitly in the Coleman-Weinberg potential and expanding the logarithm reproduces this structure precisely.

The only potentially relevant 1-loop contributions are therefore those from $\Ss$ itself and, when the flavor symmetry is gauged, from the corresponding gauge bosons. The self-contribution of $\Ss$ is controlled by its tree-level masses and therefore scales with additional powers of the tree-level quartics. As a result, for perturbative quartics it remains loop suppressed with respect to the tree-level potential and cannot qualitatively modify the vacuum structure.

By contrast, the flavor gauge boson masses scale as $m_V \sim g |\Ss|$, so their Coleman-Weinberg contribution can compete with the tree-level potential already for $g\sim \mathcal{O}(1)$, provided the tree-level quartics are of order $\lambda_{\rm tree}\sim 1/16\pi^2$. This hierarchy is radiatively stable, making the mechanism natural in our setup. We therefore focus on the gauge boson contribution in the following. In particular, we restrict attention to the $SU(2)$ contribution for two reasons. First, it is the only necessary gauge contribution that induces a nontrivial angular dependence in the vacuum structure. Second, since $SU(2)$ is anyway required by the flavor construction and is typically anomaly-free, it is natural to consider models in which it is gauged. By contrast, as discussed in \cref{sec:models}, the additional $U(1)$ factor is not strictly necessary, and anomaly-free charge assignments that allow it to be gauged are typically harder to realize.

\paragraph{$SU(2)$ Coleman-Weinberg.}
With $T^a$ the $SU(2)$ generators in the spin-$j_\Ss$ representation, the gauge boson mass matrix on the background $\Ss=|\Ss|n$ is
\begin{equation}
\bigl(M_V^2\bigr)^{ab}
=
g_F^2 |\Ss|^2 G^{ab}(n),
\qquad
G^{ab}(n)\equiv
n^\dagger \frac{\{T^a,T^b\}}{2}n.
\end{equation}
Note that $G^{ab}$ can be split in the flavor representation of spin-2 and the Casimir,
$
G=Q + j_\Ss(j_\Ss+1) \mathds{1} /3
$.
In particular, $|G|^2 = |Q| ^2 + (j_\Ss (j_\Ss+1))^2/3$ with $|Q|^2 \propto S_2$. 
The Coleman--Weinberg term can then be written as
\begin{equation}
\begin{aligned}
V_{\rm CW}^{\rm gauge}
&=
\frac{3}{64\pi^2}
\tr\!\left[
M_V^4 
\left(
\log\frac{M_V^2}{\mu^2}-\frac56
\right)
\right] =
\frac{3 g_F^4 |\Ss|^4}{64\pi^2}
\left[
A(n)\left( \log\!\left(\frac{g_F^2 |\Ss|^2}{\mu^2}\right) -\frac56 \right)
+
B(n)
\right]
\end{aligned}
\end{equation}
with 
\begin{equation}
A(n)\equiv \tr\, G(n)^2,
\qquad
B(n)\equiv \tr G(n)^2\log G(n) .
\end{equation}
Hence, the effective 1-loop potential to be minimized is now
\begin{equation}
V_{\rm eff}
=
-m^2 |\Ss|^2
+
\lambda_\text{eff} ^\text{tree} (n)|\Ss|^4
+
\frac{3 g_F^4 |\Ss|^4}{64\pi^2}
\left[
A(n)\left( \log\!\left(\frac{g_F^2 |\Ss|^2}{\mu^2}\right) -\frac56 \right)
+
B(n)
\right].
\end{equation}
This makes it explicit that the potential is no longer linear in the orbit invariants. The logarithms depend non-trivially on the orbit direction $n$, in particular through the quadrupole $Q$, and therefore the minimum is no longer expected to lie on the boundary of the orbit space. As a result, the vacuum is generically expected to have no residual stabilizer.

The gauge Coleman-Weinberg contribution should not, however, be
considered in isolation. As shown in \cref{app:CW}, whenever
$Q=0$ is accessible, the gauge contribution alone is minimized there.
This limit is itself highly special: the potential depends only on the
five independent components of the quadrupole and therefore leaves several
angular directions flat. Moreover, for $j_\Ss=5/2$, the condition $Q=0$
also implies $v_3=0$. The case $j_\Ss=3/2$ is exceptional because
$|Q|^2$ is fixed and $Q=0$ cannot be reached. Thus, the pure
gauge Coleman-Weinberg contribution does not by itself guarantee the
generic vacuum required by our construction.

The relevant effect is instead the genuinely non-linear interplay between
the Coleman-Weinberg contribution and the tree-level potential. The
perturbative analysis presented in \cref{app:CW} shows that, when
\begin{align}
    \lambda_{\rm tree}
    \sim
    \frac{g_F^4}{16\pi^2},
\end{align}
the Coleman-Weinberg contribution can destabilize the special vacuum
selected at tree level. We confirmed this numerically for
$j_\Ss=3/2$ and $j_\Ss=5/2$. The resulting vacuum
is generically free of residual stabilizers, the otherwise flat angular
directions are lifted, and the allowed spin-$1/2$ composites, including
$v_3$, become non-vanishing. This radiatively stable regime therefore
provides the desired dynamical realization of the generic spurion vacuum.
It may also lead to interesting cosmological signatures, in particular a
first-order phase transition associated with flavor symmetry breaking, as
discussed in the next section.

\section{Phenomenological implications}
\label{sec:pheno}

The phenomenology of our framework can be discussed in different layers of generality.

\paragraph{SMEFT.} The higher-representation spurions introduced to generate the Yukawas can also be used to analyze the selection rules obeyed by new physics coupled to the SM. At the lowest order, this amounts to expanding the Standard Model Effective Theory (SMEFT) Wilson coefficients in powers of the same spurions.

The flavor bounds associated with these operators typically probe very high scales, and this remains true in our framework. Operators involving only SM fields that are neutral under the flavor group are predicted to be flavor anarchic, and are therefore subject to very strong flavor bounds, up to scales of order $10^5\,\mathrm{TeV}$ \cite{deBlas:2025gyz}. Even when the fields are charged under the flavor group, however, the resulting suppressions in our framework are generally mild.

We argued how the large top Yukawa calls for $q$ and $u$ to be split as $\bm{2}\oplus\bm{1}$, while a realistic CKM hierarchy rather generically requires $q$ to transform in this way under an $SU(2)$ flavor factor. Thus, in realistic constructions, one expects a higher $SU(2)_q$ spin spurion $\Ss_q$ associated with the flavor symmetry acting on $q$.
Consider then, for instance, current bilinears involving the light doublets,
$
\bar q_\alpha \gamma^\mu X^a q_\beta,
$
where $X^a$ denotes either the identity or a generator of $SU(3)_c \times SU(2)_{\rm L}$. These bilinears are neutral under the Abelian factors and transform as $\bm{1}\oplus\bm{3}$ under $SU(2)_q$. The presence of a higher flavor spin spurion $\Ss_q$ therefore allows the generic expansion
\begin{align}
    \bar q_\alpha \gamma^\mu X^a q_\beta
    \left[
    \delta_{\alpha\beta}
    + \frac{(\Ss_q \bar \Ss_q)_{1,\alpha\beta}}{M_q^2}
    + \dots
    \right],
    \label{eq:fermionCurrentDecomp}
\end{align}
where the leading non-trivial insertion is the composite triplet $(\Ss_q\bar\Ss_q)_1$. Since in almost all the models discussed above one finds $|\Ss_q|/M_q=\eps_q\approx 0.1-0.3$, this gives only a mild $\eps_q^2$ suppression to $\Delta F=1$ transition induced by operators involving this bilinear.

An even more dramatic situation arises for four-fermion operators in which the two fermion bilinears transform under the same $SU(2)$ factor. For example, the operators \cite{Grzadkowski:2010es}
\begin{align}
    [O_{qq}^{(1,3)}]_{\alpha \beta \rho \sigma}
    =
    (\bar q_\alpha \gamma^\mu (\tau^a) q_\beta)
    (\bar q_\rho \gamma_\mu (\tau^a) q_\sigma),
    \label{eq:Oqq}
\end{align}
relevant for $\Delta F=2$ meson mixing, have light generation flavor indices transforming under $SU(2)_q$ as
$
(2)\bm{1}\oplus\bm{3}\oplus\bm{5}.
$
The spin-2 fiveplet can directly generate $\Delta F=2$ transitions, and can be contracted with the composite spin-2 spurion $(\Ss_q\bar\Ss_q)_2$. The resulting flavor-violating coefficient, therefore, scales as
\begin{align}
    [C_{qq}^{(1,3)}]_{1212}
    \sim
    \frac{\eps_q^2}{\Lambda_\text{eff}^2}.
\end{align}
Bounds from Kaon mixing require $\Lambda_\text{eff}\gtrsim 2.1\times 10^4\,\mathrm{TeV}$ in the anarchic case \cite{Silvestrini:2018dos}. Taking $\eps_q\approx 0.3$, as in most of the models in \cref{sec:models},
this is reduced only to
$
    \Lambda_\text{eff} \gtrsim 6.3\times 10^3\,\mathrm{TeV}.
$
This is an important difference from traditional models that employ only low-dimensional spurions. In standard $U(2)^5$, for instance, the triplet in \cref{eq:fermionCurrentDecomp} or the fiveplet in \cref{eq:Oqq} must be built out of the doublet $V_q$ and the bifundamentals $\Delta_{u,d}$ themselves, leading to a much stronger suppression \cite{Greljo:2022cah, Faroughy:2020ina}.

Other classes of operators may be more suppressed, depending on the specific model. Chirality-flipping bilinears, in particular, are sensitive to the detailed flavor assignments. For example, in the $U(2)_{q+\ell}$ model of \cref{sec:U2ql}, structures such as $\bar q_\alpha X d_i$ or $\bar q_\alpha X u_i$, with $X$ denoting a generic Lorentz and gauge structure, are suppressed by $\eps_{q+\ell}^3$ when they involve light generations of $q$, namely $\alpha=1,2$, and arbitrary right-handed generations $i=1,2,3$. This suppression arises from the insertion of a composite doublet that acts on $q$. By contrast, the corresponding structures involving $\bar q_3$, such as $\bar q_3 X d_i H$ or $\bar q_3 X u_i \tilde H$, are unsuppressed.
In the $U(2)_{q+e}\times U(2)_u\times SU(3)_5$ model of \cref{sec:U3_5}, the same structures are more suppressed. For instance, $\bar q_\alpha X u_\beta$ and $\bar q_\alpha X d_i$ scale respectively as $\eps_q^3\eps_u^3$ and $\eps_q^3\eps_5^3$. Chirality-flipping four-fermion operators, such as $O_{quqd}^{(1,8)}$ and $O_{\ell edq}$, follow a similar logic to the current-current operators discussed above.

Finally, in our construction, we do not assume CP to be a symmetry of the UV theory, broken spontaneously by the higher-representation spurions.\footnote{The $U(2)_{q+\ell}$ model is not compatible with spontaneous CP violation because of its rigid structure and the resulting relation between CKM elements and quark masses, as already anticipated in \cref{sec:U2ql}. Whether the other models can realize spontaneous CP violation is an interesting question that we leave for future work.} As a result, operators sensitive to flavor-conserving CP violation, such as dipoles, are not additionally suppressed.

\begin{figure}[!htb]
  \centering
  \begin{subfigure}{\textwidth}
    \centering
    \includegraphics[width=\textwidth]{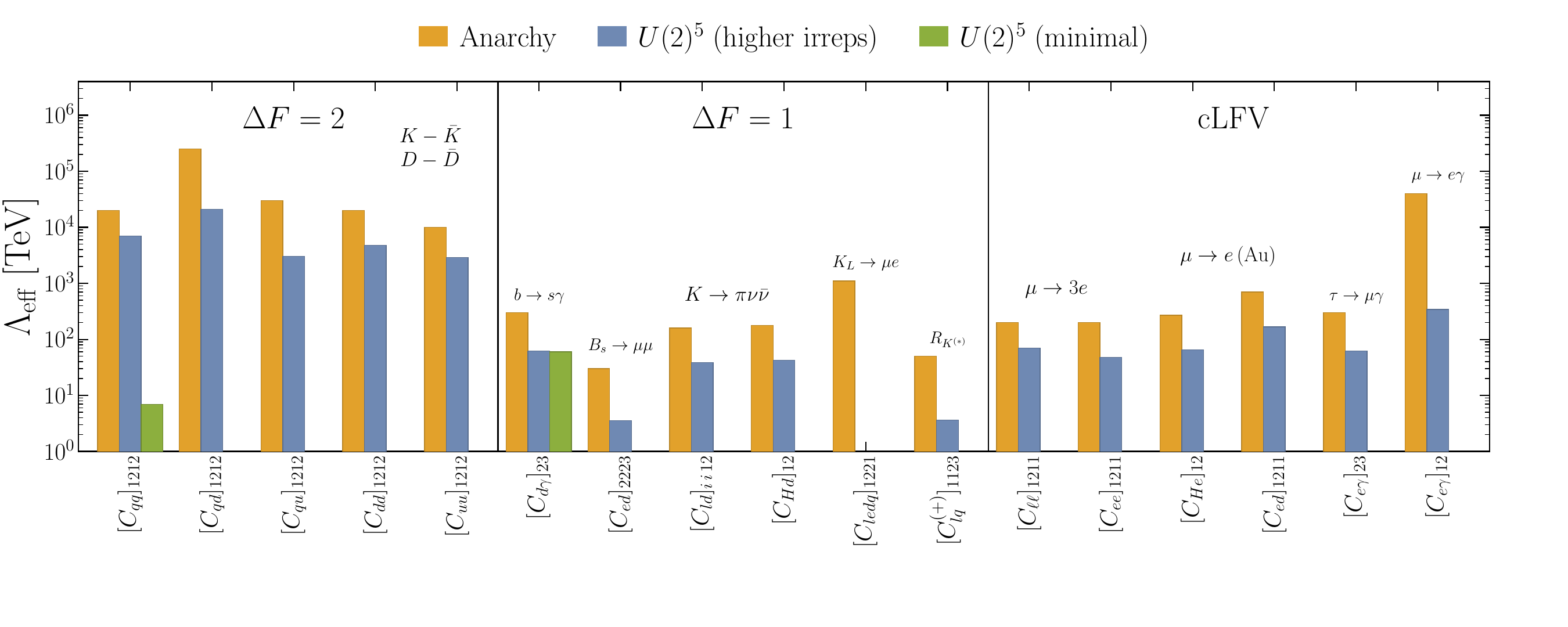}
    \caption{Bounds in the anarchic scenario, in $U(2)^5$ with higher-irrep spurions, and in traditional $U(2)^5$.}
    \label{fig:SMEFT_U2}
  \end{subfigure}
  \begin{subfigure}{\textwidth}
    \centering
    \includegraphics[width=\textwidth]{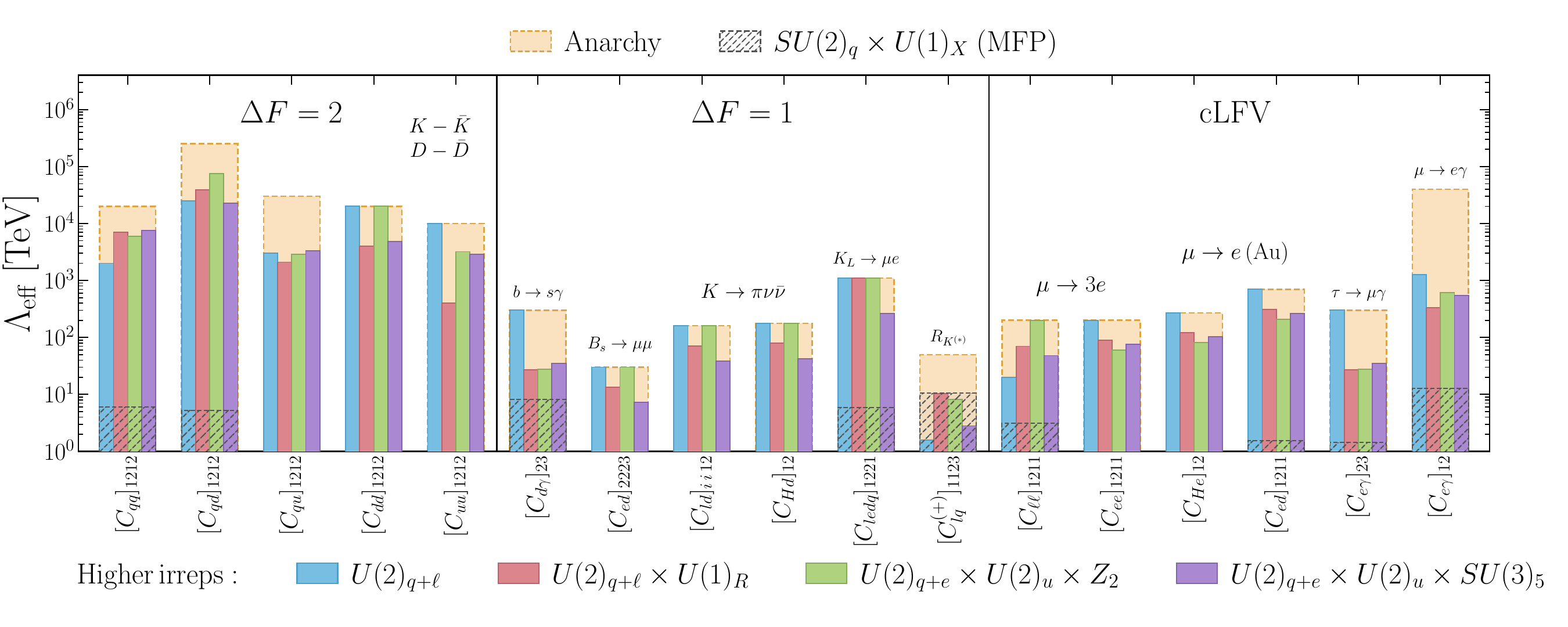}
    \caption{Bounds in the anarchic scenario, in the higher-irreps models of \cref{sec:models}, and in traditional $SU(2)_q \times U(1)_X$ (MFP).}
    \label{fig:SMEFT_models}
  \end{subfigure}
  \caption{Flavor bounds on the effective scale associated with a selected set of SMEFT operators in the anarchic scenario and with the selection rules of the models in \cref{sec:models}, compared with the traditional $U(2)^5$ setup (top) and the $SU(2)_q \times U(1)_X$ setup (bottom).}
  \label{fig:SMEFT}
\end{figure}

In summary, while the spurion structure provides some suppression, it is clearly insufficient on its own to fully address the TeV-scale flavor problem for generic SMEFT operators, at least from an EFT perspective. This is shown quantitatively in the bar plots of \cref{fig:SMEFT}, where we display the bounds from selected observables and SMEFT operators in our models, and compare them with flavor anarchy and standard $U(2)^5$ and $SU(2)_q\times U(1)_X$ MFP scenarios. In the plot, the flavor indices associated with the Wilson coefficients should be understood as indicative; for each operator, we show the least-suppressed contribution to the corresponding observable. The anarchic bounds are adapted from \cite{Greljo:2025mwj}.

\paragraph{UV signatures.}
The UV signatures are necessarily model dependent. Nevertheless, the EFT analysis above can be directly reinterpreted in terms of the mass scale of the UV states that generate the effective Yukawas. For instance, the VLF chains introduced in \cref{sec:UV} generically induce, at one loop, the same flavor-violating four-fermion operators and dipoles discussed above, with the same spurion suppressions \cite{Greljo:2024evt, Arkani-Hamed:2026wwy}. An approximate lower bound on the mediators' UV scale generating the Yukawas can therefore be inferred from these effects. Parametrically, one finds
\begin{align}
    M \gtrsim \mathcal{O}(1000)  \text{ TeV},
    \label{eq:M_UV_bound}
\end{align}
where $M$ denotes the typical VLF mass.

Operators of the kind $O_{Hf}$ can instead be generated already at tree level when the VLF chain responsible for the Yukawa ends on a fermion field $f$ that is not charged under the flavor group, such as $\bar q_3, \bar \ell_3$ or $d_i,u_i,e_i$ in the concrete example completion of $U(2)_{q+\ell}$ in \cref{sec:UV}. The corresponding bounds are typically slightly weaker than those from $\Delta F=2$ observables and, despite arising at tree level, lead to constraints qualitatively comparable to \cref{eq:M_UV_bound}.
 
Since the ratio $|\Ss|/M \approx 0.1-0.3$ is fixed by the flavor fit, the bounds on the mediator scale translate into a typical lower bound on the flavon vev,
\begin{align}
    |\Ss| \gtrsim \mathcal{O}(100) \text{ TeV}.
    \label{eq:Sbound1}
\end{align}

Additional signatures are associated with the spontaneous breaking of the flavor group by the higher-representation flavons. If the flavor symmetries are global, the corresponding ALPs have flavor-violating couplings to the SM. This typically pushes the breaking scale to very high values, up to $|\Ss|\gtrsim 10^{11}\,\mathrm{GeV}$ \cite{MartinCamalich:2020dfe, Calibbi:2025rxn, Greljo:2025ljr}. Allowing for a small explicit breaking of the flavor symmetry, for instance through soft masses for the ALPs, may lower this scale and make the corresponding bounds even weaker than those of \cref{eq:Sbound1} \cite{Greljo:2024evt}.

A different way to avoid a large number of light ALPs is to gauge the flavor group, or at least part of it. This possibility is also motivated by the vacuum dynamics discussed in \cref{sec:pot}: flavor gauge bosons can contribute to the 1-loop Coleman-Weinberg potential and help select generic higher-representation spurion vevs in the absence of residual symmetries. In this case, the model predicts the existence of massive flavor gauge bosons. Their exchange induces flavor-violating four-fermion operators, with Wilson coefficients scaling as $C\sim 1/|\Ss|^2$.
However, these operators induce flavor violation only when the two fermion currents coupled by the gauge boson exchange involve \emph{different} fields transforming under the same flavor symmetry~\cite{Darme:2023nsy, Greljo:2023bix}. Since, as argued in \cref{sec:framework}, our setup avoids charging the left- and right-handed fields entering a given Yukawa interaction under the same flavor factor, flavor violation is generated essentially only in semileptonic four-fermion operators, not in $\Delta F=2$ operators such as $(\bar q \gamma^\mu q)^2,\text{etc}$. From the bounds shown in \cref{fig:SMEFT}, one then obtains
\begin{align}
    |\Ss| \gtrsim \mathcal{O}(100) \text{ TeV}.
    \label{eq:VEVfcnc}
\end{align}
which is comparable to the bounds from the UV mediators in \cref{eq:Sbound1}. The precise numerical value is model dependent.

While all these scales are far beyond the direct kinematic reach of current high-energy colliders, future flavor precision programs at existing kaon, muon, charm, tau, and bottom factories, as well as the future FCC-ee~\cite{deBlas:2025gyz}, provide a natural context in which the indirect effects of high-scale flavor dynamics could be further tested.

\paragraph{Cosmological signatures.}

Gauged flavor symmetries based on ${SU}(2)$ or ${SU}(3)$ factors can undergo first-order phase transitions (FOPTs) under suitable conditions, leading to a stochastic gravitational-wave background (SGWB) potentially observable at future detectors. The possibility that the origin of flavor is tied to such cosmological signals has been explored in~\cite{Greljo:2019xan, Fabri:2025fsc, Chrysostomou:2025vrg}. Of particular interest is the regime where the symmetry-breaking scale lies in the range $|\Ss| \sim 100 \!- \!1000$\,TeV, which is both indirectly probed by flavor-changing neutral currents \eqref{eq:Sbound1}, \eqref{eq:VEVfcnc} and corresponds to gravitational-wave frequencies accessible to next-generation observatories such as the Einstein Telescope~\cite{ET:2019dnz} and Cosmic Explorer~\cite{Evans:2021gyd}. In this window, the spontaneous breaking of a gauged ${SU}(2)$ flavor symmetry directly links flavor experiments to gravitational-wave probes of the early universe.

However, a sufficiently strong FOPT is not generic in tree-level scalar potentials and requires relatively small quartic couplings. For breaking via a fundamental (doublet) representation~\cite{Greljo:2023bix}, one typically requires a scalar quartic of $\mathcal{O}(0.01)$~\cite{Chrysostomou:2025vrg}, with no explanation within the model. In our framework, by contrast, the generation of non-vanishing composite doublets from a higher-spin flavon \emph{requires} the vacuum to be driven away from the symmetric minima selected by the tree-level potential. As discussed in \cref{sec:additonalContrib}, this is naturally achieved when the gauge Coleman-Weinberg contribution competes with the tree-level quartics, which in turn requires the latter to be of loop size (see \cref{app:CW}). Therefore, the small quartics needed for a strong FOPT are not an independent assumption, but are tied to the same vacuum dynamics that makes the framework viable. Our setup thus further tightens the link between flavor physics and cosmological dynamics. A quantitative prediction of the SGWB from higher flavor spins is left for future work.

\section{Summary and outlook}
\label{sec:concl}

The origin of flavor remains one of the most intriguing puzzles in particle physics. In this work, inspired by the recent proposal of \cite{Banks:2025baf}, we argued that the hierarchies in the Standard Model fermion masses and mixings can naturally emerge from a small set of spurions in representations of the flavor group larger than those associated with the fermions themselves. These spurions are assumed to be completely anarchic, with entries of comparable size and no zero textures.
The hierarchies then arise from decomposing their products into doublets and triplets that can be attached to SM fermion bilinears, thereby entering the Yukawa matrices. The construction somehow resembles a non-Abelian version of the Froggatt-Nielsen mechanism.

In this setup, the flavor group should contain at least some $SU(2)$ factors, as the large top Yukawa coupling otherwise makes a spurion expansion in the up sector ill-defined. The lepton and down quark sectors, on the other hand, may be assigned to $SU(3)$ factors. In line with the philosophy of using a few spurions with generic entries, we identified a novel mechanism that applies to both cases. The essential point is that the Yukawa matrices are not generated directly as full rank matrices, but are built through successive outer products of composite doublets and triplets. At leading order in the spurion expansion, the composite $SU(2)$ doublet is unique and of order $\eps^3 = (|\Ss|/M)^3$, making the light Yukawa block naturally rank-1. This rank is then lifted by the doublets appearing at the next order, accompanied by an additional $\eps^2$ suppression. Minimally,
\begin{align}
    \big ( Y_f^{2\times 2}\big )_{\alpha i}
    \sim
    \eps^3\, v_\alpha w_i
    +
    \eps^5\, v'_\alpha w'_i
    + \dots ,
\end{align}
where here $v_\alpha$ and $v'_\alpha$ are normalized composite doublets, while $w_i$ and $w'_i$ are vectors of $\mathcal{O}(1)$ coefficients.
This naturally generates the required splitting between the second and first generations, while allowing for an unsuppressed third-generation Yukawa. The mechanism for triplets is analogous and can generate suppressions also for the bottom quark and the tau lepton. The formalization of this logic and of the general framework was given in \cref{sec:framework}.

Based on this framework, we built explicit models that realize the mechanism in \cref{sec:models}. We started by applying this philosophy to a known minimal model capable of partially addressing the flavor puzzle, based on an $SU(2)_{q+\ell}$ symmetry \cite{Greljo:2023bix, Antusch:2023shi}. In this setup, the two effective doublets required at the empirical level can be naturally generated by a single higher-representation $SU(2)$ spurion through the rank-lifting mechanism described above. We then built more refined models based on the same principles, capable of addressing the remaining small tunings at the cost of introducing additional factors in the flavor symmetry group, while still using only one higher-representation spurion per factor.
We also explained how the hierarchies of the effective spurions $V_q$ and $\Delta_{u,d,e}$ in the popular $U(2)^5$ framework \cite{Barbieri:2011ci} can arise in our construction, although with very different selection rules for new physics.

A crucial assumption of our construction is the genericity of the spurion, namely, the absence of residual symmetries and the presence of entries of comparable size in a generic basis. While this assumption seems natural at the EFT level, it is not guaranteed when the spurion is realized as the vacuum expectation value of a scalar field, since residual-symmetry configurations often arise as minima of the scalar potential. To address this issue, in \cref{sec:pot} we studied the scalar potential for a single field in detail, with half-integer flavor spin $j_\Ss \geq 3/2$. To understand the vacuum structure, we employed the geometric picture of the Majorana representation, which helps identify residual symmetries intuitively. We found that the tree-level renormalizable potential indeed tends to select vacua with discrete or $U(1)$ residual symmetries. This is dangerous for our models because the selection rules associated with these symmetries often forbid the composite doublets that are necessary ingredients of the mechanism. Among the possible ways out, the most elegant is provided by the 1-loop gauge Coleman-Weinberg potential. Its nonlinear dependence on the angular variables allows us to move the vacuum away from symmetric configurations, ensuring the genericity of the vev assumed in our construction.

The necessity of the Coleman-Weinberg contribution makes the construction particularly predictive. Indeed, spontaneous breaking of non-Abelian groups through radiative effects can lead to a strong first-order phase transition \cite{Greljo:2019xan, Fabri:2025fsc, Chrysostomou:2025vrg}, and therefore to testable gravitational-wave signals. Hence, the ingredients required by the flavor mechanism itself naturally point to interesting cosmological signatures.
Other phenomenological consequences include flavor-changing neutral currents induced by the flavor group gauge bosons and by the mediators of the UV completion, as introduced in \cref{sec:UV} and briefly addressed in \cref{sec:pheno}. These set lower bounds on the scale of the mechanism through precision flavor observables, requiring the flavor breaking scale to lie around
$$
    |\Ss| \gtrsim  \mathcal{O}(100 \text{ TeV}).
$$
This places the gravitational waves mentioned above within the range relevant to the Einstein Telescope \cite{ET:2019dnz} and Cosmic Explorer \cite{Evans:2021gyd}, making it possible to probe the framework at both the cosmological and particle physics frontiers. Further signatures arise from the different selection rules imposed on new physics coupled to the SM. We analyzed this in the SMEFT, showing that the irreducible presence of composites beyond the doublets and triplets required for the Yukawas can weaken the suppression of flavor observables. As a result, the bounds are typically stronger than in minimal MFV, $U(2)^5$, or MFP scenarios.

Overall, our results show that higher-flavor group representations provide an innovative and viable way to organize fermion hierarchies starting from generic spurions, while retaining distinctive phenomenological signatures. Several directions naturally follow from this work and deserve further exploration, including a more quantitative study of the gravitational-wave signatures discussed above, a detailed analysis of explicit UV completions, and the extension of the scalar potential and vacuum alignment analysis to higher-representation $SU(3)$ spurions.

\acknowledgments

The authors thank Matthew McCullough and Luc Darmé for useful discussions.
This work has received funding from the Swiss National Science Foundation (SNSF) program ``Swiss High Energy Physics for the FCC'' (CHEF) supported in part by SERI.

\appendix

\section{$SU(2)$}

\subsection{Counting irreps}
\label{app:countingIrreps}

Let $\Ss$ be a commuting field transforming in the spin-$j_\Ss$ irrep $V_{j_\Ss}$ of $SU(2)$, with $j_\Ss$ half-integer. Since $SU(2)$ irreps are self-dual, the conjugate spurion $\bar\Ss$ transforms in the conjugate representation $V_{j_\Ss}^*\simeq V_{j_\Ss}$. For representation counting, composite operators of the schematic form
\begin{align}
    \Ss^{p} \bar\Ss^{q}
\end{align}
therefore transform as
\begin{align}
    \Sym^{p}(V_{j_\Ss})\otimes \Sym^{q}(V_{j_\Ss}).
\end{align}
It is convenient to use the standard identification
\begin{align}
    V_{j_\Ss} \simeq \Sym^{2j_\Ss}(\mathbf 2),
\end{align}
where $\mathbf 2$ is the fundamental doublet of $SU(2)$.
The decomposition of the symmetric powers $\Sym^m(V_{j_\Ss})$ into irreducible $SU(2)$ representations, including the determination of their multiplicities, is then a standard plethysm problem. In general, this is highly non-trivial and various tools are available (for example \cite{Fonseca:2020vke}). For $SU(2)$, however, the problem simplifies considerably, since the multiplicities admit a convenient closed expression through the Cayley-Sylvester formula, see \cite{manivel2007extension} and Ref.~[6] therein:
\begin{align}
    \Sym^m\!\bigl(\Sym^n(\mathbf 2)\bigr)
    =
    \bigoplus_{w=0}^{\lfloor mn/2\rfloor}
    M_{m,n,w}\,
    \Sym^{mn-2w}(\mathbf 2),
\end{align}
with
\begin{align}
    M_{m,n,w}=[m,n,w]-[m,n,w-1].
\end{align}
Here $[m,n,w]$ denotes the number of partitions of $w$ into at most $m$ non-negative parts, each bounded by $n$. By convention, $M_{m,n,w}=0$ for $w<0$.

Setting $n=2j_\Ss$, this gives
\begin{align}
    \Sym^m(V_{j_\Ss})
    =
    \Sym^m\!\bigl(\Sym^{2j_\Ss}(\mathbf 2)\bigr)
    =
    \bigoplus_{w=0}^{\lfloor mj_\Ss\rfloor}
    M_{m,2j_\Ss,w}\,
    \Sym^{2mj_\Ss-2w}(\mathbf 2).
\end{align}
Since $\Sym^N(\mathbf 2)$ is the spin-$N/2$ irrep, this can be rewritten as
\begin{align}
    \Sym^m(V_{j_\Ss})=\bigoplus_J m^{(m)}_J\,V_J,
\end{align}
where
\begin{align}
    m^{(m)}_J = M_{m,\,2j_\Ss,\,mj_\Ss-J},
\end{align}
whenever $mj_\Ss-J\in \mathbb Z_{\ge 0}$, and vanishes otherwise.

Once the two symmetric powers are decomposed, the required multiplicity follows from the ordinary $SU(2)$ tensor product rule
\begin{align}
    V_J\otimes V_{J'}
    =
    \bigoplus_{K=|J-J'|}^{J+J'} V_K.
\end{align}

At this point, we should emphasize an important caveat. The multiplicity obtained in this way is a necessary but not sufficient criterion for the resulting composite $d-$dimensional irreps, viewed as \(d\)-dimensional vectors, to be linearly independent. Indeed, this counting refers to independent \(SU(2)\) maps, namely to the independence of the corresponding polynomial covariants as functions of the original spurion. As $d-$dimensional vectors, these covariants may still be linearly dependent.\footnote{A trivial toy example is given by $v_1=x(x,y)$ and $v_2=y(x,y)$. As polynomial covariants in $x,y$, they are independent, but as two-dimensional vectors, they are linearly dependent.} Their actual independence must therefore be checked separately. Later, we will argue how, in some cases, the multiplicity count can still be used as a sufficient condition. In all cases relevant to the models discussed in \cref{sec:models}, we performed this check explicitly. This refers particularly to the spin-0 (quartics) and 1/2 composites.

\paragraph{Spin 1/2 irreps.} A spin-$1/2$ irrep appears in $V_J\otimes V_{J'}$ iff
\begin{align}
    J'=J\pm \frac12.
\end{align}
Therefore, if
\begin{align}
    \Sym^{p}(V_{j_\Ss})=\bigoplus_J m^{(p)}_J\,V_J,
    \qquad
    \Sym^{q}(V_{j_\Ss})=\bigoplus_{J'} m^{(q)}_{J'}\,V_{J'},
\end{align}
the multiplicity of spin-$1/2$ inside
\begin{align}
    \Sym^{p}(V_{j_\Ss})\otimes \Sym^{q}(V_{j_\Ss})
\end{align}
is
\begin{align}
\begin{aligned}
    N^{(p,q)}_{1/2}
    &=
    \sum_J
    m^{(p)}_J
    \left(
        m^{(q)}_{J+1/2}
        +
        m^{(q)}_{J-1/2}
    \right)\\
    &= \sum_J
    M_{p,2j_\Ss,pj_\Ss-J}
    \left[
        M_{q,2j_\Ss,qj_\Ss-(J+1/2)}
        +
        M_{q,2j_\Ss,qj_\Ss-(J-1/2)}
    \right].
\end{aligned}
\label{eq:repMultiplicities}
\end{align}
This gives an explicit, fully algorithmic formula for the number of spin-$1/2$ irreps in arbitrary composite operators constructed from $\Ss$ and $\bar\Ss$.

We now give several explicit examples relevant to our work.
\begin{itemize}

\item  $\bm{\Ss^2 \bar \Ss}$. It contains exactly one spin-1/2 irrep:
\begin{align}
    N_{1/2}^{(2,1)}=1.
\end{align}
This is easy to understand, as it does not require all the machinery developed above. Writing it as $[(\Ss \otimes \Ss)_J \otimes \bar \Ss]_{1/2}$, from basic $SU(2)$ tensor decomposition only $J=j_\Ss \pm 1/2$ can contribute, and each gives exactly one spin-1/2 irrep when contracted with $\bar \Ss$. However, $(\Ss \otimes \Ss)_{J}$ is not null only for odd $J$, which singles out either $J=j_\Ss+1/2$ or $J=j_\Ss-1/2$. Hence, only one channel can contribute to $(\Ss^2 \bar \Ss)_{1/2}$, meaning there is exactly one spin-1/2 irrep in the decomposition.

The fact that $(\Ss \otimes \Ss)_{J}$ is non-vanishing only for odd $J$ can be understood in two equivalent ways. The first is to inspect the explicit Clebsch-Gordan coefficients: exchanging the two $\Ss$ factors in $(\Ss\Ss)_J$ produces a factor $(-1)^{2j_\Ss-J}$, so for commuting fields and half-integer $j_\Ss$ only the components with odd $J$ survive.
A more pictorial argument is the following. A spin-$j_\Ss$ representation can be viewed as the totally symmetric product of $2j_\Ss$ spin-$1/2$ representations, namely as a totally symmetric tensor with $2j_\Ss$ indices $\Ss^{a_1 a_2 \cdots a_{2j_\Ss}}$, corresponding to a Young tableau with a single row of $2j_\Ss$ boxes. In the product of two such tensors, projecting onto total spin $J$ means leaving $2J$ boxes uncontracted in a single row, while the remaining $4j_\Ss-2J$ boxes must be paired into vertical columns of two boxes, corresponding to $SU(2)$ singlet contractions. These are implemented by $\epsilon_{ab}$ contractions, and their number is therefore $2j_\Ss-J$. Upon exchanging the two $\Ss$ fields, each $\epsilon_{ab}$ contributes a minus sign because of its antisymmetry, so the total factor is again $(-1)^{2j_\Ss-J}$.

\item $\bm{\Ss^3}$. This never contains a spin-1/2 representation. From \cref{eq:repMultiplicities}, the presence of spin-1/2 requires  $2m j_\Ss -2w=6j_\Ss-2w=1$; writing $j_\Ss=(2r+1)/2$, $r\in \mathds{N}$, this means $w=3r+1$. So 
\begin{align}
    N_{1/2}^{(3,0)} = M_{3,2r+1,3r+1} = [3,2r+1,3r+1]-[3,2r+1,3r].
\end{align}
For these special arguments, one can prove that the two partition numbers are equal, $[3,2r+1,3r+1]=[3,2r+1,3r]$, so that
\begin{align}
    N^{(3,0)}_{1/2}=0.
\end{align}
A more concrete way to understand this is to return to the Young tableau picture. The product $\Ss^3$ contains $6j_\Ss$ boxes, all of which must be contracted into singlets except for one, corresponding to the free spin-$1/2$ index. Since singlet contractions are implemented through $\epsilon_{ab}$, the spin-$1/2$ representation can be written schematically as
\begin{align}
    (\Ss^3)_{1/2, i} = S_{i a_2 \cdots a_{2j_\Ss}} S_{b_1 b_2  \cdots b_{2j_\Ss}} S_{c_1 c_2  \cdots c_{2j_\Ss}} \eps^{a_2 b_1} \eps ^{a_3 c_1} \cdots \eps ^{b_{2j_\Ss} c_{2j_\Ss}} + \text{permutations}.
    \label{eq:S3expl}
\end{align}
It follows that there are $(2j_\Ss-1)/2=j_\Ss-1/2$ cross contractions between the tensor with a free index and the rest, and $(2j_\Ss+1)/2=j_\Ss+1/2$ between the inner tensors (the other possibilities are trivially identically null since they would require an $\epsilon$ contraction within the same symmetric tensor). Since $\Ss$ commutes and \cref{eq:S3expl} is completely symmetrized, including the role of the field associated to the free index, swapping any pair of tensors picks up a factor of $(-1)^{j_\Ss \pm 1/2}$ due to $\eps$ contractions. And because one among $j_\Ss \pm 1/2$ is odd, it follows that the object is identically null.

\item $\bm{\Ss^3 \bar \Ss^2}$. Using \cref{eq:repMultiplicities} together with the fact that $(\Ss \Ss)_k$ contains only odd $k$,
we obtain
\begin{align}
    N^{(3,2)}_{1/2}
    =
    \sum_J
    m^{(3)}_J\Bigl(m^{(2)}_{J+1/2}+m^{(2)}_{J-1/2}\Bigr)
    =
    \sum_{J\le 2j_\Ss+\frac12} m^{(3)}_J.
\end{align}
Indeed, for half-integer $J$, exactly one among $J\pm\tfrac12$ is an odd integer, and it contributes precisely when it lies in the allowed range $1,3,\dots,2j_\Ss$.
Using
\begin{align}
    \Sym^3(V_{j_\Ss})=\bigoplus_w M_{3,2j_\Ss,w}\,V_{3j_\Ss-w},
\end{align}
this can be rewritten as
\begin{align}
    N^{(3,2)}_{1/2}
    =
    \sum_{w\ge j_\Ss-\frac12} M_{3,2j_\Ss,w}
    =
    [3,2j_\Ss,\lfloor 3j_\Ss\rfloor]-[3,2j_\Ss,j_\Ss-3/2].
\end{align}
For example, for $j_\Ss=\tfrac32$ one finds
$ N^{(3,2)}_{1/2}
    =
    [3,3,4]-[3,3,0]
    =2$.

Note that one of the doublets is necessarily of the form $(\Ss \bar \Ss)_0 \otimes (\Ss^2 \bar \Ss)_{1/2} \sim |\Ss|^2   (\Ss^2 \bar \Ss)_{1/2}$, where $(\Ss \bar \Ss)_0 \propto |\Ss|^2$ is the unique quadratic singlet. As a 2-dimensional vector then this doublet is necessarily proportional to the unique cubic one $(\Ss^2 \bar \Ss)_{1/2}$. Since for $j_\Ss \geq 3/2$ we have $ N^{(3,2)}_{1/2} \geq 2$, it means there is at least another independent spin-1/2 composite.

\end{itemize}

\paragraph{Quartic structures.}
The quartic structures discussed in the main text, relevant for the scalar potential, are simple special cases of the more general counting described above.

\begin{itemize}
\item $\bm{\Ss^2 \bar \Ss^2}$. For the invariants $B_k = |(\Ss \Ss)_k|^2$, the situation is actually very simple. Indeed, from the standard $SU(2)$ tensor product decomposition, each spin-$k$ irrep appears exactly once in $\Ss\otimes \Ss$. Moreover,  as seen before, all structures with even $k$ vanish identically, leaving $j_\Ss+1/2$ independent tensors. Finally, a singlet can only be formed by contracting each such tensor with itself, which proves $j_\Ss+1/2$ $U(2)$ preserving quartics are independent. Using the formulas provided above leads to the same conclusion. Note that in this case, the multiplicity of singlets in the formula is exactly the number of independent quartic singlets, since we are now just counting independent polynomial maps of fixed order.

\item $\bm{\Ss^3 \bar \Ss}$. The independent singlets of type
\begin{align}
    B_{U,i} \sim (\Ss^3\bar\Ss)_0
\end{align}
are counted by the multiplicity of spin-${j_\Ss}$ inside $\Sym^3(V_{j_\Ss})$, since then this must be contracted with $\bar \Ss$ to get a singlet:
\begin{align}
    N_{B_U}
    =
    M_{3,2j_\Ss,2j_\Ss}
    =
    \left\lfloor \frac{2j_\Ss+3}{6}\right\rfloor.
\end{align}
The last identity can be obtained by explicitly computing $M_{3,2j_\Ss,2j_\Ss}=[3,n,n]-[3,n,n-1]$ for odd $n$.

\item $\bm{\Ss^4}$. Likewise,
\begin{align}
    F_{U,i} \sim (\Ss^4)_0
\end{align}
are directly counted by the multiplicity of the singlet inside $\Sym^4(V_{j_\Ss})$:
\begin{align}
    N_{F_U} =
    M_{4,2j_\Ss,4j_\Ss} = \left\lfloor \frac{2j_\Ss+3}{6}\right\rfloor.
\end{align}

\end{itemize}

The closed forms for $N_{1/2}, N_{B_U}$ and $N_{F_U}$ were cross-checked explicitly for the cases relevant to this work.

\subsection{Potential bases}
\label{app:potentialBases}

The $U(2)$ invariant quartics can be equivalently written in the $B_k$ basis, with $k=1,3,\dots 2j_\Ss$, or in the $S_k$ basis, see \cref{sec:pot}. In principle the change of basis can be derived recursively from $SU(2)$ recoupling identities, using Wigner $6j$ symbols. This quickly becomes cumbersome. A faster derivation exploits the spectral decomposition in $j\otimes j$.
Define the pure density matrix
$
\rho \equiv \ket{n}\bra{n}
$, with
$
\rho^{\otimes 2}\equiv \rho\otimes \rho
$. Then
\begin{align}
    B_k
    = |(n\otimes n)_k|^2 =
    \bra{n,n} P_k \ket{n,n}
    =
    \tr\!\left(\rho^{\otimes 2} P_k\right),
\end{align}
where $P_k$ is the projector onto the total spin-$k$ irrep inside $j_\Ss\otimes j_\Ss$. It can be written as
\begin{align}
    P_k
    =
    \prod_{k'\neq k}
    \frac{J_{\rm tot}^2-k'(k'+1)}{k(k+1)-k'(k'+1)},
\end{align}
with
\begin{align}
    J_{\rm tot}=J_1+J_2,
    \qquad
    J_{\rm tot}^2
    =
    J_1^2+J_2^2+2J_1\!\cdot\!J_2
    =
    2j_\Ss(j_\Ss+1)+2X,
\end{align}
where for later convenience we defined $
    X\equiv J_1\!\cdot\!J_2$, and used the fact that  $J_{\rm tot}^2$ acts on the spin-$k$ irrep as $k(k+1)\mathds{1}$.
It follows that $P_k$ is a polynomial in $X$, so each $B_k$ is a linear combination of the moments
\begin{align}
    \mu_m \equiv \tr\!\left(\rho^{\otimes 2}   X^m\right) = \bra{n,n} X^m \ket{n,n}.
\end{align}
These moments are combinations of expectation values of irreducible tensors built from the generators $T^a$, namely the invariants
\begin{align}
    \Phi_k &\equiv \Big|\bra n T^{(a_1}\cdots T^{a_k)}\big|_{\rm traceless}\ket n\Big|^2,
\end{align}
which are directly proportional to $S_k= |(\bar n \otimes n)_k|^2$, since they are both norms of the unique spin-$k$ bilinears. Indeed, for the first two ranks, one finds
\begin{align}
    S_1
    =
    \frac{3}{j_\Ss(j_\Ss+1)(2j_\Ss+1)}\,\Phi_1,
    \qquad
    S_2
    =
    \frac{120}{(2j_\Ss-1)(2j_\Ss)(2j_\Ss+1)(2j_\Ss+2)(2j_\Ss+3)}\,\Phi_2.
\end{align}
These factors can be obtained either by explicitly matching the bilinear $(\bar n \otimes n)_{k}$ to the matrix elements of the corresponding rank-$k$ irreducible tensor operators or, more simply, by evaluating both sides on the highest weight state $\ket{j_\Ss,j_\Ss}$ and fixing the proportionality constant.

Hence, this strategy provides a convenient way to relate $B_k$ and $S_k$ and, therefore, to switch between different bases for the potential. Note that the same relations can be used to derive analytic constraints on the orbit space by imposing $B_k\geq 0$. In practice, however, we find that the resulting bounds are generally weaker than the true physically reachable orbit space. Also, the condition $\sum_k B_k=1$ is obvious in this language.

\paragraph{$\bm{j_\Ss=3/2}$.}
For $j_\Ss=3/2$ we have only $B_{1,3}$. 
The two associated projectors read
\begin{align}
    P_1 = \frac{9}{20} - \frac{X}{5},
    \qquad
    P_3 = \frac{11}{20} + \frac{X}{5}
\end{align}
Using
\begin{align}
    \mu_1 = \bra{n,n} X\ket{n,n}= \bra{n}T_1 ^a \ket{n}\bra{n}T_1 ^a \ket{n} = \Phi_1 = 5 S_1
\end{align}
one obtains
\begin{align}
    B_1= \frac{9}{20} -  S_1
    \qquad
    B_3= \frac{11}{20} + S_1 
\end{align}

\paragraph{$\bm{j_\Ss=5/2}$.}
For $j_\Ss=5/2$ we have $B_{1,3,5}$. 
The projectors now are 
\begin{align}
\begin{aligned}
P_1 &= \frac{1}{70}X^2 - \frac{X}{20} - \frac{55}{224}, \\
P_3 &= -\frac{1}{45}X^2 - \frac{1}{30}X + \frac{155}{144}, \\
P_5 &= \frac{1}{126}X^2 +\frac{1}{12}X + \frac{341}{2016}.
\end{aligned}
\end{align}
which involve
\begin{align}
\begin{aligned}
    \mu_1 &= \bra{n,n} X\ket{n,n}= \bra{n}T ^a \ket{n}\bra{n}T ^a \ket{n} = \Phi_1 = \frac{35}{2} S_1,\\
    \mu_2 &= \bra{n,n} X^2\ket{n,n}= \bra{n}T ^a T^b \ket{n}\bra{n}T ^a T^b \ket{n} = \Phi_2 - \frac{1}{2}\Phi_1 + \frac{j_\Ss^2 (j_\Ss+1)^2}{3}\\
    &= 
    56S_2 - \frac{35}{4}S_1 + \frac{1225}{48},
\end{aligned}
\end{align}
where we used $\bra{n}T^a T^b \ket{n} =  \Phi_2 ^{ab}  + j_\Ss(j_\Ss+1) \delta^{ab}/3 + i \epsilon^{abc} T^c/2$.
Substituting these expressions, we find
\begin{align}
\begin{aligned}
    B_1 &= \frac{4}{5}S_2- S_1+\frac{5}{42},
    \\
    B_3 &= -\frac{56}{45}S_2-\frac{7}{18}S_1+\frac{55}{108},
    \\
    B_5 &= \frac{4}{9}S_2+\frac{25}{18}S_1+\frac{281}{756}.
\end{aligned}
\end{align}

\paragraph{$S_k$ identities.}
In expressing $B_k$ in terms of the $S_k$, we effectively restricted the projector identity to the symmetric subspace $\mathrm{Sym}^2(V_{j_\Ss})$, where $\ket{n,n}$ lives.
To derive instead relations among the $S_k$ themselves, it is more economical to use the minimal polynomial of
$ 
X = J_1 \cdot J_2 
$
on $\mathrm{Sym}^2(V_{j_\Ss})$. Since
\begin{align}
X=\frac12\Bigl(J_{\rm tot}^2-2j_\Ss(j_\Ss+1)\Bigr),
\end{align}
its eigenvalue on the spin-$k$ subspace is
\begin{align}
\lambda_k=\frac12\Bigl(k(k+1)-2j_\Ss(j_\Ss+1)\Bigr).
\end{align}
Hence on $\mathrm{Sym}^2(V_{j_\Ss})$ one has
\begin{align}
X^q M_{\rm sym}(X)\ket{n,n}=0,
\qquad
M_{\rm sym}(X)\equiv \prod_{k \, {\rm odd}} (X-\lambda_k), \qquad q \in \mathds{N}.
\end{align}
For $q=0,1,\dots,j_\Ss-\frac12$, these relations involve the moments
$
\mu_m = \bra{n,n}X^m\ket{n,n}
$ up to $m=2j_\Ss$, and since each of these decomposes as
\begin{align}
\mu_m=\Phi_m+\sum_{r<m} c_{mr}\Phi_r,
\end{align}
they determine recursively the higher-rank invariants $\Phi_{j_\Ss+1/2},\dots,\Phi_{2j_\Ss}$, and therefore the dependent variables
\begin{align}
S_{j_\Ss+1/2},\dots,S_{2j_\Ss},
\end{align}
in terms of
$
S_0,S_1,\dots,S_{j_\Ss-1/2}
$.

We illustrate this procedure explicitly for $j_\Ss=3/2$. Here one has
$
\lambda_1=-11/4,
\lambda_3= 9/4.
$
Therefore
\begin{align}
M_{\rm sym}(X)
&=\left(X+\frac{11}{4}\right)\left(X-\frac94\right)
= X^2+\frac12 X-\frac{99}{16},
\end{align}
and thus the recurrence relation is 
\begin{align}
\bra{n,n} X^q M_{\rm sym}(X)\ket{n,n}=0
\longrightarrow
\mu_{2+q}+\frac12\mu_{q+1}-\frac{99}{16} \mu_q=0
\end{align}
with the initial conditions $\mu_0 =1, \mu_1 =\Phi_1$. Focusing first on $q=0$, we can exploit $\mu_2 = \Phi_2- \Phi_1/2+75/16$ to obtain
\begin{align}
    \Phi_2 = \frac{3}{2} \longrightarrow S_2 = \frac{1}{4}.
\end{align}
This implies that the quadrupole size is fixed for $j_\Ss=3/2$, which is relevant for the Coleman-Weinberg potential (see \cref{app:CW}).
To obtain $S_3$ we should consider $q=1$ and explicitly decompose $\mu_3$ to obtain the $c_{3r}$ coefficients. In this case this can be avoided, because for the unit state $n$ we have 
\begin{align}
    S_0+S_1+S_2+S_3=1
\end{align}
so we get
\begin{align}
    S_3 = \frac{1}{2} - S_1.
\end{align}

For larger $j_\Ss$, the same method applies, but beyond the first step one needs the higher-moment decompositions of $\mu_m$ in order to extract explicitly the corresponding $\Phi_m$ ($S_m$).

\subsection{Residual symmetries, selection rules and Majorana representation}
\label{app:vacuumSymm}

Suppose the vacuum state $n$ has an $SU(2)$ stabilizer $H_n$, i.e., there is a subgroup $H_n \subset SU(2)$ whose action leaves it invariant up to an overall phase,
\begin{align}
    D^{(j_\Ss)}(h)\, n = e^{i \alpha_h} n \, .
\end{align}
If the original symmetry group is $U(2)$, the stabilizer $H_n$ corresponds to an exact residual symmetry group of the theory, as the overall phase can always be absorbed by $U(1) \subset U(2)$ to leave $n$ invariant. Both in the $U(2)$ and $SU(2)$ cases, however, it imposes strong constraints on the possible representations of composite operators, as we outline now.

Let us focus on the leading spin-$1/2$ composite,
\begin{align}
v_3 \equiv (n^2\bar n)_{1/2}.
\end{align}
By construction, it
must transform with the same overall phase,
\begin{align}
    D^{(1/2)}(h)\, v_3 = e^{i \alpha_h} v_3 \, .
\end{align}
This immediately implies a strong selection rule. If the phase $e^{i\alpha_h}$ induced by the action of $h$ on the vacuum cannot be realized in the spin-$1/2$ representation, then $v_3$ must vanish identically. Equivalently, if $e^{i\alpha_h}$ is an eigenvalue of $D^{(j_\Ss)}(h)$ on the vacuum $n$ but not of $D^{(1/2)}(h)$, then the spin-$1/2$ composite $v_3$ cannot be nonzero on that vacuum. The same logic also constrains whether spin-$1/2$ composites arising at different orders can be independent: all such composites must lie in the eigenspace of $D^{(1/2)}(h)$ with eigenvalue $e^{i\alpha_h}$. Note that, if this eigenspace is one-dimensional, then all nonzero spin-$1/2$ composites are necessarily proportional and therefore cannot furnish independent two-component vectors. In the present case, this is, in fact, the generic situation, since for any non-center element $h$ the spin-$1/2$ representation has two distinct eigenvalues and hence one-dimensional eigenspaces. An eigenspace of dimension larger than one can occur only for the $Z_2$ center elements $h$ for which $D^{(1/2)}(h)=- \mathds{1}_2$, and indeed the center is always a trivial stabilizer.

This selection rule provides a necessary condition for determining whether a vacuum can generate non-vanishing, independent spin-$1/2$ composites such as $v_3$. In a generic situation with no stabilizer, no such constraint is present, and all representations are, in principle, allowed and independent. By contrast, whenever the vacuum exhibits a non-trivial $H_n$, the allowed composites are strongly constrained by representation theory.
This logic applies equally to spin-$1/2$ composites with different $U(1)$ charges $(\Ss^4 \bar \Ss)_{1/2}, \dots$ or for different spin composites, for which the phase induced by the stabilizer action on the original state must be consistent with their transformation properties.\footnote{This notion can be extended straightforwardly to anti-unitary transformations too.} In particular, if we enforce $U(2)$ as in some of the models discussed in \cref{sec:models}, any vacuum with residual symmetry $H_n$ beyond the center cannot yield spin-$1/2$ composites of the same $U(1)$ charge that are independent at any order, and therefore cannot produce viable Yukawa matrices. Essentially, in these models, the residual symmetry induces an accidental $U(1)$ selection rule at the level of the spin-$1/2$ composites.

A particularly clear case is when $n$ admits a rotational stabilizer around some axis, so that for some rotation vector $\bm{\alpha}_*$ one has
$
n \to e^{-i \bm{\alpha}_* \cdot \bm{T}}\, n = e^{i\beta_*} n \, .
$
By an $SU(2)$ transformation one can always choose a basis in which this axis is aligned with $T^3$. The stabilizer condition then becomes
\begin{align}
n \;\to\; e^{-i\alpha_* T^3} n
= \sum_m e^{-i\alpha_* m} n_m \ket{j_\Ss,m}
= e^{i\beta_*} n \, .
\end{align}
This immediately constrains which components of $n$ can be simultaneously non-zero: for two non-null $n_m$ and $n_{m'}$ components, the stabilizer property implies
$
e^{-i\alpha_* m} = e^{-i\alpha_* m'} = e^{i\beta_*} \, ,
$ 
or equivalently
\begin{align}
\alpha_*(m-m') = 2\pi k \, , \qquad k\in\mathbb Z \, .
\end{align}
Since $|m|\leq j_\Ss$, this strongly restricts the set of allowed magnetic quantum numbers $m$.
Interestingly, it provides an immediate test for the existence of spin-$1/2$ composites: in the aligned basis, a spin-$1/2$ representation of $U(1)$ charge $+q$ only carries the phases $e^{\pm i \alpha_*/2}$ under the rotation, so if
\begin{align}
e^{iq\beta_*} \neq e^{\pm i\alpha_*/2},
\end{align}
it must vanish. Hence, a necessary condition for a non-vanishing spin-$1/2$ composite is that the set of $m$ satisfy
\begin{align}
m = m_0 \;\text{mod}\; \frac{2\pi}{\alpha_*}, \qquad
q m_0 = \mp \frac{1}{2} \;\text{mod}\; \frac{2\pi}{\alpha_*} \, .
\label{eq:m0condition}
\end{align}
Note that this condition is only necessary; the composite object can in principle still vanish either by accident or due to additional selection rules.

As a concrete example, a discrete $C_p$ stabilizer is generated by $\alpha_* = 2\pi/p$, so it implies $m = m_0 \text{ mod } p$
and $qm_0 = \mp 1/2 \text{ mod } p$ in order for the spin-1/2 composite to be non-identically vanishing. 
A $U(1)$ stabilizer is one for which any angle $\alpha_*$ leaves $n$ invariant up to a phase. This is possible only if all nonzero components of $n$ carry the same magnetic quantum number, so the vacuum must be of the form $\{m\}=m_0$. In this case the condition \cref{eq:m0condition} becomes exact rather than modular, and a spin-$1/2$ composite of net $U(1)$ charge $q$ can be nonzero only if
$
q m_0 = \mp 1/2
$.
This equation admits solutions only for $|q|=1$, in which case necessarily $m_0=\pm 1/2$. Therefore, in a $U(1)$ vacuum, the only potentially nonvanishing spin-$1/2$ composites are those of charge $\pm 1$, such as $v_3$, and even these require the vacuum to be aligned along $m_0=\pm 1/2$.

\paragraph{Majorana representation.} Identifying stabilizers of a generic state $n$ can be non-trivial, especially in the case of discrete symmetries. In $SU(2)$, however, this is surprisingly easy and intuitive in the \emph{Majorana representation} \cite{Majorana:1932ga}. This construction generalizes the familiar Bloch sphere picture \cite{Bloch:1946zza,Feynman:1957zz} by exploiting the fact that any pure spin-$j_\Ss$ state $n$ can be written as the fully symmetric product of $2j_\Ss$ spin-$1/2$ states $\ket{\psi_i}$,
\begin{align}
    n \sim \text{Sym} \left(\ket{\psi_1}\otimes \ket{\psi_2}\otimes \cdots \otimes \ket{\psi_{2j_\Ss}}\right).
\end{align}
Each of these spinors can in turn be associated with a point on the sphere $S^2$ through the map
\begin{align}
\ket{\psi_i} = \cos\theta/2 \ket{\uparrow} + e^{i\phi}\sin\theta/2 \ket{\downarrow}
\end{align}
where $\theta,\phi$ correspond to the polar and azimuthal angles. The angles are obtained from the roots of the Majorana polynomial associated with $n$; see \cite{RevModPhys.17.237,bengtsson2017geometry} for more details. In this way, the full state $n$ can be visualized as a constellation of $2j_\Ss$ points on the unit sphere, transforming under $SU(2)$ via the natural map $SU(2)\to SO(3)$.\footnote{
The group isomorphism is $SU(2)/Z_2 \simeq SO(3)$. The $Z_2$ quotient is the center of $SU(2)$, acting as $n\to (-1)^{2j_\Ss}n$. Hence, it acts only by a sign on each Majorana spinor and has no effect on the constellation or on the selection rules discussed above.} States of the form $n = \ket{j_\Ss,m}$ correspond to $j_\Ss + m$ points at the north pole and $j_\Ss - m$ points at the south pole in our convention.

This representation makes the identification of stabilizers of $n$ particularly clear: they are simply the geometric symmetries of the constellation.\footnote{This follows from the fact that $n$ is the fully symmetric product of the Majorana states $\ket{\psi_i}$. A stabilizer of the constellation acts by permuting these spinors, possibly up to an overall phase, and therefore leaves the full state $n$ invariant up to a global phase.} Furthermore, $SU(2)$ flavor transformations act as global rotations of the sphere, so that one can always choose a basis aligned with some symmetry axis of the constellation, in which the selection rules outlined earlier are evident. In particular, since for half-integer $j_\Ss$ the number of points is odd, the only allowed symmetries are axial ones, namely $SO(2) \simeq U(1)$, as well as cyclic and dihedral groups $C_p$ and $D_p$ ($p \leq 2j_\Ss$), for which it is always possible to identify a preferred symmetry axis \cite{Bacry:1974hc}.

Interestingly, in this language, the $S_k$ invariants admit an even more intuitive interpretation: they are proportional to the quantum multipole moments of a set of spin-$1/2$ particles with identical charges. This makes it very natural to identify which constellations maximize or minimize a given $S_k$.\footnote{Note that quantum and classical multipoles can differ for $k\geq 2$, due to interference effects that are absent in the classical picture. However, the correspondence at order $k$ becomes exact whenever all lower moments up to order $k-1$ vanish, since in that case the interference contributions are absent.}

\section{Potential minimization}

\subsection{$j_\Ss=3/2$}
\label{app:genericSU2_3/2}

For $j_\Ss=3/2$, there is one independent invariant of each type,
$B_u$ and $F_u$. We choose the convention
\begin{align}
    B_u
    &=
    \left[
        (n\otimes n)_1
        \otimes
        (n\otimes\bar n)_1
    \right]_0,
    &
    F_u
    &=
    \left[
        (n\otimes n)_1
        \otimes
        (n\otimes n)_1
    \right]_0,
\end{align}
where any other choice is equivalent up to a rescaling of the associated
quartic couplings. Using $S_0=1/4$ and a phase rotation of $\Ss$ to make
$\lambda_{B_u}$ real, the most general $SU(2)$-invariant angular potential
can be written as
\begin{align}
    \lambda_{\rm eff}(n)
    =
    \frac{1}{4}\lambda_0
    +\lambda_1 S_1
    +\lambda_{B_u}\Re B_u
    +\Re\lambda_{F_u}\Re F_u
    -\Im\lambda_{F_u}\Im F_u ,
    \label{eq:j32GeneralPotential}
\end{align}
with $\lambda_0$, $\lambda_1$, and $\lambda_{B_u}$ real. The corresponding
orbit space is four-dimensional and can be parametrized by
$(S_1,\Re B_u,\Re F_u,\Im F_u)$. In the Majorana representation, the state
is described by three points on the sphere and can possess
$U(1)$, $C_3$ or $D_3$, and $C_2$ stabilizers.

\paragraph{$U(2)$ case.}
In the $U(2)$-invariant limit,
$\lambda_{B_u}=\lambda_{F_u}=0$, and the orbit space reduces to the interval
parametrized by the dipole invariant $S_1$. The two extrema and their
representative states are given in \cref{eq:j32U2minima}. Their geometrical
interpretation is particularly simple in the Majorana representation.

\begin{figure}[!ht]
    \centering
    \begin{subfigure}[t]{0.48\textwidth}
        \centering
        \includegraphics[width=\textwidth]
        {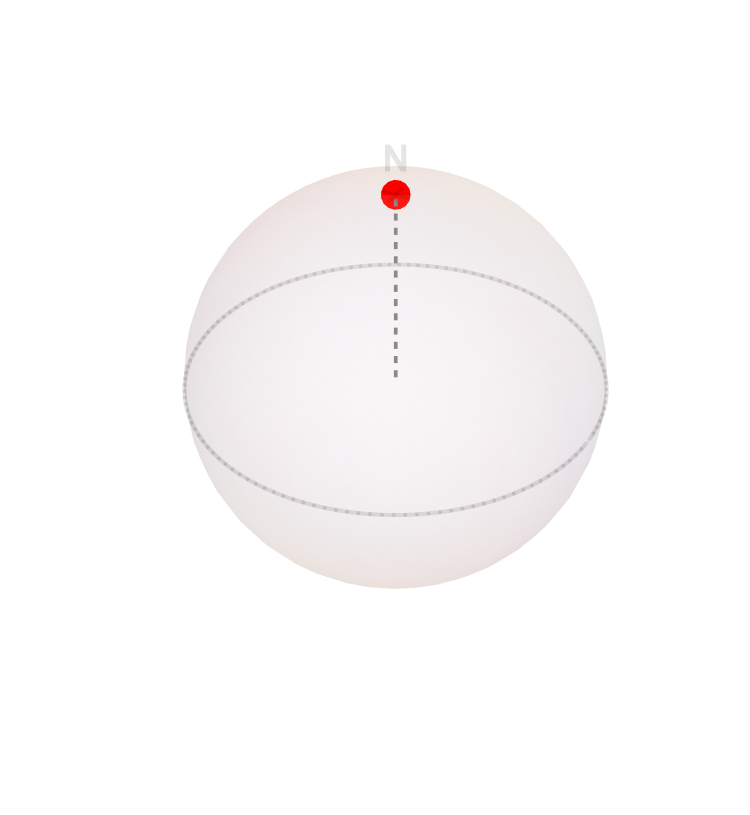}
        \caption{The maximum $S_1=9/20$, for which the three points
        coincide at a pole and the stabilizer is $U(1)_{3/2}$.}
    \end{subfigure}
    \hfill
    \begin{subfigure}[t]{0.48\textwidth}
        \centering
        \includegraphics[width=\textwidth]
        {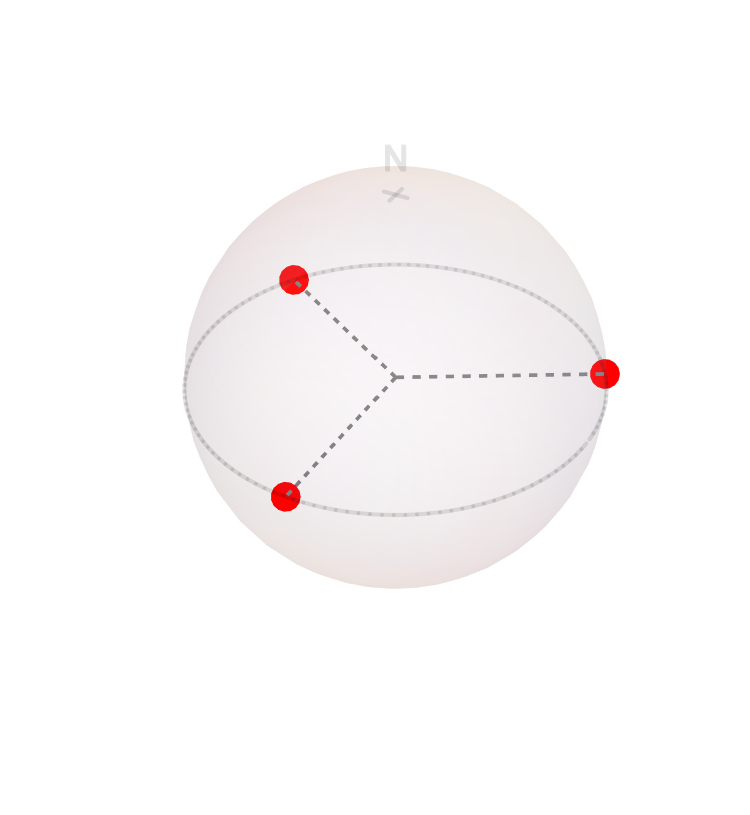}
        \caption{The minimum $S_1=0$, for which the three points form an
        equilateral triangle and the stabilizer is $D_3$.}
    \end{subfigure}
    \caption{Majorana constellations associated with the two extrema of
    the $U(2)$-invariant potential for $j_\Ss=3/2$.}
    \label{fig:3/2_constellation}
\end{figure}

For $\lambda_1<0$, the dipole is maximized by placing all three Majorana
points at the same pole. This gives the state
$n=\ket{3/2,\pm3/2}$ and preserves a continuous $U(1)_{3/2}$ stabilizer.
For $\lambda_1>0$, the dipole vanishes when the three points form an
equilateral triangle on the equator. The corresponding state
\begin{align}
    n
    =
    \frac{1}{\sqrt{2}}
    \left(
        \ket{3/2,-3/2}
        +
        \ket{3/2,3/2}
    \right)
\end{align}
preserves a $D_3$ stabilizer. The two constellations are shown in
\cref{fig:3/2_constellation}.

The corresponding selection rules make the obstruction transparent. For
the $U(1)_{3/2}$ vacuum, the vacuum charge cannot be matched by a
spin-$1/2$ composite. For the $D_3$ vacuum, the $C_3$ subgroup restricts
the state to magnetic components $m=\pm3/2$, which likewise forbids the
required spin-$1/2$ composites. Thus, both extrema give $v_3=0$, together
with vanishing higher-order spin-$1/2$ composites of the relevant charges.

\paragraph{$SU(2)$ case.}
The invariants $B_u$ and $F_u$ enlarge the orbit space to four dimensions,
which cannot be visualized directly. We first consider the
$Z_2$/CP-conserving limit,
\begin{align}
    \Im\lambda_{F_u}=0,
\end{align}
for which only the three orbit variables
$(S_1,\Re B_u,\Re F_u)$ enter the potential.

\begin{figure}[!ht]
    \centering
    \includegraphics[width=0.8\linewidth]
    {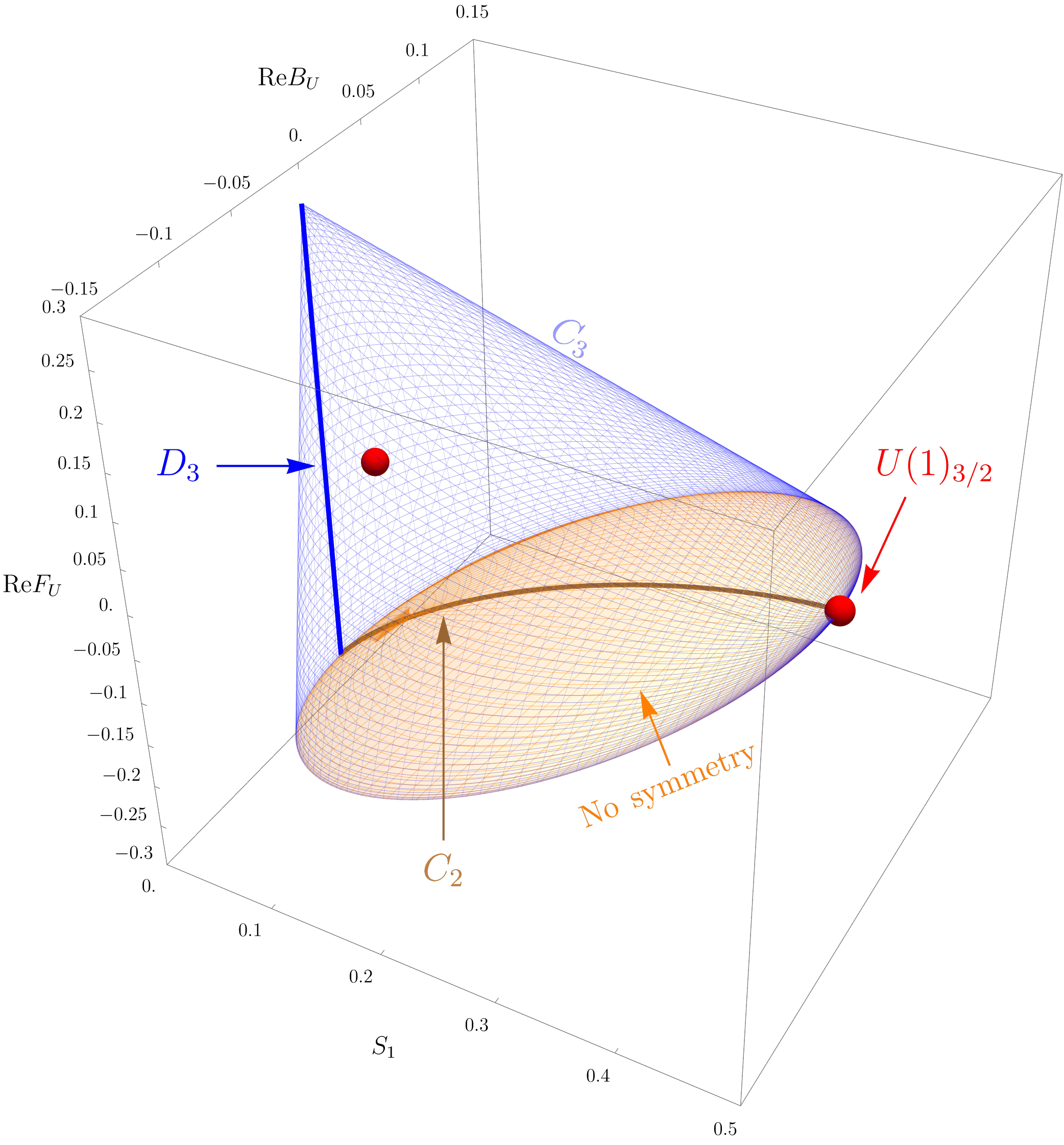}
    \caption{Orbit space for $j_\Ss=3/2$ in the $Z_2$/CP-conserving
    $SU(2)$ case, parametrized by
    $(S_1,\Re B_u,\Re F_u)$. The outer conical surface is generated by
    the $C_3$ family, while its special limits exhibit enhanced
    $D_3$ and $U(1)_{3/2}$ stabilizers. The concave base contains
    configurations with $v_3\neq0$, including a $C_2$ contour, but cannot
    generically support the minimum of a linear potential. The
    $U(1)_{1/2}$ configuration lies in the interior of the orbit space.}
    \label{fig:3/2_3d}
\end{figure}

We determine the boundary by combining a Monte Carlo scan of normalized
states with the boundary-search algorithm of
\cite{Kim:1997wf,Kim:1998qy}. Symmetric ans\"atze suggested by the
Majorana representation then provide an analytical parametrization of the
different boundary components. The resulting orbit space is shown in
\cref{fig:3/2_3d}.
The convex outer boundary is generated by the $C_3$ family
\begin{align}
    n
    =
    e^{i\gamma}
    \left(
        \cos\alpha\,\ket{3/2,-3/2}
        +
        \sin\alpha\,\ket{3/2,3/2}
    \right),
\end{align}
for which
\begin{align}
\begin{cases}
    S_1
    =
    \frac{9}{20}\cos^2(2\alpha),
    \\
    B_u
    =
    B_0 e^{2i\gamma}\sin(4\alpha),
    \\
    F_u
    =
    2B_0 e^{4i\gamma}\sin^2(2\alpha),
\end{cases}
\qquad
    B_0\equiv -\frac{3\sqrt{3}}{40},
\label{eq:3/2_C3fam}
\end{align}
and $v_3=0$. The endpoints $\alpha=0,\pi/2$ correspond to the
$U(1)_{3/2}$ configurations, while at $\alpha=\pi/4$ the stabilizer is
enhanced to $D_3$. The remaining part of the boundary is a concave base,
which contains configurations with $v_3\neq0$ and, along a special contour,
a $C_2$ stabilizer. Since the potential is linear in the orbit variables,
a generic minimum is selected on a convex or protruding component of the
boundary and cannot lie on this concave base. The latter can contribute
only for tuned orientations of the equipotential hyperplane, in which case
its minima are degenerate with configurations on the $C_3$ surface.
Therefore, already in the $Z_2$/CP-conserving limit, the global minimum can
always be chosen on the $C_3$ family and has $v_3=0$.

The three-dimensional orbit space provides a useful geometrical picture,
but the conclusion is not restricted to the $Z_2$/CP-conserving limit.
For a general complex $\lambda_{F_u}$, the orbit space is genuinely
four-dimensional. Nevertheless, the minimization can be performed
analytically by deriving inequalities among $S_1$, $B_u$, and $F_u$.
We now show that, at each fixed value of $S_1$, the lower bound on the
remaining part of the potential is saturated by the same $C_3$ family.

Introduce $T = (n\otimes n)_1$, $J = (\bar n \otimes n)_1$, and the associated Cartesian tensors $T_c$, $J_c$ ($J_c = J_c^*$ by construction).\footnote{The relation between the spherical and Cartesian components of a spin-1 tensor is \(X_{c,x}=(X_{-1}-X_{+1})/\sqrt{2}\), \(X_{c,y}=i(X_{+1}+X_{-1})/\sqrt{2}\), \(X_{c,z}=X_0\). Since the transformation between spherical and Cartesian is unitary, the norm is preserved, \( |X|^2 \equiv \sum_{m=-1}^{1} |X_m|^2 = \sum_{i=x,y,z} |X_{c,i}|^2 \equiv |X_c|^2 \).} Note that $F_u = (TT)_0 = -T_c \cdot T_c/\sqrt{3}$ and $B_u = (TJ)_0 = - T_c \cdot J_c/\sqrt{3}$. Split $T_c$ into a component proportional to $J_c$ and an orthogonal one $U_c$:
\begin{align}
    T_c= \alpha J_c + U_c.
\end{align}
Then simple manipulations lead to
\begin{align}
    \begin{cases}
        T_c \cdot J_c = \alpha J_c \cdot J_c \longrightarrow \alpha = -\sqrt{3} B_u/S_1\\
        |T_c|^2 = |\alpha|^2 |J_c|^2 + |U_c|^2 \longrightarrow |U_c|^2 = B_1 - 3 |B_u|^2/S_1 \\
        T_c \cdot T_c = \alpha^2 J_c \cdot J_c + U_c \cdot U_c \longrightarrow U_c \cdot U_c = - \sqrt{3} F_u - 3 B_u^2/S_1
    \end{cases}
\end{align}
which, together with the inequality $|U_c \cdot U_c| \leq |U_c|^2$, lead to 
\begin{align}
    | \sqrt{3} F_u S_1 + 3 B_u^2| \leq S_1 B_1- 3 |B_u|^2.
\end{align}
From this one also separately obtains, using $B_1 = 9/20-S_1$ (see \cref{app:potentialBases}) and introducing $s_1 = S_1/S_1 ^\text{max}$,
\begin{align}
    |F_u| \leq \frac{3 \sqrt{3}}{20} (1-s_1), \qquad |B_u| \leq  \frac{3 \sqrt{3}}{20} \sqrt{s_1(1-s_1)}
\end{align}
which immediately imply $|F_u|^\text{max} = 3 \sqrt{3}/20$ and $|B_u|^\text{max} = 3 \sqrt{3}/40$.
If we introduce the normalized variables
\begin{align}
    u = \frac{B_u}{\sqrt{S_1 B_1/3}}, \qquad v = \frac{\sqrt{3} F_u}{B_1},
\end{align}
the inequalities read
\begin{align}
    |u|\le 1, \qquad |v| \leq 1, 
    \qquad
    |v+u^2|\le 1-|u|^2 .
\end{align}
Crucially, on the $C_3$ family, these are saturated. This family corresponds to the curve $(u,v)=(-e^{2i\gamma} \,  \text{sign}(\sin 4\alpha), -e^{4i\gamma} ) \equiv (z,-z^2)$ with $|z|=1$.

Now define
\begin{align}
    a \equiv \lambda_{B_u} \sqrt{\frac{S_1 B_1}{3}} \in \mathbb R,
    \qquad
    b \equiv \frac{\lambda_{F_u} B_1}{\sqrt{3}} \in \mathbb C,
\end{align} 
so that the effective quartic can be written as
\begin{align}
    \lambda_{\rm eff} = \lambda_1 S_1 +  a\, \Re u + \Re (b v) \, .
\end{align}
The mixed inequality implies that, at fixed $u$, the allowed values of $v$ form the disk
\begin{align}
    v = -u^2 + (1-|u|^2) w,
    \qquad
    |w|\le 1.
\end{align}
Since the potential is linear in $v$, for fixed $u$ it is minimized on the boundary of this disk, namely for $w = - e^{- i \arg b}$, so that
\begin{align}
    \lambda_{\rm eff}
    \ge
    \lambda_1 S_1 + \Re(a u - b u^2) - |b| (1-|u|^2).
    \label{eq:appLambdaEff}
\end{align}
At this point one can minimize over $u$. The function to minimize is
\begin{align}
    F(u)
    =
    a\,\mathrm{Re}\,u
    + |b|\,|u|^2
    - \mathrm{Re}(b u^2),
    \qquad |u|\leq 1 .
\end{align}
Let first $b\neq0$ and write $b=|b|e^{-2i\eta}$. We define real variables $X,Y$ by
\begin{align}
    e^{-i\eta}u = X+iY ,
    \qquad X^2+Y^2\leq 1 .
\end{align}
Then 
\begin{align}
    F(X,Y)
    =
    a(X\cos\eta-Y\sin\eta)+2|b|Y^2,
    \qquad X^2+Y^2\leq 1 .
\end{align}
For fixed $Y$, the dependence on $X$ is linear. If $a\cos\eta\neq0$, no point with
$X^2+Y^2<1$ can be a minimum, since $X$ can be varied at fixed $Y$ in the direction
that lowers $F$. If $a\cos\eta=0$, the function is independent of $X$, so any interior
minimum is degenerate with a boundary point at the same value of $Y$. Hence, for any
value of $Y$, the minimum can be chosen with $X^2+Y^2=1$, namely with
\begin{align}
    |u|=1 .
\end{align}
If $b=0$, the function reduces to $F(u)=a\,\mathrm{Re}\,u$, and the same conclusion is immediate for $a\neq0$. For $a=0$ too, the function is fully degenerate, and a boundary
representative is as good as any other. Thus, for generic values of the parameters the minimum lies on $|u|=1$, except at nongeneric degenerate points where additional interior minimizers may occur, always degenerate with boundary ones. The mixed inequality then forces
\begin{align}
    |v+u^2|=0
    \qquad\Longrightarrow\qquad
    v=-u^2,
\end{align}
hich is precisely the normalized $C_3$ family. Hence, for every fixed
$S_1$, the lower bound in \cref{eq:appLambdaEff} is saturated by a physical
configuration on the $C_3$ branch. Since this branch spans the full interval
$0\leq S_1\leq S_1^{\rm max}$, the global minimum of the full tree-level
potential can always be chosen on the $C_3$ family. The final minimization
over $S_1$ only fixes the latitude $\alpha$. At nongeneric degenerate points,
additional non-symmetric minima may coexist with the $C_3$ representative.
Thus, also in the CP-violating case, the tree-level potential always admits
a global minimum with $v_3=0$.

We therefore conclude that the most general renormalizable
$SU(2)$-invariant potential for $j_\Ss=3/2$ always admits a global minimum
on the $C_3$ family, for which $v_3=0$. Non-symmetric minima with
$v_3\neq0$ may occur only at nongeneric degenerate points and are then
degenerate with a $C_3$ representative.

\subsection{$j_\Ss=5/2$}
\label{app:genericSU2_5/2}

For $j_\Ss=5/2$, the angular potential contains both the dipole and
quadrupole invariants. There is again one independent invariant of each
type, $B_u$ and $F_u$, and, after using a phase rotation of $\Ss$ to make
$\lambda_{B_u}$ real, the most general $SU(2)$-invariant potential reads
\begin{align}
    \lambda_{\rm eff}(n)
    =
    \frac{1}{6}\lambda_0
    +\lambda_1 S_1
    +\lambda_2 S_2
    +\lambda_{B_u}\Re B_u
    +\Re\lambda_{F_u}\Re F_u
    -\Im\lambda_{F_u}\Im F_u .
    \label{eq:j52GeneralPotential}
\end{align}
The corresponding orbit space is five-dimensional. In the Majorana representation,
$n$ is described by five points on the sphere and can support continuous
$U(1)$ stabilizers as well as $C_5$, $D_5$, $C_4$, $C_3$, $D_3$, and
$C_2$ stabilizers.

\begin{figure}[!htb]
    \centering
    \includegraphics[width=1.1\linewidth]
    {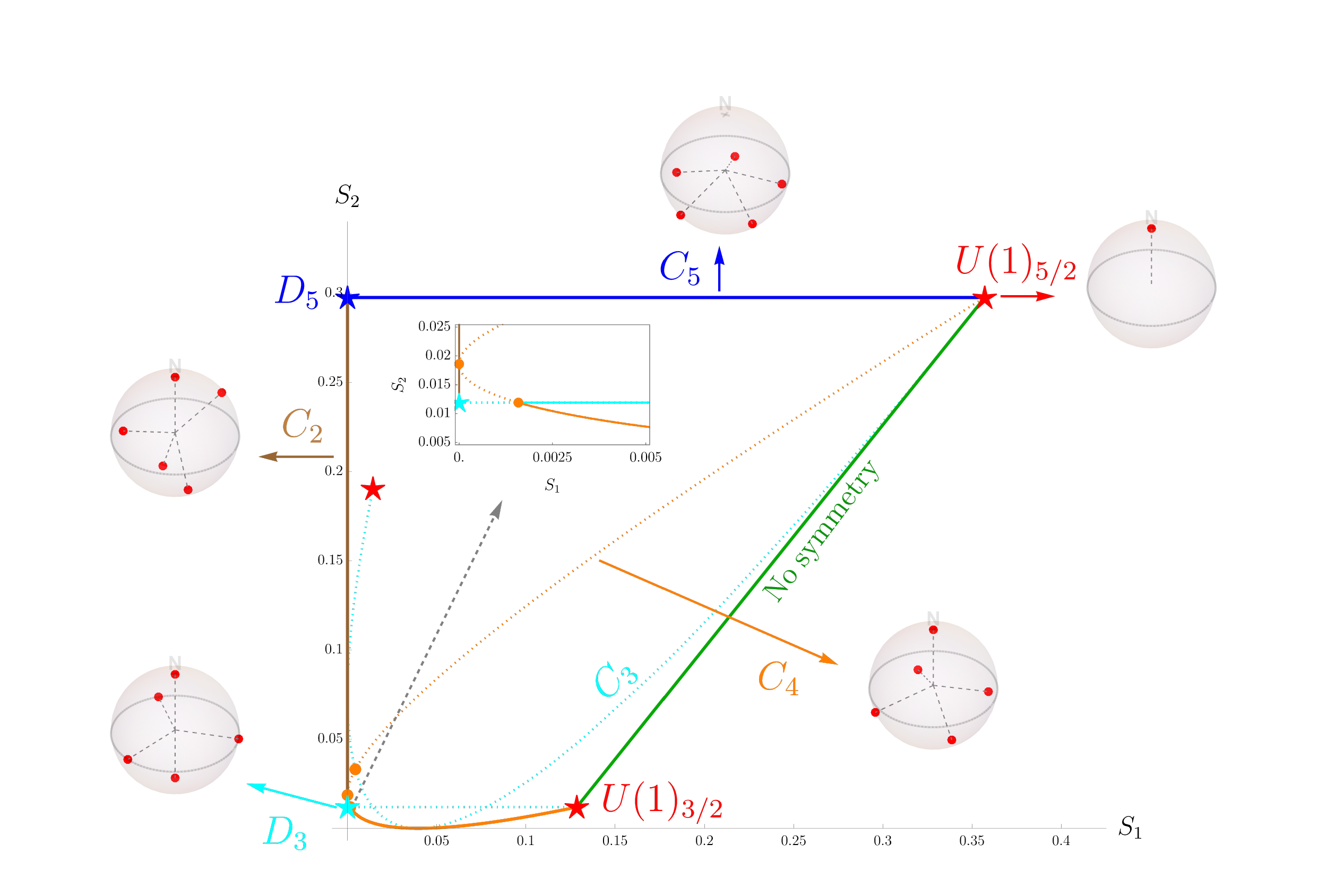}
    \caption{Detailed boundary of the orbit space for $j_\Ss=5/2$ in the
    $U(2)$-invariant case. Representative Majorana constellations are
    displayed next to the corresponding boundary components. The inset
    highlights the non-trivial crossing near the $D_3$ limit, while the
    dotted lines denote symmetric configurations lying in the interior of
    the orbit space.}
    \label{fig:5/2_domain_majorana}
\end{figure}

\paragraph{$U(2)$ case.}
In the $U(2)$-invariant limit, $\lambda_{B_u}=\lambda_{F_u}=0$, and the
orbit space reduces to the two-dimensional domain parametrized by
$(S_1,S_2)$. We determine its boundary by combining a numerical scan of
normalized states with symmetric ans\"atze suggested by the Majorana
representation. The resulting domain is shown in
\cref{fig:5/2_domain_majorana}.

Up to $SU(2)$ rotations, the relevant symmetric families are characterized
by the following non-vanishing magnetic components:
\begin{itemize}
    \item $U(1)$: $n=\ket{5/2,m}$, corresponding to Majorana points at
    the north and south poles.
    \item $C_5$: $m\in\{-5/2,5/2\}$, corresponding to a regular pentagon.
    \item $C_4$: $m\in\{\mp5/2,\pm3/2\}$, corresponding to a square and
    one point at a pole.
    \item $C_3$: $m\in\{\mp5/2,\pm1/2\}$ or
    $m\in\{-3/2,3/2\}$, corresponding respectively to an equilateral
    triangle with two points at a pole, or with one point at each pole.
    \item $C_2$: $m\in\{\mp5/2,\mp1/2,\pm3/2\}$, corresponding to a
    constellation invariant under a rotation by $\pi$.
\end{itemize}
The $D_5$ and $D_3$ stabilizers arise at special points where the cyclic
symmetry is enhanced by additional reflections. Among these symmetric
families, only the first $C_3$ family, with
$m\in\{\mp5/2,\pm1/2\}$, and the $C_2$ family can support a non-vanishing
$v_3$.\footnote{Although the $C_4$ stabilizer forces $v_3=0$, it allows
spin-$1/2$ composites with charge $+3$, such as
$(\Ss^4\bar\Ss)_{1/2}$, while $(\Ss^3)_{1/2}$ vanishes identically.}

The $C_3$ family with $v_3\neq0$ is represented by the horizontal cyan
dotted line in \cref{fig:5/2_domain_majorana}. Since it lies in the
interior of the orbit space, it cannot be selected as an isolated minimum
of a linear potential. Equipotential configurations are represented by
straight lines in the $(S_1,S_2)$ plane, and a generic orientation selects
the protruding or convex components of the boundary, associated with
$U(1)_{5/2}$, $D_5$, and $C_4$ configurations. All these vacua have
$v_3=0$.
The only boundary branches with $v_3\neq0$ are the non-symmetric and
$C_2$ branches. They are exposed only for the tuned relations given in
the main text. Any departure from these relations tilts the equipotential
line toward a branch with $v_3=0$. Moreover, at the tuned points, the
$v_3\neq0$ configurations are degenerate with other vacua for which
$v_3=0$.

\paragraph{$SU(2)$ case.}
Once $B_u$ and $F_u$ are included, the orbit space becomes
five-dimensional, or four-dimensional in the $Z_2$/CP-invariant limit,
and can no longer be visualized directly. Unlike for $j_\Ss=3/2$, we do
not have an analytical argument forcing the minimum onto a single
symmetric family.

The origin of this difference can be understood from the inequalities
used in \cref{app:genericSU2_3/2}. For $j_\Ss=5/2$, the norm $B_1$ entering
those inequalities depends on both independent $U(2)$ orbit variables,
\begin{align}
    B_1
    =
    \frac{4}{5}S_2-S_1+\frac{5}{42},
    \label{eq:j52B1}
\end{align}
as derived in \cref{app:potentialBases}. An analytical argument analogous
to the $j_\Ss=3/2$ case would therefore require a family of physical
configurations that both saturates the relevant inequalities and spans the
full two-dimensional $(S_1,S_2)$ domain. No symmetric family with these
properties exists.

The $C_5$ family and the $C_3$ family with
$m\in\{-3/2,3/2\}$ do saturate the inequalities, but each traces only a
one-dimensional curve in the $(S_1,S_2)$ plane. The $C_5$ family is
particularly competitive because it spans the full allowed range of
$S_1$, while $S_2$ remains fixed at its maximum. It is consequently
favored when $\lambda_2$ is sufficiently negative. For more general
quartic couplings, however, it competes with the $C_3$ family and with
configurations that do not saturate the inequalities but have greater
freedom to optimize the $(S_1,S_2)$-dependent part of the potential.

We therefore study the general potential numerically by sampling the
quartic-coupling space and minimizing over normalized field
configurations. The $C_5$ vacuum is selected particularly often, while
$C_4$ and other symmetric configurations also occur. Within our scan, we
find no non-symmetric global minima and no minima with $v_3\neq0$.
Although this does not constitute an analytical proof analogous to the
$j_\Ss=3/2$ case, it provides strong evidence that the most general
renormalizable $SU(2)$-invariant potential for $j_\Ss=5/2$ also fails to
produce a viable vacuum generically.

\subsection{Coleman-Weinberg}
\label{app:CW}

The $SU(2)$ gauge Coleman-Weinberg term can be written as
\begin{equation}
V_{\rm CW}^{\rm gauge}
=
\frac{3 g_F^4 x^2}{64\pi^2}
\left[
A(n)\left( \log\!\left(\frac{g_F^2 x}{\mu^2}\right) -\frac56 \right)
+
B(n)
\right]
\label{eq:appCW}
\end{equation}
with $x=|\Ss|^2$, and 
\begin{equation}
A(n)\equiv \tr\, G(n)^2,
\qquad
B(n)\equiv \tr \bigl(G(n)^2\log G(n)\bigr).
\end{equation}
Since \(G\) is a real symmetric \(3\times 3\) matrix, it can always be diagonalized in adjoint space,
\begin{equation}
G = O^T \,\mathrm{diag}(\rho_1,\rho_2,\rho_3)\, O,
\qquad
\rho_i \geq 0,
\qquad
\sum_{i=1}^3 \rho_i = \tr G = j_\Ss(j_\Ss+1) \equiv S \text{ (Casimir)},
\end{equation}
where $\rho_i = \rho_i(n)$.
Therefore the pure gauge CW depends only on the three positive eigenvalues \(\rho_i\), subject to fixed sum:
\begin{equation}
V_{\rm CW}^{\rm gauge}
=
\frac{3 g_F^4 x^2}{64\pi^2}
\sum_{i=1}^3
f(\rho_i),
\qquad
f(\rho)=\rho^2\left(\log\frac{g_F^2 x\,\rho}{\mu^2}-\frac56\right).
\end{equation}
To determine the minimum at fixed $x$, one may extremize the function
\begin{align}
    \sum_{i=1}^3 f(\rho_i)
\end{align}
under the constraint $\sum_i \rho_i=S $. Introducing a Lagrange multiplier $\lambda$,
\begin{align}
    \mathcal L = \sum_{i=1}^3 f(\rho_i) - \lambda \left(\sum_{i=1}^3 \rho_i - S\right),
\end{align}
the stationarity conditions read
\begin{align}
    f'(\rho_i)=\lambda,
    \qquad i=1,2,3.
\end{align}
Since the problem is symmetric in the three eigenvalues, the isotropic point
$
    \rho_1=\rho_2=\rho_3=S/3
$
is always a stationary point.
To determine whether it is a minimum, note that
\begin{align}
    f''(\rho)
    &=2\log\frac{g_F^2 x\,\rho}{\mu^2}+\frac43.
\end{align}
Hence $f$ is convex whenever
\begin{align}
    \frac{g_F^2 x\,\rho}{\mu^2}\ge e^{-2/3}.
\end{align}
If this condition holds throughout the physically allowed range of $\rho$, then by convexity
\begin{align}
    \frac13\sum_{i=1}^3 f(\rho_i)\ge
    f\!\left(\frac{\rho_1+\rho_2+\rho_3}{3}\right)
    =
    f\!\left(\frac{S}{3}\right),
\end{align}
with equality if and only if $\rho_1=\rho_2=\rho_3=S/3$. In that case, the pure gauge CW potential is globally minimized by the isotropic configuration
\begin{align}
    \rho_1=\rho_2=\rho_3=\frac{j_\Ss(j_\Ss+1)}{3},
    \qquad\Longleftrightarrow\qquad
    Q=0.
\end{align}

Now for half-integer $j_\Ss$, each $\rho_i$ is an expectation value of a squared spin component and therefore satisfies $\rho_i\ge 1/4$. A sufficient condition for global convexity is then
\begin{align}
    1 \le \frac{e^{2/3}}{4}  \frac{ g_F^2 x}{\mu^2} \approx 0.49  \frac{ g_F^2 x}{\mu^2} 
\end{align}
This condition is sufficient but not necessary. A weaker and more directly relevant requirement is obtained by evaluating the convexity condition at the isotropic point itself,
\begin{align}
    f''(\rho_*) = 2 \log \!\left(\frac{g_F^2 x\,S}{3\mu^2}\right) + \frac{4}{3}.
\end{align}
Hence the isotropic configuration is a local minimum whenever
\begin{align}
    \frac{g_F^2 x }{\mu^2 } > \frac{3 e^{-2/3}}{S}.
\end{align}
For the natural choice $\mu\sim g_F\sqrt{x}$ this condition is comfortably satisfied in the cases of interest. Indeed, $3 e^{-2/3}/S < 1$ for $j_\Ss \geq 3/2$.
Thus, for the natural scale choice, the isotropic point is certainly a local minimum of the pure gauge CW potential. Proving that it is also the global minimum requires the stronger global convexity condition above.

\paragraph{Flat directions.}
Whenever the isotropic configuration $Q=0$ is selected, the pure gauge
Coleman--Weinberg potential necessarily possesses several massless angular
directions. To see this, let $\theta_i$, with
$i=1,\ldots,4j_\Ss+1$, denote real coordinates on the space of normalized
configurations $n^\dagger n=1$. At fixed $x$, the angular dependence of
the potential can be expressed entirely in terms of the five real
components $Q_A$ of the quadrupole,
\begin{align}
    V_{\rm CW}^{\rm gauge}(x,n)
    =
    V_{\rm CW}^{\rm gauge}\bigl(x,Q_A(n)\bigr),
    \qquad
    A=1,\ldots,5.
\end{align}
At $Q=0$, rotational invariance implies
\begin{align}
    \left.
    \frac{\partial V_{\rm CW}^{\rm gauge}}{\partial Q_A}
    \right|_{Q=0}
    =0.
\end{align}
The angular Hessian therefore takes the form
\begin{align}
    \left(M_{\rm ang}^2\right)_{ij}
    &=
    \left.
    \frac{\partial^2V_{\rm CW}^{\rm gauge}}
         {\partial Q_A\partial Q_B}
    \right|_{Q=0}
    \frac{\partial Q_A}{\partial\theta_i}
    \frac{\partial Q_B}{\partial\theta_j},
    \label{eq:CWHessianRank}
\end{align}
where the term proportional to
$\partial V_{\rm CW}^{\rm gauge}/\partial Q_A$ vanishes at the isotropic point.
It follows that
\begin{align}
    \operatorname{rank}M_{\rm ang}^2\leq 5,
\end{align}
and hence that the number of massless real angular directions satisfies
\begin{align}
    N^{\rm ang}
    \geq
    \bigl[2(2j_\Ss+1)-1\bigr]-5
    =
    4j_\Ss-4.
    \label{eq:CWNumberFlatDirections}
\end{align}
If $SU(2)$ is completely broken, three of these directions are eaten by the gauge fields, leaving at least
\begin{align}
    N^{\rm moduli} \geq 4j_\Ss-7
\end{align}
physical massless moduli at this level. If a continuous subgroup remains unbroken,
fewer Goldstone modes are eaten and the number of moduli is correspondingly larger.

\paragraph{$\bm{j_\Ss=3/2}$ perturbations.}
To see how the CW shifts the vacuum away from the tree-level minimum, let us consider the simplest example, so the  $j_\Ss=3/2$ in the $U(2)$ limit. Recall here $\lambda_\text{eff}^\text{tree} = \lambda_0 + \lambda_1 S_1$ and the minima were either $U(1)$ conserving ($\lambda_1<0$, $n= \ket{3/2,\pm 3/2}$) or $D_3$ conserving  ($\lambda_1>0$, $n= ( \ket{3/2, -3/2} + \ket{3/2, 3/2}\sqrt{2}$).
Let's focus on the $U(1)$ vacuum
\begin{equation}
n_{U(1)}= (0,0,0,1).
\end{equation}
A perturbation in the $m=1/2$ direction is a pure infinitesimal $SU(2)$ rotation at leading order and therefore not a physical deformation of the orbit, while along $m=-3/2$ leads exactly to the $C_3$ family.\footnote{$U(\alpha)n \approx i \alpha^a T^a n$ and $T^- \ket{3/2,3/2}\propto \ket{3/2,1/2}$.} Hence, the first non-trivial and interesting physical perturbation is along the $m=-1/2$ direction:
\begin{equation}
n(\epsilon)= \left(
0,
\eps,
0,
\sqrt{1-\eps^2} \right) ,
\qquad
\epsilon\in \mathbb R.
\end{equation}
This family is not completely generic: it still preserves a $C_2$ symmetry, since the occupied magnetic quantum numbers differ by $2$, but allows a non-null $v_3 \sim (\epsilon, 0)$.
Now for $j_\Ss=3/2$ one has
\begin{equation}
S_1=\frac{1}{4}\, |\langle \vec T\rangle|^2.
\end{equation}
Along the above deformation,
\begin{equation}
\langle T^1\rangle=\langle T^1\rangle=0,
\langle T^3\rangle=\frac32-2\epsilon^2 \longrightarrow  S_1(\epsilon)=\frac{9}{20}-\frac{6}{5}\epsilon^2+\mathcal{O}(\epsilon^4).
\end{equation}
Therefore, the tree-level potential expands as
\begin{equation}
V_{\rm tree}(\epsilon)=
V_{\rm tree}(0)
+
-\frac{6}{5}\lambda_1\epsilon^2 x^2
+
\mathcal{O}(\epsilon^4).
\end{equation}
The dimensionless gauge mass matrix is diagonal,
\begin{equation}
G(\epsilon)=
\begin{pmatrix}
\frac34+\epsilon^2+\sqrt3\,\epsilon\sqrt{1-\epsilon^2} & 0 & 0\\
0 & \frac34+\epsilon^2-\sqrt3\,\epsilon\sqrt{1-\epsilon^2} & 0\\
0 & 0 & \frac94-2\epsilon^2
\end{pmatrix},
\end{equation}
and expanding the gauge Coleman-Weinberg term gives
\begin{equation}
V_{\rm CW}^{\rm gauge}(\epsilon)=
V_{\rm CW}^{\rm gauge}(0)
+
\frac{3 g_F^4 x^2}{64\pi^2}
\left(6-9\log 3\right)\epsilon^2
+
\mathcal{O}(\epsilon^4).
\end{equation}
The total quadratic coefficient is therefore
\begin{equation}
x^2
\left[
-\frac{6}{5}\lambda_1
+
\frac{3 g_F^4}{64\pi^2}\left(6-9\log 3\right)
\right].
\end{equation}
and hence the aligned tree vacuum becomes unstable when
\begin{equation}
-\frac{6}{5}\lambda_1
+
\frac{3 g_F^4}{64\pi^2}\left(6-9\log 3\right)
<0 \longrightarrow |\lambda_1|
<
\frac{5}{6}\,
\frac{3 g_F^4}{64\pi^2}\,
\left(9\log 3-6\right).
\end{equation}
Numerically this is 
$|\lambda_1|
\lesssim 1.5\times 10^{-2}\, g_F^4$, 
which is indeed approximately a 1-loop factor.
This perturbative analysis shows that the gauge boson CW term does not induce a linear shift away from the old high-symmetry vacuum, as is obvious from symmetry arguments. It, however, modifies the curvature in physical directions. If the tree-level quartic is sufficiently small, the 1-loop gauge contribution can make the old vacuum unstable and drive the system toward a new minimum. The perturbation considered above preserves a residual $C_2$ symmetry, so a full numerical minimization is required to determine whether the true new vacuum is this or fully generic, and whether the corresponding $v_3$ is indeed nonzero.
A similar analysis can be carried out for the $C_3$ minimum as well.

\paragraph{$\bm{j_\Ss=3/2}$ minimization.}
For completeness, we report the full minimization of the Coleman-Weinberg potential alone. Recall that for $j_\Ss=3/2$ the size of the quadrupole is fixed, $|Q|^2=\Phi_2 = 3/2$, see \cref{app:potentialBases}. As a result, in \cref{eq:appCW} one has $A(n)=\text{const}$, and the angular dependence of the potential is entirely controlled by the minimization of $B(n)$, subject to the constraints $\Phi_2 = 3/2$ and $\tr G = \sum_i \rho_i = 15/4$. In particular, the first condition already implies that the isotropic configuration $G\propto \mathds{1}$ is not accessible in this case.

\begin{figure}[!htb]
    \centering
    \includegraphics[width=0.5\linewidth]{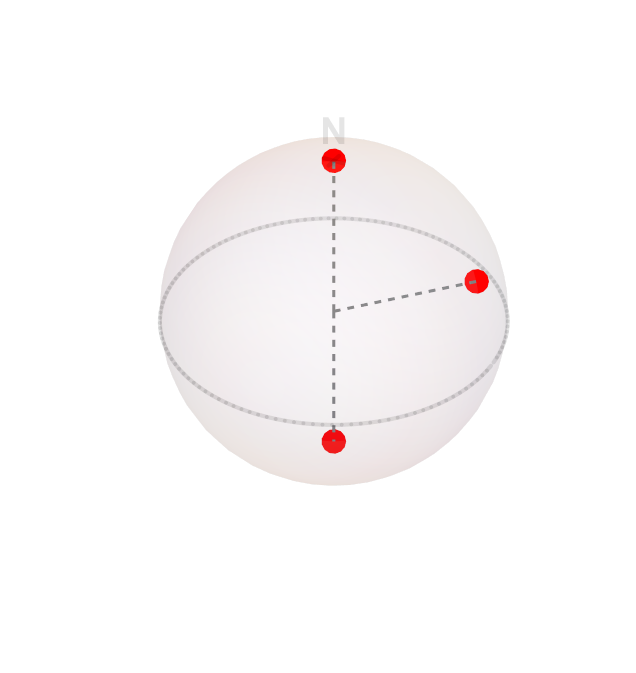}
    \caption{Majorana constellation associated to the state minimizing the CW potential alone for $j_\Ss=3/2$. Here $\alpha = \pi/3$ ($\gamma$ is uninfluential).}
    \label{fig:app_majoCW}
\end{figure}

Together, these two constraints fix two of the three eigenvalues $\rho_i$, so that the minimization of $B(n)$ reduces to a simple one-variable problem. One finds the minimum
\begin{align}
    G=
    \begin{pmatrix}
        \frac{1}{4} & 0 & 0 \\
        0 & \frac{7}{4} & 0 \\
        0 & 0 & \frac{7}{4}
    \end{pmatrix}.
\end{align}
This minimum is realized by the ansatz
\begin{align}
n = e^{i \gamma} \left( \cos \alpha \ket{3/2,-1/2}+ \sin \alpha \ket{3/2,1/2} \right),
\end{align}
for which $G(n)$ is already diagonal and takes precisely the form above. The Majorana constellation associated to this state is depicted in \cref{fig:app_majoCW}.

A few remarks are in order. First, for generic values of $\alpha$ and $\gamma$ the minimum does not preserve any special symmetry, and therefore leads to non-vanishing spin-$1/2$ composites, in agreement with our general expectations; in particular $v_3 \propto (-\cos \alpha, \sin \alpha)$. At the endpoints $\alpha=0,\pi/2$, the vacuum exhibits a residual $U(1)_{1/2}$ symmetry, which still allows spin-$1/2$ irreps. For $\alpha = \pi/4$ it exhibits a $Z_2$ along axes orthogonal to $\hat z$; indeed for this value the state is $T^1$ eigenstate.

Second, the matrix $G(n)$, and therefore the Coleman-Weinberg potential itself, is independent of both $\alpha$ and $\gamma$. These two parameters thus correspond to accidental flat directions, not fixed by the minimization, and are exactly massless at this level. Indeed, a straightforward calculation shows that among the eight real scalar degrees of freedom of the $j_\Ss=3/2$ multiplet, only three acquire a mass. The remaining five are massless: three correspond to the Goldstone modes associated with the breaking of $SU(2)$ and are eaten by the gauge bosons, while the other two correspond to the phase $\gamma$, Goldstone boson of the accidental global $U(1)$, and to the accidental modulus $\alpha$. The tree-level potential \cref{eq:potentialSU2} lifts the moduli.

\bibliographystyle{JHEP}
\bibliography{biblioPaper.bib}

@article{Antusch:2023shi,
    author = "Antusch, Stefan and Greljo, Admir and Stefanek, Ben A. and Thomsen, Anders Eller",
    title = "{U(2) Is Right for Leptons and Left for Quarks}",
    eprint = "2311.09288",
    archivePrefix = "arXiv",
    primaryClass = "hep-ph",
    reportNumber = "KCL-PH-TH/2023-64",
    doi = "10.1103/PhysRevLett.132.151802",
    journal = "Phys. Rev. Lett.",
    volume = "132",
    number = "15",
    pages = "151802",
    year = "2024"
}

@article{Fuentes-Martin:2019mun,
    author = "Fuentes-Mart{\'\i}n, Javier and Isidori, Gino and Pag{\`e}s, Julie and Yamamoto, Kei",
    title = "{With or without U(2)? Probing non-standard flavor and helicity structures in semileptonic B decays}",
    eprint = "1909.02519",
    archivePrefix = "arXiv",
    primaryClass = "hep-ph",
    reportNumber = "ZU-TH-42/19",
    doi = "10.1016/j.physletb.2019.135080",
    journal = "Phys. Lett. B",
    volume = "800",
    pages = "135080",
    year = "2020"
}

@article{Greljo:2025ljr,
    author = "Greljo, Admir and Palavri{\'c}, Ajdin and Tunja, Mirsad and Zupan, Jure",
    title = "{Expanding the landscape of exotic muon decays}",
    eprint = "2510.08674",
    archivePrefix = "arXiv",
    primaryClass = "hep-ph",
    doi = "10.1103/cb52-r75c",
    journal = "Phys. Rev. D",
    volume = "113",
    number = "7",
    pages = "075022",
    year = "2026"
}

@article{MartinCamalich:2020dfe,
    author = "Martin Camalich, Jorge and Pospelov, Maxim and Vuong, Pham Ngoc Hoa and Ziegler, Robert and Zupan, Jure",
    title = "{Quark Flavor Phenomenology of the QCD Axion}",
    eprint = "2002.04623",
    archivePrefix = "arXiv",
    primaryClass = "hep-ph",
    doi = "10.1103/PhysRevD.102.015023",
    journal = "Phys. Rev. D",
    volume = "102",
    number = "1",
    pages = "015023",
    year = "2020"
}

@article{Evans:2021gyd,
    author = "Evans, Matthew and others",
    title = "{A Horizon Study for Cosmic Explorer: Science, Observatories, and Community}",
    eprint = "2109.09882",
    archivePrefix = "arXiv",
    primaryClass = "astro-ph.IM",
    reportNumber = "CE-P2100003-v7, Cosmic Explorer technical report CE-P2100003-v6",
    month = "9",
    year = "2021"
}

@article{ET:2019dnz,
    author = "Maggiore, Michele and others",
    collaboration = "ET",
    title = "{Science Case for the Einstein Telescope}",
    eprint = "1912.02622",
    archivePrefix = "arXiv",
    primaryClass = "astro-ph.CO",
    doi = "10.1088/1475-7516/2020/03/050",
    journal = "JCAP",
    volume = "03",
    pages = "050",
    year = "2020"
}

@article{Fabri:2025fsc,
    author = "Fabri, Noemi and Isidori, Gino and Racco, Davide",
    title = "{Probing Flavour Deconstruction via Primordial Gravitational Waves}",
    eprint = "2509.12414",
    archivePrefix = "arXiv",
    primaryClass = "hep-ph",
    reportNumber = "ZU-TH 55/25",
    month = "9",
    year = "2025"
}

@article{Greljo:2019xan,
    author = "Greljo, Admir and Opferkuch, Toby and Stefanek, Ben A.",
    title = "{Gravitational Imprints of Flavor Hierarchies}",
    eprint = "1910.02014",
    archivePrefix = "arXiv",
    primaryClass = "hep-ph",
    reportNumber = "CERN-TH-2019-162",
    doi = "10.1103/PhysRevLett.124.171802",
    journal = "Phys. Rev. Lett.",
    volume = "124",
    number = "17",
    pages = "171802",
    year = "2020"
}

@article{Chrysostomou:2025vrg,
    author = "Chrysostomou, Anna and Cornell, Alan S. and Darm{\'e}, Luc and Deandrea, Aldo and Demartini, Thibault",
    title = "{Gravitational waves from flavoured SU(2) early-universe phase transitions}",
    eprint = "2512.02148",
    archivePrefix = "arXiv",
    primaryClass = "hep-ph",
    month = "12",
    year = "2025"
}

@article{Barbieri:2011ci,
    author = "Barbieri, Riccardo and Isidori, Gino and Jones-Perez, Joel and Lodone, Paolo and Straub, David M.",
    title = "{$U(2)$ and Minimal Flavour Violation in Supersymmetry}",
    eprint = "1105.2296",
    archivePrefix = "arXiv",
    primaryClass = "hep-ph",
    doi = "10.1140/epjc/s10052-011-1725-z",
    journal = "Eur. Phys. J. C",
    volume = "71",
    pages = "1725",
    year = "2011"
}

@article{Alonso:2013nca,
    author = "Alonso, R. and Gavela, M. B. and Isidori, G. and Maiani, L.",
    title = "{Neutrino Mixing and Masses from a Minimum Principle}",
    eprint = "1306.5927",
    archivePrefix = "arXiv",
    primaryClass = "hep-ph",
    reportNumber = "CERN-PH-TH-2013-147",
    doi = "10.1007/JHEP11(2013)187",
    journal = "JHEP",
    volume = "11",
    pages = "187",
    year = "2013"
}

@article{DAmbrosio:2002vsn,
    author = "D'Ambrosio, G. and Giudice, G. F. and Isidori, G. and Strumia, A.",
    title = "{Minimal flavor violation: An Effective field theory approach}",
    eprint = "hep-ph/0207036",
    archivePrefix = "arXiv",
    reportNumber = "CERN-TH-2002-147, IFUP-TH-2002-17",
    doi = "10.1016/S0550-3213(02)00836-2",
    journal = "Nucl. Phys. B",
    volume = "645",
    pages = "155--187",
    year = "2002"
}

@article{Alonso:2011yg,
    author = "Alonso, R. and Gavela, M. B. and Merlo, L. and Rigolin, S.",
    title = "{On the scalar potential of minimal flavour violation}",
    eprint = "1103.2915",
    archivePrefix = "arXiv",
    primaryClass = "hep-ph",
    reportNumber = "FTUAM-11-39, IFT-UAM-CSIC-11-09, TUM-HEP-796-11, DFPD-11-TH-2",
    doi = "10.1007/JHEP07(2011)012",
    journal = "JHEP",
    volume = "07",
    pages = "012",
    year = "2011"
}

@article{Arkani-Hamed:2026wwy,
    author = "Arkani-Hamed, Nima and Figueiredo, Carolina and Hall, Lawrence J. and Manzari, Claudio Andrea",
    title = "{Generating the fermion mass hierarchy at the TeV scale}",
    eprint = "2602.17754",
    archivePrefix = "arXiv",
    primaryClass = "hep-ph",
    month = "2",
    year = "2026"
}

@article{Cornella:2023zme,
    author = "Cornella, Claudia and Curtin, David and Neil, Ethan T. and Thompson, Jedidiah O.",
    title = "{Mapping and probing Froggatt-Nielsen solutions to the quark flavor puzzle}",
    eprint = "2306.08026",
    archivePrefix = "arXiv",
    primaryClass = "hep-ph",
    reportNumber = "MITP-23-026",
    doi = "10.1103/PhysRevD.111.015042",
    journal = "Phys. Rev. D",
    volume = "111",
    number = "1",
    pages = "015042",
    year = "2025"
}

@article{Greljo:2024evt,
    author = "Greljo, Admir and Smolkovi{\v{c}}, Aleks and Valenti, Alessandro",
    title = "{Froggatt-Nielsen ALP}",
    eprint = "2407.02998",
    archivePrefix = "arXiv",
    primaryClass = "hep-ph",
    doi = "10.1007/JHEP09(2024)174",
    journal = "JHEP",
    volume = "09",
    pages = "174",
    year = "2024"
}

@article{Fedele:2020fvh,
    author = "Fedele, Marco and Mastroddi, Alessio and Valli, Mauro",
    title = "{Minimal Froggatt-Nielsen textures}",
    eprint = "2009.05587",
    archivePrefix = "arXiv",
    primaryClass = "hep-ph",
    reportNumber = "UCI-TR-2020-13",
    doi = "10.1007/JHEP03(2021)135",
    journal = "JHEP",
    volume = "03",
    pages = "135",
    year = "2021"
}

@article{Leurer:1993gy,
    author = "Leurer, Miriam and Nir, Yosef and Seiberg, Nathan",
    title = "{Mass matrix models: The Sequel}",
    eprint = "hep-ph/9310320",
    archivePrefix = "arXiv",
    reportNumber = "RU-93-43, WIS-93-93-PH",
    doi = "10.1016/0550-3213(94)90074-4",
    journal = "Nucl. Phys. B",
    volume = "420",
    pages = "468--504",
    year = "1994"
}

@article{Leurer:1992wg,
    author = "Leurer, Miriam and Nir, Yosef and Seiberg, Nathan",
    title = "{Mass matrix models}",
    eprint = "hep-ph/9212278",
    archivePrefix = "arXiv",
    reportNumber = "RU-92-59, WIS-92-94-PH",
    doi = "10.1016/0550-3213(93)90112-3",
    journal = "Nucl. Phys. B",
    volume = "398",
    pages = "319--342",
    year = "1993"
}

@article{Froggatt:1978nt,
    author = "Froggatt, C. D. and Nielsen, Holger Bech",
    title = "{Hierarchy of Quark Masses, Cabibbo Angles and CP Violation}",
    reportNumber = "CERN-TH-2519",
    doi = "10.1016/0550-3213(79)90316-X",
    journal = "Nucl. Phys. B",
    volume = "147",
    pages = "277--298",
    year = "1979"
}

@article{Feruglio:2015jfa,
    author = "Feruglio, Ferruccio",
    title = "{Pieces of the Flavour Puzzle}",
    eprint = "1503.04071",
    archivePrefix = "arXiv",
    primaryClass = "hep-ph",
    doi = "10.1140/epjc/s10052-015-3576-5",
    journal = "Eur. Phys. J. C",
    volume = "75",
    number = "8",
    pages = "373",
    year = "2015"
}

@article{Altmannshofer:2024hmr,
    author = "Altmannshofer, Wolfgang and Greljo, Admir",
    title = "{Recent Progress in Flavor Model Building}",
    eprint = "2412.04549",
    archivePrefix = "arXiv",
    primaryClass = "hep-ph",
    doi = "10.1146/annurev-nucl-121423-100950",
    journal = "Ann. Rev. Nucl. Part. Sci.",
    volume = "75",
    number = "1",
    pages = "201--322",
    year = "2025"
}

@article{Huang:2020hdv,
    author = "Huang, Guo-yuan and Zhou, Shun",
    title = "{Precise Values of Running Quark and Lepton Masses in the Standard Model}",
    eprint = "2009.04851",
    archivePrefix = "arXiv",
    primaryClass = "hep-ph",
    doi = "10.1103/PhysRevD.103.016010",
    journal = "Phys. Rev. D",
    volume = "103",
    number = "1",
    pages = "016010",
    year = "2021"
}

@article{Greljo:2025mwj,
    author = "Greljo, Admir and Palavri{\'c}, Ajdin and Stefanek, Ben A.",
    title = "{Minimal Flavor Protection for TeV-scale New Physics}",
    eprint = "2512.04159",
    archivePrefix = "arXiv",
    primaryClass = "hep-ph",
    month = "12",
    year = "2025"
}

@article{Banks:2025baf,
    author = "Banks, Hannah and Crawford, Graeme and McCullough, Matthew and Sutherland, Dave",
    title = "{Flavour, Accidentally}",
    eprint = "2510.03403",
    archivePrefix = "arXiv",
    primaryClass = "hep-ph",
    reportNumber = "CERN-TH-2025-157",
    month = "10",
    year = "2025"
}

@article{manivel2007extension,
  title={An extension of the Cayley--Sylvester formula},
  author={Manivel, Laurent},
  journal={European Journal of Combinatorics},
  volume={28},
  number={6},
  pages={1839--1842},
  year={2007},
  publisher={Elsevier}
}

@article{Greljo:2023bix,
    author = "Greljo, Admir and Thomsen, Anders Eller",
    title = "{Rising through the ranks: flavor hierarchies from a gauged SU(2) symmetry}",
    eprint = "2309.11547",
    archivePrefix = "arXiv",
    primaryClass = "hep-ph",
    doi = "10.1140/epjc/s10052-024-12556-5",
    journal = "Eur. Phys. J. C",
    volume = "84",
    number = "2",
    pages = "213",
    year = "2024"
}

@techreport{Michel:121951,
      author        = "Michel, L",
      title         = "{Minima of Higgs-Landau polynomials}",
      institution   = "CERN",
      reportNumber  = "CERN-TH-2716",
      address       = "Geneva",
      year          = "1979",
      url           = "https://cds.cern.ch/record/121951",
}

@article{Kim:1998qy,
    author = "Kim, J. S. and Toledano, J. C. and Toledano, P.",
    title = "{Monte Carlo optimization applied to symmetry breaking}",
    doi = "10.1016/S0010-4655(97)00139-2",
    journal = "Comput. Phys. Commun.",
    volume = "109",
    pages = "207--226",
    year = "1998"
}

@article{Kim:1997wf,
    author = "Kim, Jai Sam and Toledano, J. C. and Toledano, P.",
    title = "{Monte Carlo minimization of the Higgs-Landau potentials}",
    eprint = "cond-mat/9708084",
    archivePrefix = "arXiv",
    month = "8",
    year = "1997"
}

@article{Kim:1983mc,
    author = "Kim, Jai Sam",
    title = "{Orbit Spaces of Low Dimensional Representations of Simple Compact Connected Lie Groups and Extrema of a Group Invariant Scalar Potential}",
    reportNumber = "CALT-68-1017",
    doi = "10.1063/1.526347",
    journal = "J. Math. Phys.",
    volume = "25",
    pages = "1694",
    year = "1984"
}

@article{Abud:1981tf,
    author = "Abud, M. and Sartori, G.",
    title = "{The Geometry of Orbit Space and Natural Minima of Higgs Potentials}",
    reportNumber = "IFPD 45/81",
    doi = "10.1016/0370-2693(81)90578-5",
    journal = "Phys. Lett. B",
    volume = "104",
    pages = "147--152",
    year = "1981"
}

@article{Abud:1983id,
    author = "Abud, M. and Sartori, G.",
    title = "{The Geometry of Spontaneous Symmetry Breaking}",
    reportNumber = "Print-83-0073 (PADUA)",
    doi = "10.1016/0003-4916(83)90017-9",
    journal = "Annals Phys.",
    volume = "150",
    pages = "307",
    year = "1983"
}

@article{Coleman:1973jx,
    author = "Coleman, Sidney R. and Weinberg, Erick J.",
    title = "{Radiative Corrections as the Origin of Spontaneous Symmetry Breaking}",
    doi = "10.1103/PhysRevD.7.1888",
    journal = "Phys. Rev. D",
    volume = "7",
    pages = "1888--1910",
    year = "1973"
}

@article{Martin:2001vx,
    author = "Martin, Stephen P.",
    title = "{Two Loop Effective Potential for a General Renormalizable Theory and Softly Broken Supersymmetry}",
    eprint = "hep-ph/0111209",
    archivePrefix = "arXiv",
    reportNumber = "FERMILAB-PUB-01-348-T",
    doi = "10.1103/PhysRevD.65.116003",
    journal = "Phys. Rev. D",
    volume = "65",
    pages = "116003",
    year = "2002"
}

@article{Greljo:2024zrj,
    author = "Greljo, Admir and Thomsen, Anders Eller and Tiblom, Hector",
    title = "{Flavor hierarchies from SU(2) flavor and quark-lepton unification}",
    eprint = "2406.02687",
    archivePrefix = "arXiv",
    primaryClass = "hep-ph",
    doi = "10.1007/JHEP08(2024)143",
    journal = "JHEP",
    volume = "08",
    pages = "143",
    year = "2024"
}

@article{Barbieri:1995uv,
    author = "Barbieri, Riccardo and Dvali, G. R. and Hall, Lawrence J.",
    title = "{Predictions from a U(2) flavor symmetry in supersymmetric theories}",
    eprint = "hep-ph/9512388",
    archivePrefix = "arXiv",
    reportNumber = "LBL-38065, UCB-PTH-95-44",
    doi = "10.1016/0370-2693(96)00318-8",
    journal = "Phys. Lett. B",
    volume = "377",
    pages = "76--82",
    year = "1996"
}

@article{Barbieri:1996ww,
    author = "Barbieri, Riccardo and Hall, Lawrence J. and Raby, Stuart and Romanino, Andrea",
    title = "{Unified theories with U(2) flavor symmetry}",
    eprint = "hep-ph/9610449",
    archivePrefix = "arXiv",
    reportNumber = "IFUP-TH-61-96, LBL-39488, OHSTPY-HEP-T-96-033, UCB-PTH-96-45",
    doi = "10.1016/S0550-3213(97)00134-X",
    journal = "Nucl. Phys. B",
    volume = "493",
    pages = "3--26",
    year = "1997"
}

@article{Barbieri:1997tu,
    author = "Barbieri, Riccardo and Hall, Lawrence J. and Romanino, Andrea",
    title = "{Consequences of a U(2) flavor symmetry}",
    eprint = "hep-ph/9702315",
    archivePrefix = "arXiv",
    reportNumber = "IFUP-TH-4-97, LBL-39946, LBNL-39946, UCB-PTH-97-06",
    doi = "10.1016/S0370-2693(97)00372-9",
    journal = "Phys. Lett. B",
    volume = "401",
    pages = "47--53",
    year = "1997"
}

@article{Barbieri:2012uh,
    author = "Barbieri, Riccardo and Buttazzo, Dario and Sala, Filippo and Straub, David M.",
    title = "{Flavour physics from an approximate $U(2)^3$ symmetry}",
    eprint = "1203.4218",
    archivePrefix = "arXiv",
    primaryClass = "hep-ph",
    doi = "10.1007/JHEP07(2012)181",
    journal = "JHEP",
    volume = "07",
    pages = "181",
    year = "2012"
}

@article{Greljo:2022cah,
    author = "Greljo, Admir and Palavri{\'c}, Ajdin and Thomsen, Anders Eller",
    title = "{Adding Flavor to the SMEFT}",
    eprint = "2203.09561",
    archivePrefix = "arXiv",
    primaryClass = "hep-ph",
    doi = "10.1007/JHEP10(2022)005",
    journal = "JHEP",
    volume = "10",
    pages = "010",
    year = "2022"
}

@article{Faroughy:2020ina,
    author = "Faroughy, Darius A. and Isidori, Gino and Wilsch, Felix and Yamamoto, Kei",
    title = "{Flavour symmetries in the SMEFT}",
    eprint = "2005.05366",
    archivePrefix = "arXiv",
    primaryClass = "hep-ph",
    doi = "10.1007/JHEP08(2020)166",
    journal = "JHEP",
    volume = "08",
    pages = "166",
    year = "2020"
}

@article{Linster:2018avp,
    author = "Linster, Matthias and Ziegler, Robert",
    title = "{A Realistic $U(2)$ Model of Flavor}",
    eprint = "1805.07341",
    archivePrefix = "arXiv",
    primaryClass = "hep-ph",
    reportNumber = "CERN-TH-2018-121, TTP18-019",
    doi = "10.1007/JHEP08(2018)058",
    journal = "JHEP",
    volume = "08",
    pages = "058",
    year = "2018"
}

@article{Bloch:1946zza,
    author = "Bloch, F.",
    title = "{Nuclear Induction}",
    doi = "10.1103/PhysRev.70.460",
    journal = "Phys. Rev.",
    volume = "70",
    pages = "460--474",
    year = "1946"
}

@article{Feynman:1957zz,
    author = "Feynman, Richard P. and Vernon, Jr., Frank L. and Hellwarth, Robert W.",
    title = "{Geometrical Representation of the Schrodinger Equation for Solving Maser Problems}",
    doi = "10.1063/1.1722572",
    journal = "J. Appl. Phys.",
    volume = "28",
    pages = "49--52",
    year = "1957"
}

@article{Majorana:1932ga,
    author = "Majorana, Ettore",
    title = "{Oriented atoms in a variable magnetic field}",
    doi = "10.1007/BF02960953",
    journal = "Nuovo Cim.",
    volume = "9",
    pages = "43--50",
    year = "1932"
}

@article{RevModPhys.17.237,
  title = {Atoms in Variable Magnetic Fields},
  author = {Bloch, F. and Rabi, I. I.},
  journal = {Rev. Mod. Phys.},
  volume = {17},
  issue = {2-3},
  pages = {237--244},
  numpages = {0},
  year = {1945},
  month = {Apr},
  publisher = {American Physical Society},
  doi = {10.1103/RevModPhys.17.237},
  url = {https://link.aps.org/doi/10.1103/RevModPhys.17.237}
}

@book{bengtsson2017geometry,
  title={Geometry of Quantum States: An Introduction to Quantum Entanglement},
  author={Bengtsson, I. and {\.Z}yczkowski, K.},
  isbn={9781107026254},
  lccn={2016059634},
  url={https://books.google.it/books?id=sYswDwAAQBAJ},
  year={2017},
  publisher={Cambridge University Press}
}

@article{Bacry:1974hc,
    author = "Bacry, H.",
    title = "{Orbits of the rotation group on spin states}",
    doi = "10.1063/1.1666525",
    journal = "J. Math. Phys.",
    volume = "15",
    pages = "1686--1688",
    year = "1974"
}

@article{Espinosa:2012uu,
    author = "Espinosa, Jose R. and Fong, Chee Sheng and Nardi, Enrico",
    title = "{Yukawa hierarchies from spontaneous breaking of the $SU(3)_L\times SU(3)_R$ flavour symmetry?}",
    eprint = "1211.6428",
    archivePrefix = "arXiv",
    primaryClass = "hep-ph",
    doi = "10.1007/JHEP02(2013)137",
    journal = "JHEP",
    volume = "02",
    pages = "137",
    year = "2013"
}

@article{deMedeirosVarzielas:2025byb,
    author = "de Medeiros Varzielas, Ivo and Liu, Ming-Shau and Sengupta, Amartya and Talbert, Jim",
    title = "{Residual Symmetries and Scalar Multiplet Vacuum Alignment in Non-Abelian Flavour Models}",
    eprint = "2512.19789",
    archivePrefix = "arXiv",
    primaryClass = "hep-ph",
    reportNumber = "LA-UR-25-31854",
    month = "12",
    year = "2025"
}

@article{Fonseca:2020vke,
    author = "Fonseca, Renato M.",
    title = "{GroupMath: A Mathematica package for group theory calculations}",
    eprint = "2011.01764",
    archivePrefix = "arXiv",
    primaryClass = "hep-th",
    doi = "10.1016/j.cpc.2021.108085",
    journal = "Comput. Phys. Commun.",
    volume = "267",
    pages = "108085",
    year = "2021"
}

@article{Feruglio:2019ybq,
    author = "Feruglio, Ferruccio and Romanino, Andrea",
    title = "{Lepton flavor symmetries}",
    eprint = "1912.06028",
    archivePrefix = "arXiv",
    primaryClass = "hep-ph",
    doi = "10.1103/RevModPhys.93.015007",
    journal = "Rev. Mod. Phys.",
    volume = "93",
    number = "1",
    pages = "015007",
    year = "2021"
}

@article{Procesi:1985hr,
    author = "Procesi, C. and Schwarz, G. W.",
    title = "{THE GEOMETRY OF ORBIT SPACES AND GAUGE SYMMETRY BREAKING IN SUPERSYMMETRIC GAUGE THEORIES}",
    doi = "10.1016/0370-2693(85)90620-3",
    journal = "Phys. Lett. B",
    volume = "161",
    pages = "117--121",
    year = "1985"
}

@article{Talamini:2006wd,
    author = "Talamini, Vittorino",
    title = "{Affine-P-matrices in orbit spaces and invariant theory}",
    eprint = "hep-th/0607165",
    archivePrefix = "arXiv",
    doi = "10.1088/1742-6596/30/1/005",
    journal = "J. Phys. Conf. Ser.",
    volume = "30",
    pages = "30",
    year = "2006"
}

@incollection{Cabibbo:1970rza,
      author         = "Cabibbo, N. and Maiani, L.",
      title          = "{Weak interactions and the breaking of hadron
                        symmetries}",
      editor         = "Conversi, Marcello",
      booktitle      = "Evolution of particle physics: A volume dedicated to
                        Edoardo Amaldi in his sixtieth birthday",
      pages          = "50-80",
      year           = "1970",
      SLACcitation   = "%%CITATION = INSPIRE-913267;%%"
}

@article{deBlas:2025gyz,
    author = "de Blas, Jorge and others",
    title = "{Physics Briefing Book: Input for the 2026 update of the European Strategy for Particle Physics}",
    eprint = "2511.03883",
    archivePrefix = "arXiv",
    primaryClass = "hep-ex",
    reportNumber = "CERN--2025-008, CERN-ESU-2025-001",
    doi = "10.23731/CYRM-2025-008",
    month = "11",
    year = "2025"
}

@article{Silvestrini:2018dos,
    author = "Silvestrini, Luca and Valli, Mauro",
    title = "{Model-independent Bounds on the Standard Model Effective Theory from Flavour Physics}",
    eprint = "1812.10913",
    archivePrefix = "arXiv",
    primaryClass = "hep-ph",
    reportNumber = "CERN-TH-2018-279, UCI-TR-2018-24",
    doi = "10.1016/j.physletb.2019.135062",
    journal = "Phys. Lett. B",
    volume = "799",
    pages = "135062",
    year = "2019"
}

@article{Grzadkowski:2010es,
    author = "Grzadkowski, B. and Iskrzynski, M. and Misiak, M. and Rosiek, J.",
    title = "{Dimension-Six Terms in the Standard Model Lagrangian}",
    eprint = "1008.4884",
    archivePrefix = "arXiv",
    primaryClass = "hep-ph",
    reportNumber = "IFT-9-2010, TTP10-35",
    doi = "10.1007/JHEP10(2010)085",
    journal = "JHEP",
    volume = "10",
    pages = "085",
    year = "2010"
}

@article{Calibbi:2025rxn,
    author = "Calibbi, Lorenzo and Yi, Jiangyi",
    title = "{Phenomenology of Non-Abelian Gauge and Goldstone Bosons in a U(2) Flavor Model}",
    eprint = "2511.10468",
    archivePrefix = "arXiv",
    primaryClass = "hep-ph",
    month = "11",
    year = "2025"
}

@article{Jetzer:1983ij,
    author = "Jetzer, P. and Gerard, J. M. and Wyler, D.",
    title = "{Possible Symmetry Breaking Patterns With Totally Symmetric and Antisymmetric Representations}",
    reportNumber = "CERN-TH-3697",
    doi = "10.1016/0550-3213(84)90206-2",
    journal = "Nucl. Phys. B",
    volume = "241",
    pages = "204--220",
    year = "1984"
}

@article{Hisano:2013sn,
    author = "Hisano, Junji and Tsumura, Koji",
    title = "{Higgs boson mixes with an SU(2) septet representation}",
    eprint = "1301.6455",
    archivePrefix = "arXiv",
    primaryClass = "hep-ph",
    reportNumber = "IPMU13-0022",
    doi = "10.1103/PhysRevD.87.053004",
    journal = "Phys. Rev. D",
    volume = "87",
    pages = "053004",
    year = "2013"
}

@article{Heikinheimo:2017nth,
    author = {Heikinheimo, Matti and Kannike, Kristjan and Lyonnet, Florian and Raidal, Martti and Tuominen, Kimmo and Veerm{\"a}e, Hardi},
    title = "{Vacuum Stability and Perturbativity of SU(3) Scalars}",
    eprint = "1707.08980",
    archivePrefix = "arXiv",
    primaryClass = "hep-ph",
    reportNumber = "HIP-2017-19-TH",
    doi = "10.1007/JHEP10(2017)014",
    journal = "JHEP",
    volume = "10",
    pages = "014",
    year = "2017"
}

@article{Moultaka:2020dmb,
    author = "Moultaka, Gilbert and Peyran{\`e}re, Michel C.",
    title = "{Vacuum stability conditions for Higgs potentials with $SU(2)_L$ triplets}",
    eprint = "2012.13947",
    archivePrefix = "arXiv",
    primaryClass = "hep-ph",
    doi = "10.1103/PhysRevD.103.115006",
    journal = "Phys. Rev. D",
    volume = "103",
    number = "11",
    pages = "115006",
    year = "2021"
}

@article{Brummer:2023znr,
    author = {Br{\"u}mmer, Felix and Ferrante, Giacomo and Frigerio, Michele and Hambye, Thomas},
    title = "{Accidentally light scalars from large representations}",
    eprint = "2307.10092",
    archivePrefix = "arXiv",
    primaryClass = "hep-ph",
    reportNumber = "CERN-TH-2023-247",
    doi = "10.1007/JHEP01(2024)075",
    journal = "JHEP",
    volume = "01",
    pages = "075",
    year = "2024"
}

@article{Jurciukonis:2024bzx,
    author = "Jur{\v{c}}iukonis, Darius and Lavoura, Lu{\'\i}s",
    title = "{On the extension of the SM through a scalar quadruplet}",
    eprint = "2406.01628",
    archivePrefix = "arXiv",
    primaryClass = "hep-ph",
    month = "6",
    year = "2024"
}

@article{Milagre:2025vcx,
    author = "Milagre, Andr{\'e} and Jur{\v{c}}iukonis, Darius and Lavoura, Lu{\'\i}s",
    title = "{Vacuum Stability Conditions for New $SU(2)$ Multiplets}",
    eprint = "2505.05272",
    archivePrefix = "arXiv",
    primaryClass = "hep-ph",
    doi = "10.1093/ptep/ptaf105",
    journal = "PTEP",
    volume = "2025",
    pages = "093",
    month = "5",
    year = "2025"
}

@article{Jurciukonis:2025cyy,
    author = "Jur{\v{c}}iukonis, Darius and Lavoura, Lu{\'\i}s",
    title = "{On the addition of an $SU(2)$ quadruplet of scalars to the Standard Model}",
    eprint = "2507.17870",
    archivePrefix = "arXiv",
    primaryClass = "hep-ph",
    doi = "10.21468/SciPostPhysCore.9.2.022",
    journal = "SciPost Phys. Core",
    volume = "9",
    pages = "022",
    year = "2026"
}

@article{Esteban:2024eli,
    author = "Esteban, Ivan and Gonzalez-Garcia, M. C. and Maltoni, Michele and Martinez-Soler, Ivan and Pinheiro, Jo{\~a}o Paulo and Schwetz, Thomas",
    title = "{NuFit-6.0: updated global analysis of three-flavor neutrino oscillations}",
    eprint = "2410.05380",
    archivePrefix = "arXiv",
    primaryClass = "hep-ph",
    reportNumber = "IFT-UAM/CSIC-24-140, YITP-SB-2024-24, IPPP/24/64, IPPP/24/64, IFT-UAM/CSIC-24-140, YITP-SB-2024-24",
    doi = "10.1007/JHEP12(2024)216",
    journal = "JHEP",
    volume = "12",
    pages = "216",
    year = "2024"
}

@article{Darme:2023nsy,
    author = "Darm{\'e}, Luc and Deandrea, Aldo and Mahmoudi, Farvah",
    title = "{Gauge SU(2)$_{f}$ flavour transfers}",
    eprint = "2307.09595",
    archivePrefix = "arXiv",
    primaryClass = "hep-ph",
    reportNumber = "CERN-TH-2023-139",
    doi = "10.1007/JHEP05(2024)313",
    journal = "JHEP",
    volume = "05",
    pages = "313",
    year = "2024"
}

@article{Choi:1992xp,
    author = "Choi, Ki-woon and Kaplan, David B. and Nelson, Ann E.",
    title = "{Is CP a gauge symmetry?}",
    eprint = "hep-ph/9205202",
    archivePrefix = "arXiv",
    reportNumber = "UCSD-PTH-92-11",
    doi = "10.1016/0550-3213(93)90082-Z",
    journal = "Nucl. Phys. B",
    volume = "391",
    pages = "515--530",
    year = "1993"
}

@article{Dine:1992ya,
    author = "Dine, Michael and Leigh, Robert G. and MacIntire, Douglas A.",
    title = "{Of CP and other gauge symmetries in string theory}",
    eprint = "hep-th/9205011",
    archivePrefix = "arXiv",
    reportNumber = "SCIPP-92-16",
    doi = "10.1103/PhysRevLett.69.2030",
    journal = "Phys. Rev. Lett.",
    volume = "69",
    pages = "2030--2032",
    year = "1992"
}

@article{Michel:1971th,
    author = "Michel, L. and Radicati, L. A.",
    title = "{Properties of the breaking of hadronic internal symmetry}",
    doi = "10.1016/0003-4916(71)90079-0",
    journal = "Annals Phys.",
    volume = "66",
    pages = "758--783",
    year = "1971"
}

\end{document}